\documentclass[aps,pre,onecolumn,showpacs]{revtex4}
\expandafter\let\csname equation*\endcsname\relax 
\expandafter\let\csname endequation*\endcsname\relax 
\usepackage{amsmath}
\usepackage{amssymb, cancel}
\usepackage{graphicx}% Include  files
\usepackage{bm}% bold math
\usepackage{epstopdf}
\usepackage{color}
\usepackage[colorlinks=true, pdfstartview=FitV, linkcolor=blue, citecolor=blue, urlcolor=blue]{hyperref}
\definecolor{shadecolor}{rgb}{0.95, 0.95, 0.86}

\def\blue#1{\textcolor[rgb]{0,0,1}{#1}}

\newcommand{\eps}{\varepsilon}

\newcommand{\sech}{\hbox{sech}}
\newcommand{\bos}{\boldsymbol}

\def\ra{\rightarrow}
\newcommand{\bt}{\beta}

\renewcommand{\k}{\varkappa}

\renewcommand{\o}{\omega}
\newcommand{\g}{\gamma}
\newcommand{\la}{\lambda}

\def\bt{\begin{theorem}}
\def\et{\end{theorem}}
\def\bc{\begin{corollary}}
\def\ec{\end{corollary}}
\def\bx{\begin{example}\small}
\def\ex{\end{example}}
\def\bxr{\begin{exercise}\small}
\def\exr{\end{exercise}}
\def\bl{\begin{lemma}}
\def\el{\end{lemma}}
\def\bd{\begin{definition}}
\def\ed{\end{definition}}
\def\bp{\begin{proposition}}
\def\ep{\end{proposition}}

\def\br{\begin{remark}}
\def\er{\end{remark}}

\def\be{\begin{equation}}
\def\ee{\end{equation}}
\def\&{\hspace{-15pt}&}
\def\bea{\begin{eqnarray}}
\def\eea{\end{eqnarray}}

\def \part{\partial}

\def\R{{\mathbb R}}
\def\N{{\mathbb N}}

\def\a{\alpha}
\def\b{\beta}
\def\g{\gamma}
\def\gt{\hat\gamma}

\def\l{\lambda}

\def\1{{\bf 1}}

\def\th{ {\theta}}

\def\z{\zeta}

\def\ri{\right}
\def \ggg{{\mathfrak g}}

\newcommand{\Rscr}{\mathcal R}

\begin{document}

\title{Dam break problem for the focusing nonlinear Schr\"odinger equation  and  the generation of rogue waves }

\author{G.A. El$^{1}$}
\author{E.G. Khamis$^{2,3}$}
\author{A. Tovbis$^{4}$}

\address{
$^{1}$ Department of Mathematical Sciences, Loughborough University, Loughborough LE11 3TU, UK \\
$^{2}$ Instituto de F\'isica, Universidade de S\~ao Paulo, 05508-090 S\~ao Paulo, Brazil \\
$^{3}$ Center for Weather Forecasting and Climate Studies-CPTEC, National Institute for
       Space Research (INPE), Cachoeira Paulista, S\~ao Paulo, Brazil \\
 $^{4}$ Department of Mathematics, University of Central Florida, USA  }

\date{\today}

\
\begin{abstract}
We propose a novel,  analytically tractable, scenario of the rogue wave formation in the framework of the small-dispersion focusing nonlinear Schr\"odinger (NLS) equation with the initial condition in the form of  a rectangular barrier (a ``box''). We use the Whitham modulation theory combined with the nonlinear steepest descent for the  semi-classical inverse scattering transform, to describe the evolution and interaction of two counter-propagating  nonlinear wave trains --- the dispersive dam break flows  --- generated in the NLS box problem.  We show that the interaction dynamics results in the emergence of modulated large-amplitude quasi-periodic breather lattices whose amplitude profiles are closely approximated by the Akhmediev and Peregrine breathers within certain space-time domain. Our semi-classical analytical results are shown to be in excellent agreement   with the results of direct numerical simulations of  the small-dispersion focusing NLS equation.
\end{abstract}

\maketitle

\section{Introduction}

There has been much interest over the last two decades in the modelling rogue wave formation in the framework of the one-dimensional focusing nonlinear Schr\"odinger   (NLS) equation,
 \begin{equation}\label{nls1}
i  \eps \psi_{t} +  \frac12 \eps^2 \psi_{xx} + |\psi|^2 \psi = 0 \, ,
\end{equation}
where $\psi$ is the complex wave envelope and $\eps$ is a free parameter defining the modulation scale;  all variables are dimensionless. Rogue waves represent the waves of unusually large amplitude $|\psi|_m$, whose appearance statistics deviates from the Gaussian distribution.  The conventional ``rogue wave criterion'' is  
$|\psi|_m/|\psi|_s > 2$, where $|\psi|_s$ is the significant wave height computed as the average wave height of the largest 1/3 of waves  (see e.g. \cite{kharif_book}, \cite{onorato_etal2013}, \cite{agaf_zakh2014}).   For the random wave field with Gaussian statistics one has $|\psi|_s^2= 2 |\psi_0|^2$, where $\psi_0$ is the background (mean field) amplitude, leading to the  criterion $|\psi|_m^2  >8 |\psi_0|^2$.

As is clear from the above discussion, the description of rogue waves is a two-sided problem: it is concerned  both with the dynamics of their formation and evolution, and with the statistics of their occurrence. In this paper we shall be concerned only with certain dynamical aspects of the rogue wave generation.
The NLS equation (\ref{nls1}) has a number of relatively simple exact solutions which  are widely considered as prototypes of rogue waves, the principal representatives being the Akhmediev, the Kuznetsov-Ma and the Peregrine breathers (see, e.g., \cite{dysthe99}). The main physical contexts for rogue waves are oceanography and nonlinear optics (see \cite{kharif_book}, \cite{osborne_book}, \cite{rogue_nat2007} and references therein).  

Finding controllable ways to excite rogue waves  has been one of the central topics of the ``deterministic'' rogue wave research (see e.g. \cite{ akhmediev_etal2009} and references therein). Various mechanisms have been proposed in the framework of the NLS equation and its generalisations (see e.g. \cite{onorato_etal2013}, \cite{turitsyn2013},  \cite{dudley_nat2014} and references therein). Many of them relate the rogue wave appearance to the development of modulational instability of the plane wave  due to small perturbations (see, e.g., \cite{shrira}, \cite{agaf_zakh2014}, \cite{zakharov_gelash2014}) or  large-scale initial modulations \cite{grim_tovbis}.  Other proposed mechanisms  involve nonlinear wave interactions: e.g., the interactions of individual solitons \cite{dudley2010} or the interaction of solitons with the plane wave \cite{zakharov_gelash2013}. 
One should note, however,  that, while there have been many papers developing the methods for finding particular rogue wave solutions (Darboux transformation, $\bar \partial$-method and oth. --- see, e.g., \cite{akhmediev_PRE09}, \cite{zakharov_gelash2013} and references therein), an analytical description of their formation from reasonably generic inital data remains a challenging, and to a large degree unsolved problem. In most cases numerical simulations remain the only available resort.

One can distinguish two contrasting general classes of initial-boundary value problems associated with equation (\ref{nls1}). The first one is concerned with the evolution of rapidly decaying potentials. The result of such an evolution generically represents a combination of fundamental solitons and some dispersive radiation with no rogue waves at the output.  A comprehensive description of this process is achieved in the framework of  the Inverse Scattering Transform (IST)  \cite{ZS}.
The second class of problems deals with the NLS equation with non-decaying boundary conditions,  and is much less explored analytically.  One of the most interesting and physically relevant problems of this kind  is  the  evolution of various perturbations of the plane wave  (the ``condensate''),
\begin{equation}\label{pw}
\psi= q e^{iq^2 t/\eps},
\end{equation}
where the amplitude $q > 0$. Small harmonic perturbations $\sim e^{i(kx-\omega t)}$ of (\ref{pw}) satisfy  the dispersion relation $\omega(k) = \pm \frac12 k[(\eps k)^2 - 4q^2]^{1/2}$, which implies modulational instability for  sufficiently long waves with the wavenumbers $k < k_c =2q/\eps$.  The  description of the nonlinear stage of the development of modulation instability has been the subject of many papers including some very recent developments  \cite{zakharov_gelash2014}, \cite{agaf_zakh2014}, \cite{biondini_ist_2015}, \cite{biondini_prl_2016}.  It has been proposed in \cite{agaf_zakh2014} that the  long-time evolution of this process, in the case of random initial conditions, results in the establishment  of a  complex, globally incoherent strongly nonlinear wave state which can be associated with ``integrable turbulence'', the notion introduced by Zakharov in \cite{zakharov2009}. It has also been shown that the rogue wave formation plays an important role in the characterisation of the  early stage of the integrable turbulence development from a randomly  perturbed plane wave \cite{agaf_zakh2014},
but also  has a  noticeable effect on the power spectrum of the established  integrable turbulence developing from  random initial conditions which are not small perturbations of a plane wave \cite{suret2014}.

In this paper, we consider the problem that combines some of the key features of the both above contrasting fundamental mathematical and physical  set-ups (NLS with decaying vs. non-decaying boundary conditions). 
Specifically, we consider the evolution of a {\it large-scale} (compared to the medium's coherence length) decaying data in the form of a rectangular barrier  (a `box' ) of finite height $q>0$  and the length $2L$: 
\begin{equation}\label{ic0}
\psi(x,0) =\left\{
\begin{array}{ll}
q  &\quad \hbox{for} \quad |x| <L,\\
 0& \quad \hbox{for} \quad |x|>L \, .
\end{array}
\right.
\end{equation}
In the dimensionless variables of (\ref{nls1}) we assume that $L=O(1)$, and the dispersion parameter $\eps \ll 1$. We shall refer to the problem (\ref{nls1}), (\ref{ic0}) as the small-dispersion NLS box problem.  It is expected that the evolution (\ref{nls1}), (\ref{ic0}) at times $t \ll \eps^{-1}$ will model some features of the nonlinear stage of the development of modulational instability, in particular, the formation of rogue waves. 

The initial evolution of the box data for the small-dispersion focusing NLS equation  can be viewed as a combination of two ``dam break'' problems, which are known to  lead to the instantaneous formation  {\it single-phase} nonlinear modulated wave trains regularising the discontinuities at the opposite edges of the initial profile \cite{kamch97},  \cite{kenjen14}. These wave trains have the structure similar to  dispersive shock waves (DSWs) \cite{dsw_review}, or undular bores,  extensively studied in the defocusing (hyperbolic) NLS theory \cite{GK87, eggk95, kodama99, ha06}. There are, however, important differences, which we emphasise by using the term  {\it dispersive dam break flow} rather than DSW.   The reason for using this term in the context of the focusing NLS equation is that the regularisation of  discontinuous data via a single-phase modulated wave train in the focusing NLS occurs {\it only} if the upstream constant state is the vacuum, which the key feature of the classical shallow-water  dam break problem \cite{whitham_book}.  However, the shallow-water ``dry bottom'' dam break problem does not involve the formation of a shock \cite{whitham_book} so  the regularising dispersive wave trains in the  NLS box  problem do not have classical shock counterparts in the dispersionless hyperbolic case.  In contrast, the ``genuine'' focusing DSW analog is expected to have  a multi-phase structure \cite{bikbaev}.

The dispersive dam break flows regularising the box data (\ref{ic0}) expand inside the interval $-L <x<L$ and, after a certain moment of time, $t_0=L/(2\sqrt{2}q)$, start to interact resulting in the formation of a region  filled with 
a two-phase, $x$-$t$ quasi-periodic  wave, which we term the {\it breather lattice} due to the characteristic shape of the individual oscillations resembling standard breather solutions of the NLS equation.
Indeed, we show  that  the wave form (the amplitude profile) of the oscillations in this lattice at each given $t$ is quite well approximated by that of the Akhmediev breather with the spatial period depending on the value of $t$. Towards the end of the interaction region the spatial and temporal periods of the breather lattice increase so that locally, the oscillations  are closely approximated by the  Peregrine solitons with the amplitude $3q$.  Our numerical simulations show that further in time, the evolution leads to the generation of multiphase regions in the $x$-$t$ plane with the number of oscillatory phases  $g$ (the ``genus'' of the solution) at any particular point $-L <x<L$ growing with time, the expected asymptotic behaviour being $g \sim t $ for $t \gg 1$.   Assuming the existence of the long-time asymptotics for the solution $\psi(x,t, \eps)$ one can associate it with the `breather gas' (see \cite{ekmv01, el03, ek05, ekpz2011} for the description of the counterpart soliton gas in the KdV theory). One of the numerically observed features of the regions with $g \ge 4$ is the presence of the higher-order rogue  waves with the maximum wave height $4q <|\psi|_m  < 5q$.

We now outline the analytical approach adopted in this paper. 
Although the NLS box problem  admits  exact  analytical description via the IST \cite{ZS}, such a description  becomes not feasible in practical terms when $\eps  \ll 1$ as the number of (global) degrees of freedom in the IST solution is then $ O (\eps^{-1}) \gg 1$.  In that case, the natural analytical framework is  the {\it semi-classical approximation}, which enables one to asymptotically reduce the complicated exact IST construction to a more manageable description of a  modulated  multi-phase  NLS solution approximating the exact solution and involving just a few (local) degrees of freedom. Unlike in the exact solution, the IST spectrum of the approximate, semi-classical solution consists of finite number of bands which slowly evolve in space-time.  There are two complementary mathematical approaches to the construction of such slowly modulated multiple-scale  solutions to integrable nonlinear dispersive equations. The first one is based on the Whitham averaging procedure \cite{whitham65, whitham_book} leading to a system of quasilinear equations governing \blue{the} slow evolution of the endpoints of spectral bands \cite{ffm, flee86, pavlov87}. For the focusing NLS, the Whitham modulation system is elliptic, implying modulational instability of the underlying nolinear periodic wave  \cite{flee86}.
(One should be clear that, for the NLS equation, which is a modulation equation itself, the Whitham equations describe the `super-modulations', i.e.  the modulations of the nonlinear periodic or quasi-periodic solutions. See \cite{newell}, \cite{el_dispersive_2016} for the discussion of the relation between the NLS equation and the Whitham theory.)
 
While the type (hyperbolic vs. elliptic) of the Whitham system yields the essential information about stability/instability of the underlying periodic wave \cite{whitham_book},  the  {\it modulation solution}  provides the detailed information about the evolution of \blue{a} slowly modulated wave train.  Vast majority of papers on the integration of the Whitham equations deal with the hyperbolic case, most notably in the DSW theory  (see \cite{dsw_review} and references therein).  In contrast, the solutions of  elliptic Whitham equations are far less explored, especially in the context of applications. The existing applied results are restricted to the simplest self-similar solutions in the single-phase case (see e.g. \cite{egkk93},  \cite{bikbaev_sharapov94}, \cite{ekk96}, \cite{kamch97}, \cite{abdullaev}). 
 
The second approach to the semi-classical NLS equation is the  nonlinear steepest descent method by Deift and Zhou \cite{DZ1} involving the  Riemann-Hilbert problem (RHP) formulation of the  IST.  
As a matter of fact, the two approaches are consistent, the modulation {\it solution}  directly arising as part of the more general 
and mathematically rigorous (but also more technically involved) RHP analysis. 
In this paper, we use an appropriate combination of the two methods to construct a relatively simple analytic solution describing the $x$-$t$ evolution of the approximate, semi-classical, IST specrum in the NLS box problem in the region of interaction of two dispersive dam break flows.  This solution describes slow modulations of the two-phase breather lattice and   enables us to predict the formation of rogue waves. 
 We note that rigorous RHP analysis of the initial stage of the box evolution  involving genus zero and genus one solutions was done  in  the recent work \cite{kenjen14}.  Part of the results obtained in \cite{kenjen14} appear in the earlier papers \cite{egkk93}, \cite{kamch97}, where they were derived via the Whitham modulation theory. 

The semi-classical focusing NLS  has only started to be explored as an analytical platform for the rogue wave research.  We mention two recent papers \cite{grim_tovbis}  and    
\cite{hoefer_tovbis}  demonstrating the relevance of the semi-classical NLS scaling to the deep water ocean dynamics and the experimentally achievable configurations of Bose-Einstein condensates (BECs). Both  above-mentioned works use the rigorous  (RHP) mathematical results of  \cite{BT1}  establishing the role of the Peregrine solitons in the regularisation of the focussing gradient catastrophe  during the semi-classical NLS evolution of 
a certain family of analytic initial data that includes $\psi(x,0)=\sech(x)$.   Our paper continues this emerging line of research by  putting forward and studying a novel scenario of the rogue wave formation via the interaction of two modulationally stable nonlinear wavetrains. Integrability of the NLS equation (\ref{nls1}) and the semi-classical approximation play the key roles in the mathematical description of the proposed scenario.  The proposed mechanism of the rogue wave formation can be realised in fibre optics experiments. Our analytical results are favourably compared with direct numerical simulations of the small-dispersion NLS box problem.  In this regard we note that, although the semi-classical focusing NLS has been the subject of many numerical investigations (see e.g. \cite{bronski_kutz1999, cai2002, ceniceros_tian2002, Clarke-Miller} and more recent papers \cite{lyng_vankova2012, lee_lyng2013} and references therein), we are  aware only of a few examples, such as  \cite{LM}, \cite{BT1}   of the previous work undertaking quantitative comparison of the semi-classical analytical solutions for $g \ge 1$ with the direct numerical simulations of the NLS. 

The paper is organised as follows. In Section 2 we present an account of the necessary results from finite-gap theory of the focusing NLS equation and the associated Whitham modulation theory.
Along with the well known results, this section contains a new general compact representation (\ref{Vjexp}) for the characteristic speeds of the multiphased NLS-Whitham modulation equations. In Section 3 the modulation solution of the dam break problem is constructed following earlier analytical results of \cite{egkk93}, \cite{kamch97}, and then compared with numerical simulations. We  show that, despite generic modulational instability in the system, this solution has the enhanced stability properties due to vanishing of the imaginary parts of the nonlinear characteristic speeds. Section 4 is central and 
is devoted to the analysis of the dispersive dam break flow interaction in the semi-classical NLS box problem.  Using a combination of the Whitham modulation theory, the generalised hodograph transform \cite{tsarev85} and elements of the RHP techniques we construct and analyse the modulation solution describing the interaction region, and then use it for the prediction of the rogue wave appearance. 
In Section 5 we numerically consider  effects of small perturbations on the qualitative structure of the small-dispersion NLS box problem solution. We consider both perturbation of the initial conditions and perturnbations of the equation itself.  In Section 6 we draw conclusions from our study and outline directions of future research.
Appendix A contains an outline of  the RHP calculations used in the construction of the modulation solution, and Appendix B is a brief description of the numerical method used for the solution of the NLS equation with small dispersion parameter. 

\section{Multi-phase solutions, rogue waves and modulation equations}

As is widely appreciated (see e.g. \cite{tracy_chen}, \cite{osborne_book}), one of the natural mathematical frameworks for the description of the development of modulational instability is the finite-gap theory \cite{belokolos_etal1994} which is a counterpart of the IST for the NLS periodic problem \cite{abma}.  The general finite-gap (or, better, finite-band) solution of (\ref{nls1})  is given by
\begin{equation}
\label{fingap}
\psi_g=q\frac{\Theta_g(x/\eps,  t/\eps;  \boldsymbol{\nu}^0_-)}{\Theta_g(x/\eps, t/\eps;\boldsymbol{\nu}_+^0)} e^{{iq^2 t/\eps}},
\end{equation}
where $\Theta_g$ is the Riemann theta-function associated with the hyperelliptic Riemann surface $\Gamma_g$
of genus $g$ given by
\begin{equation}
\label{rsurf}
\Rscr_g (\la; \bos{\a}, \bos{\bar \a})=  \sqrt{(\la - \a_0)(\la - \bar \a_0) \dots (\la - \a_{g})(\la - \bar \a_{g})} ,
\end{equation}
where $\la$ is the complex spectral parameter. The branch points $\bos{\a}=(\a_0, \dots, \a_{g})$ and  c.c. are the points of simple spectrum of the periodic  non-self-adjoint Zakharov-Shabat scattering operator (see \cite{belokolos_etal1994}, \cite{tracy_chen}, \cite{osborne_book}).
The phases $\boldsymbol{\nu}^0_{\pm} \in \mathbb{R}^{g}$ in (\ref{fingap}) are defined by the initial conditions. The plane wave solution (\ref{pw}) corresponds to the zero-genus spectral surface specified by Eq. (\ref{rsurf}) with $\a_0=iq$, i.e. $\Rscr_0 (\la; \bos{\a}, \bos{\bar \a}) =  \sqrt{(\la - iq)(\la + i q)}$. Thus the spectral portrait of the plane wave is a vertical branch cut between the simple spectrum points $\a_0=iq$ and $\bar \a_0=-iq$. (Note that, by associating a spectral portrait with a particular branchcut configuration we do not imply that the actual spectral bands (the {\it spines} in the terminology of \cite{tracy_chen}, \cite{osborne_book})  are necessarily located exactly along the branchcuts.)
%Thus, in terms of the spectrum characterisation 
%the branchcuts chosen in such a way are analogous to  {\it spines} introduced in \cite{tracy_chen} (see also \cite{osborne_book})}.  

For $g \ge 1$  the theta-solution (\ref{fingap}) is a quasi-periodic function depending on $g$ nontrivial oscillatory phases $\eps^{-1 }\eta_j(x,t)$, so that $\psi_g( \dots \eps^{-1}\eta_j + 2 \pi, \dots) = \psi_g(\dots \eps^{-1}\eta_j, \dots)$ for all $j=1, \dots, g$. The phases  are given by  $\eta_j =k_j x + \omega_j t + \eta_j^0$, $j=1,\dots, g$. Here the (normalised by $\eps$) wavenumbers $ k_j$ and the frequencies $ \omega_j$  are defined in terms of the branch points $\alpha_j$;  and $\eta_j^0$ are arbitrary initial phases.   Also, associated with the `carrier' plane wave is an extra, trivial phase $\eta_0=q^2t$.

We present here the expressions for the wavenumbers $k_j$  and the frequencies $\omega_j$ \cite{flee86}, \cite{tracy_chen},  \cite{osborne_book} which will be needed in what follows,
\begin{equation}
\label{kj}
k_j=-4\pi i  \k_{j, 1} \, , \qquad \omega_j= -4\pi i  \left[  \tfrac12  \sum \limits_{k=0}^{g} (\a_k + \bar \a_k ) \  \k_{j,1} + \k_{j, 2} \right], \quad j=1, \dots, g,
\end{equation}
where $\k_{j,k}(\bos{\a}, \bos{\bar \a})$ are found from the system
\begin{equation}
\label{holonorm}
\sum \limits_{i=1}^{g}\k_{j,i} \oint \limits_{\gt_k} \frac{\z^{g-i}}{\Rscr_g (\z; \bos{\a}, \bos{\bar \a})} d\z = \delta_{jk} \, , \quad j,k = 1, \dots, g.
\end{equation}
Here $\delta_{ik}$ is the Kronkecker symbol and  $\gt_k$ is a  clockwise oriented loop around the  branchcut connecting $\bar \a_k$ and $\a_k$.  Explicit expressions for $\k_{i,j}$ for the case $g=2$ can be found in Appendix A (see (\ref{coeffs})).

For $g=1$  the solution (\ref{fingap}) is periodic and can be expressed in terms of the Jacobi elliptic functions. We introduce  the notation $\alpha_j=a_j+ib_j$. Then for the intensity we have (see e.g. \cite{kamch97})
\begin{equation}
\label{cnoidal}
|\psi|^2 = (b_0+b_1)^2 - 4b_0b_1 \, \hbox{sn}^2 \left(\sqrt{(a_0-a_1)^2+(b_0+b_1)^2} \, (x-Ut + \xi_0)\eps^{-1}; m \right),
\end{equation}
where the phase velocity $U$ and the modulus $0 \le m \le 1$ are given by
\begin{equation}
\label{Um}
U= \tfrac12 (\a_0 +\bar \a_0 + \a_1 + \bar \a_1)= a_0+a_1, \quad m=\frac{( \alpha_0 - \bar \alpha_0)( \alpha_1 - \bar \alpha_1)}{(\alpha_1 - \bar \alpha_0)(\alpha_0 - \bar \alpha_1)}=\frac{4b_0 b_1}{(a_0 - a_1)^2+(b_0+ b_1)^2}\, ,
\end{equation}
and $\xi_0$ is an arbitrary initial phase.
For the wavenumber  $k $  of the cnoidal wave (\ref{cnoidal}) we have
\begin{equation}
\label{k1}
 k= \frac{\pi  \sqrt{(\alpha_1 - \bar \alpha_0)(\alpha_0 - \bar \alpha_1)}}{K(m)} = \frac{\pi }{K(m)} \sqrt{(a_0-a_1)^2+(b_0+b_1)^2}\, ,
\end{equation}
where $K(m)$ is the complete elliptic integral of the first kind. Note that (\ref{k1}) can be obtained from the general representation (\ref{kj}), (\ref{holonorm}) by setting $g=1$. The wave frequency $\omega=kU$  then follows from the second expression (\ref{kj}), where  $\k_{1,1} \equiv 0$.

In the harmonic limit, $m=0$, the spectral portrait  of the solution is a double point (either $\a_1=\bar \a_1$ or $\a_0=\bar \a_0$) on the real axis -- this corresponds to a {\it stable} linear modulation of the plane wave (\ref{pw}); while the soliton limit, $m=1$,  coresponds to two complex conjugate double points:  $\a_1 = \a_0$ and  c.c. on the imaginary axis.  This limit of (\ref{fingap}) describes a  fundamental soliton riding (or resting) on a zero background.

The `classical' prototypes of rogue waves (Akhmediev,  Kuznetsov-Ma and Peregrine breathers) represent unsteady solitary wave modes on the finite background $\psi=q$ and are described  by special limits of the genus two (two-phase)  solution (Eq. (\ref{fingap}) with $g=2$) \cite{osborne_book}.
The Akhmediev breather solution of (\ref{nls1})  has the form (see e.g. \cite{dysthe99})
 \begin{equation}
\label{AB}
\psi_A=qe^{iq^2t/\eps} \frac{\cosh(\Omega t \eps^{-1}- 2i\phi) - \cos\phi \cos(px\eps^{-1})}{\cosh(\Omega t \eps^{-1}) - \cos \phi \cos(px\eps^{-1})} ,
\end{equation}
where
\begin{equation}
\label{pphi}
p=2\sin \phi, \quad \hbox{and} \quad \Omega=2\sin(2\phi),
\end{equation}
for $\phi$ real. Solution (\ref{AB}) is periodic in space with the period $P=2\pi \eps/p$, and tends to the plane wave solution (\ref{pw}) in the limits $t \to \pm \infty$.
The largest modulation occurs at $t=0$ with the maximum envelope at $x=0$,
\begin{equation}
\label{ ABamp}
|\psi_A|= 1+ 2 \cos \phi.
\end{equation}

The   Peregrine breather can be obtained from the Akhmediev breather by letting $p \to 0$. It is described by the rational solution of (\ref{nls1})
\begin{equation}
\label{per}
\psi_P= qe^{iq^2t/\eps} \left[ 1- \frac{4(1+2it\eps^{-1})}{1 +  4\eps^{-2}(x^2 + t^2)} \right],
\end{equation}
which is localised both in space and time around $x=0, t=0$. It can also be obtained directly from the finite-band solution (\ref{fingap}) with $g=2$ by setting $\a_3=\alpha_2=\alpha_1=iq$ (and the c.c. expressions), i.e the spectral portrait of the Peregrine breather consists of two complex conjugate double points placed at the endpoints $\pm iq$ of the basic branchcut. The wave (\ref{per}) has the maximum height  $|\psi_P|_{\max} = 3q$ and represents a homoclinic solution starting from the plane wave (\ref{pw}) at $t \to -\infty$ and returning to the same state at $t \to + \infty$  (see \cite{shrira} for  the discussion of the special role of the Peregrine breather in the theory of rogue waves).

The Madelung transformation
\begin{equation}\label{madelung}
\psi=\sqrt{\rho}e ^{i \frac{\phi}{\eps}}, \qquad \phi_x=u
\end{equation}
maps the NLS equation (\ref{nls1}) to the dispersive hydrodynamics-like system with the negative classical pressure $p=-\rho^2/2$,
\begin{equation}\label{dh}
\begin{split}
    \rho_t+(\rho u)_x=0,\\
 u_t+uu_x - \rho_x - \eps^2\left(\frac{\rho_x^2}{8\rho^2}
   -\frac{\rho_{xx}}{4\rho}\right)_x=0 \, .
   \end{split}
\end{equation}
Here  $\rho \ge 0$ and $u$  are the density and velocity respectively, of the ``fluid''. 

The prominent feature of the small-dispersion NLS evolution for decaying potentials is that, although it is globally described by the IST solution with a very large ($ \sim \eps^{-1} \gg 1$) number of degrees of freedom, the semi-classical asymptotics for $t=O(1)$  is locally (i.e.  for $\Delta x, \Delta t \sim \eps$) described by the  finite-band formula (\ref{fingap}) with {\it just a few degrees of freedom}, but with slowly varying parameters.  Specifically,  the  approximate solution (\ref{fingap})  of the semi-classical NLS equation (\ref{nls1})  is associated  with the  Riemann surface (\ref{rsurf}) of genus  $g=O(1)$,  whose slow (i.e. occurring on the scale much larger than $\eps$) spatiotemporal deformations  are governed by the Whitham modulation equations for the  branch points $\a_j(x,t)$:
\be \label{whitham-g}
(\a_j)_t  = V_j^{(g)}({\bos{\a}},\bos{\bar\a})(\a_j)_x, \quad (\bar \a_j)_t  = \overline V_j^{(g)}(\bos{\a},\bos{\bar \a})(\bar \a_j)_x, \qquad j=0,\dots, g \, .
\ee

Remarkably, $\a_j$ are {\it Riemann invariants} of the Whitham modulation system (\ref{whitham-g}), whose characteristic speeds  $V_j^{(g)}(\bos{\a}, \bos{\bar \a})$  can be expressed in terms of  Abelian differentials \cite{flee86}.  Another compact and physically insightful representations for $V_j$'s as nonlinear group velocities is (see \cite{el96}, \cite{ekv2001})
\begin{equation}
\label{Vjg}
V_j^{(g)} =  \frac{\partial \omega_i}{\partial \a_j} / \frac{\partial k_i}{\partial \a_j} ,    \qquad \hbox{for any} \quad i=1, \dots, g. 
\end{equation}
Formula (\ref{Vjg}) follows from the consideration of   the system  of $g$ wave conservation equations 
\begin{equation}
\label{wncl}
\frac{\partial}{\partial t} k_j (\bos{\a}, \bos{ \bar \a}) =  \frac{\partial}{\partial x} \omega_j (\bos{\a}, \bos{\bar \a}), \qquad  j=1, \dots, g \, ,
\end{equation} 
as a consequence of the diagonal system (\ref{whitham-g}).  Equations (\ref{wncl}) represent the consistency conditions in the formal averaging procedure  \cite{whitham65}, leading to the Whitham equations (\ref{whitham-g}). In this procedure, the local conservation laws of the NLS equation (\ref{nls1}) are averaged over the finite-band solutions (\ref{fingap}) (see \cite{flee86, pavlov87}) respectively. On the other hand, equations (\ref{wncl}) require that the fast phases $\eps^{-1}\eta_j(x,t)$ in the modulated finite-band solution (\ref{fingap}) must be consistent with the  generalised definitions of $k_j$ and $\omega_j$ as the local wave number and local frequency respectively  \cite{whitham_book}, \cite{dn89}, 
\be \label{kom_local}
k_j=(\eta_j)_x,  \quad \omega_j=(\eta_j)_t, \quad j=1, \dots, g.
\ee
Substitution of $\eta_j=k_jx+\omega_j t + \eta_j^0$ in  (\ref{kom_local})  yields the expressions for the initial phases $\eta_j^0$,  which are also subject to slow modulations and  are expressed in terms of the spectrum branch points, $\eta_j^0(x,t)=\Upsilon_j(\boldsymbol {\a, \bar \a, }), \ j=1, \dots, g$ \cite{el_dispersive_2016}. As we shall show, the {\it modulation phase functions} $\Upsilon_j(\boldsymbol {\a, \bar \a, })$  play the key role in the  description of the rogue wave formation due to the interaction of  the ``regular''  single-phase modulated wave trains.
 
 Using (\ref{Vjg}), (\ref{kj}), (\ref{holonorm}) one obtains the explicit compact representation for the characteristic speeds (from now on we omit the subscript $g$ in $\Rscr_g$) 
 \be\label{Vjexp}
V_j^{(g)} = \hbox{Re} \left[\sum_{k=0}^g\a_k \right]+
\dfrac{\sum_{k=1}^g\k_{k,2}\oint_{\gt_k}\frac{d\z}{(\z-\a_j)\Rscr(\z)}d\z}{\sum_{k=1}^g\k_{k,1}\oint_{\gt_k}\frac{d\z}{(\l-\a_j)\Rscr(\z)}d\z} \, , \ee
where the parameters $\k_{k,1}$, $\k_{k,2}$ are defined by (\ref{holonorm}). The derivation of the expression (\ref{Vjexp}) for $g=2$ is presented in Appendix A (see (\ref{velos})).

The characteristic speeds   (\ref{Vjexp}) are  generally complex-valued, i.e. the modulation system is elliptic implying modulational instability of the underlying nonlinear periodic wave \cite{whitham_book}  (see \cite{zakh-ostr} for the historical account and \cite{bronski2015} for recent advances in the mathematical understanding of the predictions of Whitham's theory).   However, it turns out that, for $g \ge 1$ some characteristic speeds (\ref{Vjexp}) can undergo a degeneracy and assume real values (see  (\ref{vm01}), (\ref{vm02})  below), so stable (or weakly unstable) configurations of nonlinear modulated waves in the focusing NLS dynamics are possible despite generic modulational instability.

In the genus zero case, $g=0$, the NLS modulation system (\ref{whitham-g}) has the form
\begin{equation}
\label{whitham0}
(\alpha_0)_t = (\tfrac32 \a_0+ \tfrac12 \bar\a_0)(\alpha_0)_x, \qquad   (\bar \alpha_0)_t = (\tfrac 32 \bar \a_0 + \tfrac 12 \a_0)(\bar\alpha_0)_x\, ,
\end{equation}
and  is equivalent to the  dispersionless limit of  (\ref{nls1})
\begin{equation}\label{dless}
 \rho_t+(\rho u)_x=0, \quad 
 u_t+uu_x - \rho_x =0 \, ,
\end{equation}
where the Riemann invariants and characteristic speeds in (\ref{whitham-g}) are expressed in terms of the hydrodynamic density and velocity as
\begin{equation}
\label{alpha}
 \alpha_0 = -(\frac{u}{2} +  i \sqrt{\rho}), \qquad V_0^{(0)}= \tfrac32 \a_0+ \tfrac12 \bar\a_0 =-(u + i \sqrt{\rho}) \, .
 \end{equation}
 One can see that the characteristics of (\ref{dless}) are complex unless $\rho=0$ implying nonlinear modulational instability of the NLS equation (\ref{nls1}) in the long-wave limit, which agrees with the linearised theory result for the plane wave (\ref{pw}).  Since for $\rho>0$ system (\ref{whitham0}) is elliptic, the initial-value problem for  (\ref{whitham0}) is ill-posed for all but analytical initial data.

For $g=1$ the characterstic speeds $V_{0,1}^{(1)}$ (\ref{Vjg}) can be explicitly represented in terms of  the complete elliptic integrals $K(m)$  and $E(m)$ of the first and the second kind respectively \cite{pavlov87}

\begin{equation}
\label{V12}
\begin{split}
V_0^{(1)} = U + \frac{( \alpha_0 - \alpha_1)(\alpha_0- \bar \alpha_0)}{( \alpha_0 - \alpha_1)+ (\alpha_1-\bar \alpha_0)E(m)/K(m)} , \\
V_1^{(1)} = U + \frac{(\alpha_1-  \alpha_0)( \alpha_1 -  \bar  \alpha_1)}{( \alpha_1 - \alpha_0) +(\alpha_0- \bar \alpha_1)E(m)/K(m)} .
\end{split}
\end{equation}
Here the modulus $m$ and the phase velocity $U$ are given by (\ref{Um}).
Of particular interest are behaviours of the characteristic speeds (\ref{V12}) for $m \to 0$ (linear limit) and $m \to 1$ (soliton limit). The linear limit can be achieved in one of the two ways (see (\ref{Um})): either via  $\a_1 \to \bar \a_1$ or via $\a_0 \to \bar \a_0$.  In the first case we have
\begin{equation}
\label{vm02}
\a_1 \to \bar \a_1:  \qquad V_0^{(1)}=  \frac32 \a_0 +\frac12 \bar \a_0\,  , \quad V_1^{(1)}=\bar V_1^{(1)}=  2a_1+\frac{b_0^2}{a_1-a_0} \,  .
\end{equation}
Similarly,
\begin{equation}
\label{vm01}
\a_0 \to \bar \a_0:  \qquad V_0^{(1)}=\bar V_0^{(1)}= 2a_0 + \frac{b_1^2}{a_0-a_1} , \quad V_1= \frac32 \a_1 +\frac12 \bar \a_1\,  .
\end{equation}
One can see that in both cases one complex conjugate pair of the characteristic velocities degenerates into  a single real value while the other pair transforms into the pair of characteristic speeds of the genus zero  system (\ref{whitham0}).

In the soliton limit we have
\begin{equation}
\label{vm1}
m \to 1: \qquad \a_0 = \a_1, \quad V_0^{(1)}=  V_1^{(1)}= U=  2 a_1 \, ,
\end{equation}
i.e.  in this limit all the characteristic speeds are real, which is the modulation theory expression of stability of fundamental NLS solitons.

\section{Dispersive dam-break flows in the focusing NLS equation}

Within the modulation theory framework the initial conditions for the  semi-classical NLS (\ref{nls1})  are considered to be specified on the genus zero Riemann surface with added double points.  The location of the double points in the complex $\lambda$-plane is determined by the initial potential. The solution genus changes during the evolution implying the emergence of new oscillatory phases. In the spectral plane this process is the opening of the double points into spectral bands. The phase transition lines in the $x$-$t$ plane, where the genus changes are often called the  {\it breaking curves}.
\begin{figure}[ht]
\includegraphics[height=1.1in]{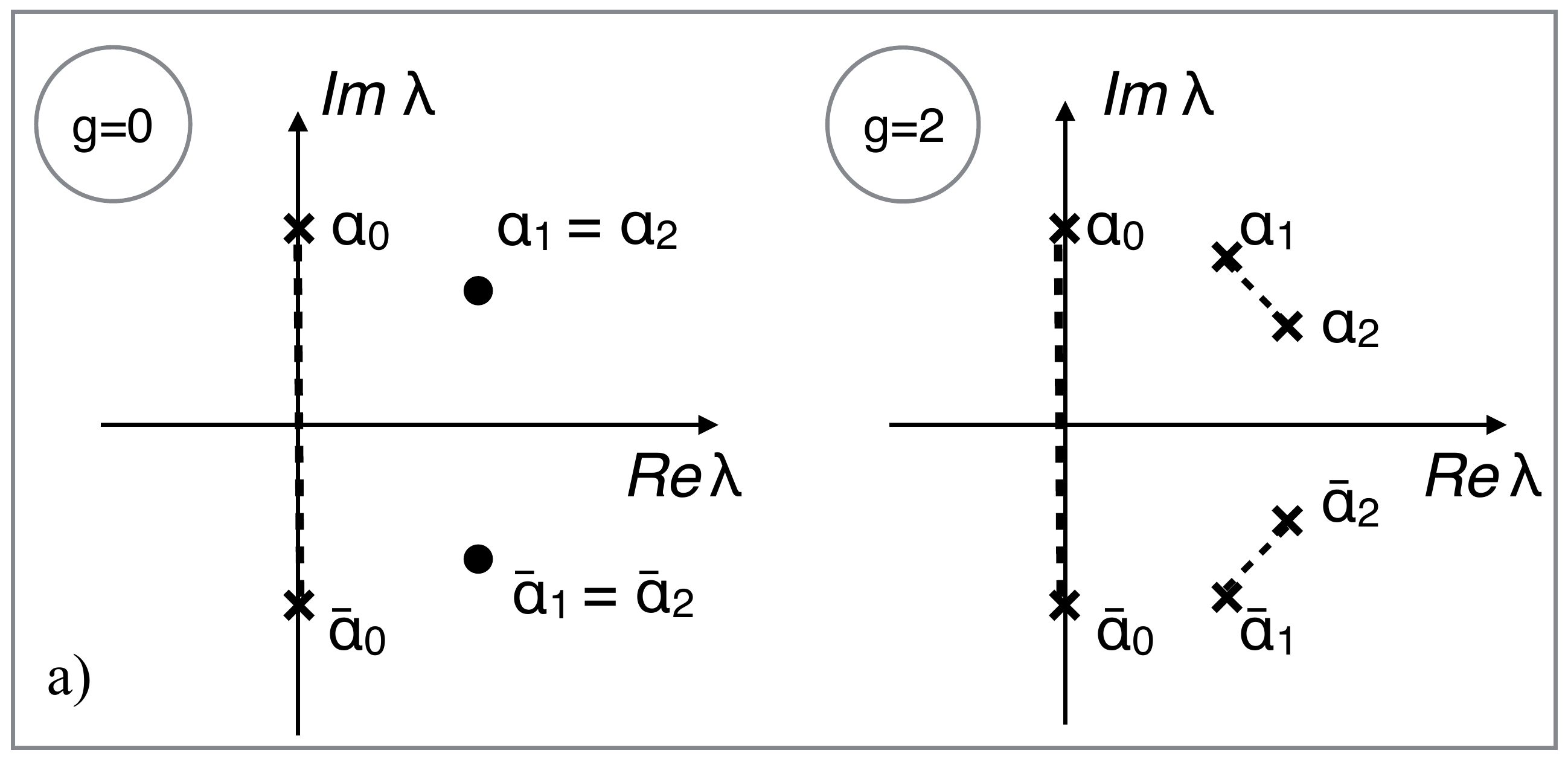}  \qquad \includegraphics[height=1.1in]{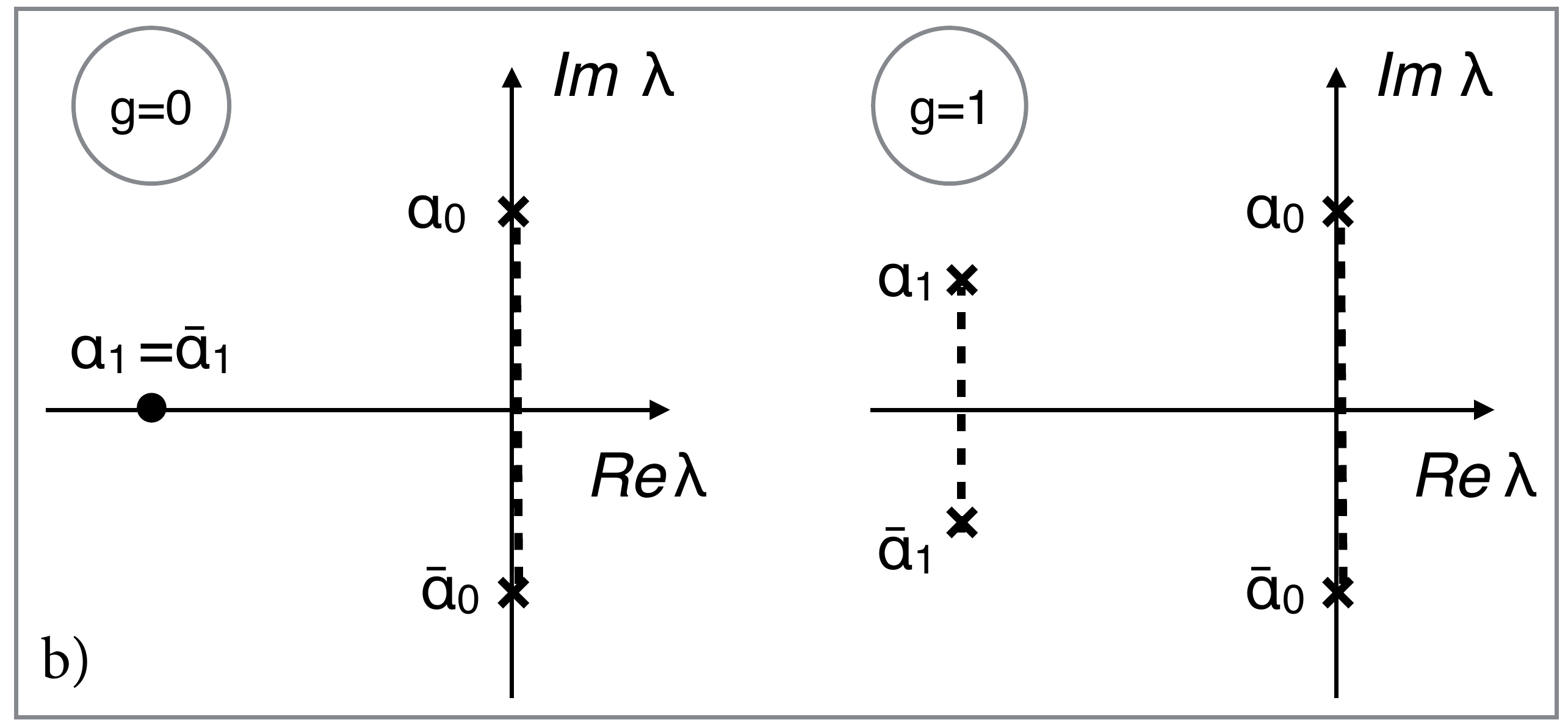}
\caption{Two contrasting generic types of the  spectrum modification across the first breaking curve: a) transition  $(g=0) \to (g=2)$; b) transition  $(g=0) \to (g=1)$}
 \label{fig1}
 \end{figure}

For a class of analytic  initial data sufficiently rapidly decaying at infinity the semiclassical NLS evolution is initially defined on the genus zero Riemann surface and is governed by the elliptic system (\ref{whitham0}). This evolution leads to the onset of a gradient catastrophe, which is dispersively regularised via the generation of rapid, $\eps$-scaled, oscillations. These oscillations within certain neighbourhood of the gradient catastrophe point $(x_0, t_0)$ are asymptotically described by the finite-band solution (\ref{fingap}) with $g=2$ and slowly varying parameters $\a_j$ \cite{TVZ}
(see the proof of universality of such an asymptotic behaviour  for analytic decaying potentials in \cite{BT2},  \cite{BT1}). The  spectrum modification across  the first breaking curve in the above scenario consists in the opening of two Schwartz-symmetric bands (see Fig. ~\ref{fig1}a) so the solution in the region just above this curve describes a nonlinear {\it two-phase} wave which is prominently manifested in the appearance of the Peregrine soliton right beyond the point of the gradient catastrophe at $x=0$ \cite{BT1}.  The exact mechanism of the genus modification is described within the RHP approach in\cite{TVZ},  \cite{BT1} (see also \cite{el_tovbis_2016}) but the necessity to introduce the genus two solution (rather than the genus one solution as it usually happens for  ``hyperbolic'' dispersive systems like the defocusing NLS, and involves the DSW formation \cite{eggk95}, \cite{kodama99}) can be qualitatively explained as follows. The genus zero system (\ref{whitham0})  is elliptic for all $\rho>0$.  One can  readily see that this system does not support  solutions involving variations of both variables $\rho$ and $u$ (or, equivalently $\a_0$ and $\bar \a_0$) along  a single characteristic direction (in hyperbolic quasilinear systems such solutions are called simple waves).    As a result, a generic gradient catastrophe   in (\ref{whitham0}) necessarily occurs for
$\a$ and $\bar \a$ at the same point  in the $x$-$t$ plane,  and so, its regularisation requires the introduction of two new, Schwartz-symmetric, spectral bands and thus,  involves the change of the genus of the Riemann surface $\Gamma(x,t)$  from zero to two.

There is another class of problems, where the phase transition across the first breaking curve occurs via the opening of a {\it single} spectral band emerging from the double point on the real axis (see Fig.~ \ref{fig1}b), so the oscillatory solution $(\rho(x,t), u(x,t))$ asymptotically represents a modulated  expanding {\it single-phase} wave train connecting the vacuum state $(0,0)$ upstream with the plane wave $(q^2,0)$ downstream. This type of dynamics resembles the behavour of  a DSW and is more in line with what one would expect in the hyperbolic case. As we shall see, this similarity is for a reason.

Consider the `dam-break' problem for the focusing dispersive hydrodynamics (\ref{dh}):
\begin{equation}\label{db}
\rho (x,0) =\left\{
\begin{array}{ll}
q^2>0 &\quad \hbox{for} \quad x <0,\\
 0& \quad \hbox{for} \quad x>0,
\end{array}
\right.
\qquad  u(x,0)=0 \, .
\end{equation}

In the stable, defocusing, case the regularisation  of the initial dam break (\ref{db}) occurs via a smooth rarefaction wave and does not involve the generation of a nonlinear dispersive wave except for small ($\sim \eps$)  oscillations regularising the weak discontinuity at the rarefaction wave corner \cite{eggk95} ( see also \cite{kodama_wabnitz}). In other words, the dam break problem solution to the defocusing NLS equation is  asymptotically ($\eps \to 0$)  equivalent to the  classical dam break problem solution for shallow-water waves \cite{whitham_book}.  The dam break solution for the focusing NLS equation is of  drastically different nature and  involves dispersive regularisation via a finite-amplitude modulated single-phase wavetrain defined inside an expanding transition region between two disparate states. In many respects this dynamic transition is analogous to a dispersive shock wave (DSW) \cite{dsw_review} with the important difference that the `hyperbolic' DSWs, similar to classical shocks, cannot provide transition from a vacuum state  (see e.g. \cite{whitham_book}).

The  modulation solution for the focusing dispersive dam break flow is constructed by noticing that one of the Riemann invariant c.c. pairs must be fixed to provide matching with the plane wave (\ref{pw})  downstream --- this pair is $\alpha_0=iq$, $\bar \a_0 = -iq$. The modulation solution for the second pair of the invariants $\alpha_1(x,t)$, $\bar \alpha_1(x,t)$ must depend on $s=x/t$ alone due to the scaling invariance of the problem (both initial conditions (\ref{db}) and the modulation equations (\ref{whitham-g}) are invariant with respect to the transformation $x \to Cx$, $t \to Ct$, where $C$ is an arbitrary constant). As a result, the required modulation is given by a centred  characteristic fan of the {\it genus one} Whitham equations (\ref{whitham-g}) \cite{egkk93},  \cite{bikbaev_sharapov94}, \cite{kamch97}:
\begin{equation}
\label{g1}
\alpha_0= i q, \quad \hbox{Re} [V_1(iq, -iq, \alpha_1, \bar \a_1)]=-\frac{x}{t},  \quad \hbox{Im}[ V_1(iq, -iq, \alpha_1, \bar \a_1)] =0\, .
\end{equation}
(here and below in this section we omit the superscript $^{(1)}$ for the characteristic speeds denoting the genus of the associated Riemann surface). Due to the symmetry in the expressions (\ref{V12}) for the characteristic speeds and the cnoidal solution (\ref{cnoidal}) the modulation (\ref{g1}) could be obtained in terms of $\alpha_0$, $\bar \a_0$ using the characteristic speed $V_0$ --  in that case one would need to set constant the other pair of invariants: $\a_1$, $\bar \a_1$.

We recall the notation $\alpha_j=a_j+ib_j$. Then, using  (\ref{V12})  we obtain from (\ref{g1}) the explicit expressions
\begin{equation}
\label{simplew}
\begin{split}
a_0=0, \qquad b_0=  q,  \\
\quad \frac{a_1^2+(q-b_1)^2}{a_1^2 -b_1^2+ q^2}=\frac{E(m)}{K(m)}\, , \quad m=\frac{4qb_1}{a_1^2+(q+b_1)^2}, \\
-\frac{x}{t}=2a_1 + \frac{q^2 - b_1^2}{a_1}.\\
\end{split}
\end{equation}
Since the real part $a_1$ of the Riemann invariant  enters (\ref{simplew}) only as a square, the solution (\ref{simplew}) defines two possible modulations of the cnoidal wave (\ref{cnoidal}) corresponding to the right- and left-propagating dispersive dam break flows. 
To single out the modulation corresponding to the particular initial-value problem (\ref{db})  it is instructive to represent solution (\ref{simplew}) in terms of a single parameter $0 \le m \le 1$ \cite{egkk93}:
\begin{equation}\label{abm}
a_1= \pm  \frac{2q}{m \mu(m)}\sqrt{(1-m)[\mu^2(m)+m -1]} ,  \qquad  b_1 = \frac{q}{m \mu(m)}[(2-m)\mu(m) -2(1-m)] \, ,  
\end{equation}
\begin{equation}
\label{mxt}
\frac{x}{t}= \pm \frac{2q}{m \mu(m)}\sqrt{(1-m)(\mu^2(m)+m-1)} \left ( 1+ \frac{(2-m)\mu(m) -2(1-m)}{\mu^2(m) +m - 1}\right) \, ,
\end{equation}
where $\mu(m)=E(m)/K(m)$. Now one has to choose the lower sign  in (\ref{abm}), (\ref{mxt})  to provide correct matching with the plane wave (\ref{pw})
downstream, corresponding to the initial condition (\ref{db}). The upper sign corresponds to the dam break flow of the opposite orientation, i.e. propagating upstream.

It follows from (\ref{abm}) that for the chosen (leftward) propagation direction,
\begin{equation}
\label{limitsimple}
m \to 0 \Longrightarrow  b_1 =  0, \ a_1 =  -q/\sqrt{2}; \qquad \hbox{and} \qquad m \to 1 \Longrightarrow a_1 = 0, \  b_1  = q \, .
\end{equation}
Then,  considering solution (\ref{simplew}) in the limits $m\to 0$ and $m \to 1$ we obtain the speeds of the trailing (leftmost) and leading (rightmost) edges to be $-2\sqrt{2}q t$ and $0$ respectively.
Thus, the modulated wave train defined by (\ref{cnoidal}), (\ref{simplew}) is confined to the expanding region $-2\sqrt{2}q t\le x \le 0$, where the modulus gradually varies from $m=0$ (harmonic, leftmost,  edge) to $m=1$ (soliton, rightmost, edge).  The wave amplitude $A=2\sqrt{b_1q}$ varies from $A=0$ at the harmonic edge to $A=2q$ at the soliton edge  where the wave form is described by the formula  (assuming zero phase shift $\xi_0$ in \eqref{cnoidal})
\begin{equation}
\label{sol}
\psi_S= 2q \ \sech (2q x/\eps) e^{i 4q^2 t/\eps} \, .
\end{equation}

 At the harmonic edge $x=-2\sqrt{2}qt$  the wavenumber (\ref{k1}) assumes the value $k_0 \equiv k(m=0)=\sqrt{6}q/ \eps>2q/\eps$ (stable modulation).  We note that  the harmonic edge velocity $s_-$ coincides with the linear group velocity of the left-going wave
 $\omega'(k)= -((\eps k)^2 -2q^2)((\eps k)^2 - 4 q^2)^{-1/2}$ evaluated at $k=k_0$.

Remarkably,  solution (\ref{simplew}) is essentially a {\it simple-wave solution} of the  modulation system (\ref{whitham-g}) with $g=1$, for which two of the Riemann invariants $(\a_0, \bar \a_0)$ are constant, and the remaining two ($\a_1, \bar \a_1$) vary along the same double characteristics family $V_2=x/t$. As we mentioned earlier, the genus zero system (\ref{whitham0}) does not support solutions of this type. 
%\red{It has been shown in a very recent IST-based  study \cite{biondini_prl_2016} that the the single-phase modulated solution (\ref{cnoidal}), (\ref{abm}), (\ref{mxt})  provides a  unversal description of the nonlinear stage of  the development of modulational instability induced by a localised perturbation of the plane wave.}

\begin{figure}[h]
\centerline{\includegraphics[height=2.5in]{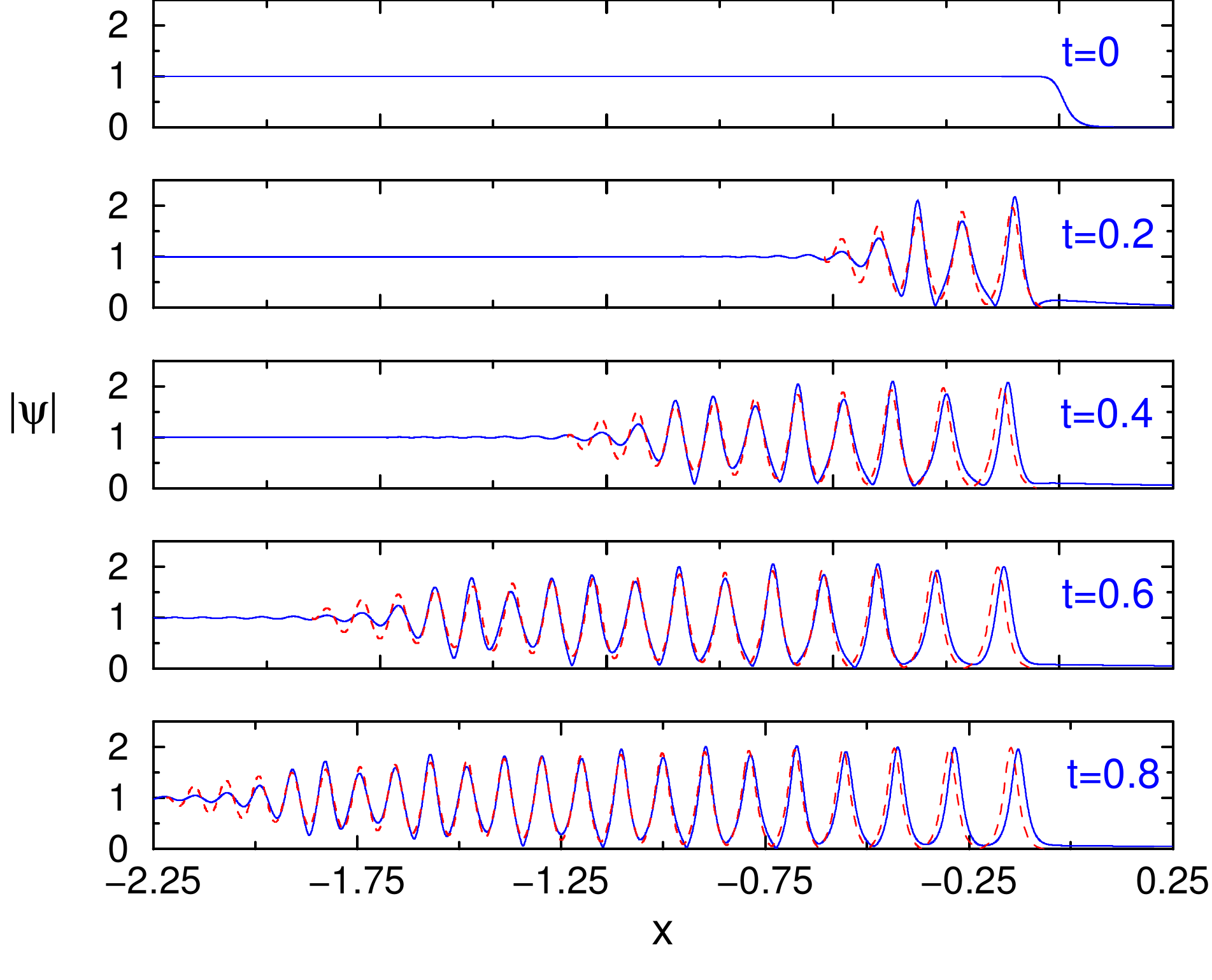} }
 \caption{(Color online) Dispersive regularisation of the dam break (\ref{db}) in the focusing NLS eq. (\ref{dh}): numerical  (solid line) vs analytical (modulation theory,  dashed line) solution for $|\psi|=\sqrt{\rho}$. The parameter values used  are: $q=1$, $\eps=1/33$. }
 \label{fig2}
 \end{figure}

Comparison of the modulated wave train described by (\ref{cnoidal}), (\ref{simplew}) with the numerical simulation of the dam break problem for the focusing NLS equation is presented in Fig.~\ref{fig2} and shows excellent agreement.   We note that an adjustment in the position of the wavetrain was required due to the  use of smoothed initial step data in the numerical simulations. 
Such a long-time persistence of the self-similar  modulation dynamics in the smoothed step problem is  well established for the  stable, defocusing, case (see \cite{biondini_kodama_2006}), but is not obvious and quite remarkable in the present, focusing, case, and deserves further analytical justification.

\begin{figure}[h]
\includegraphics[height=2in]{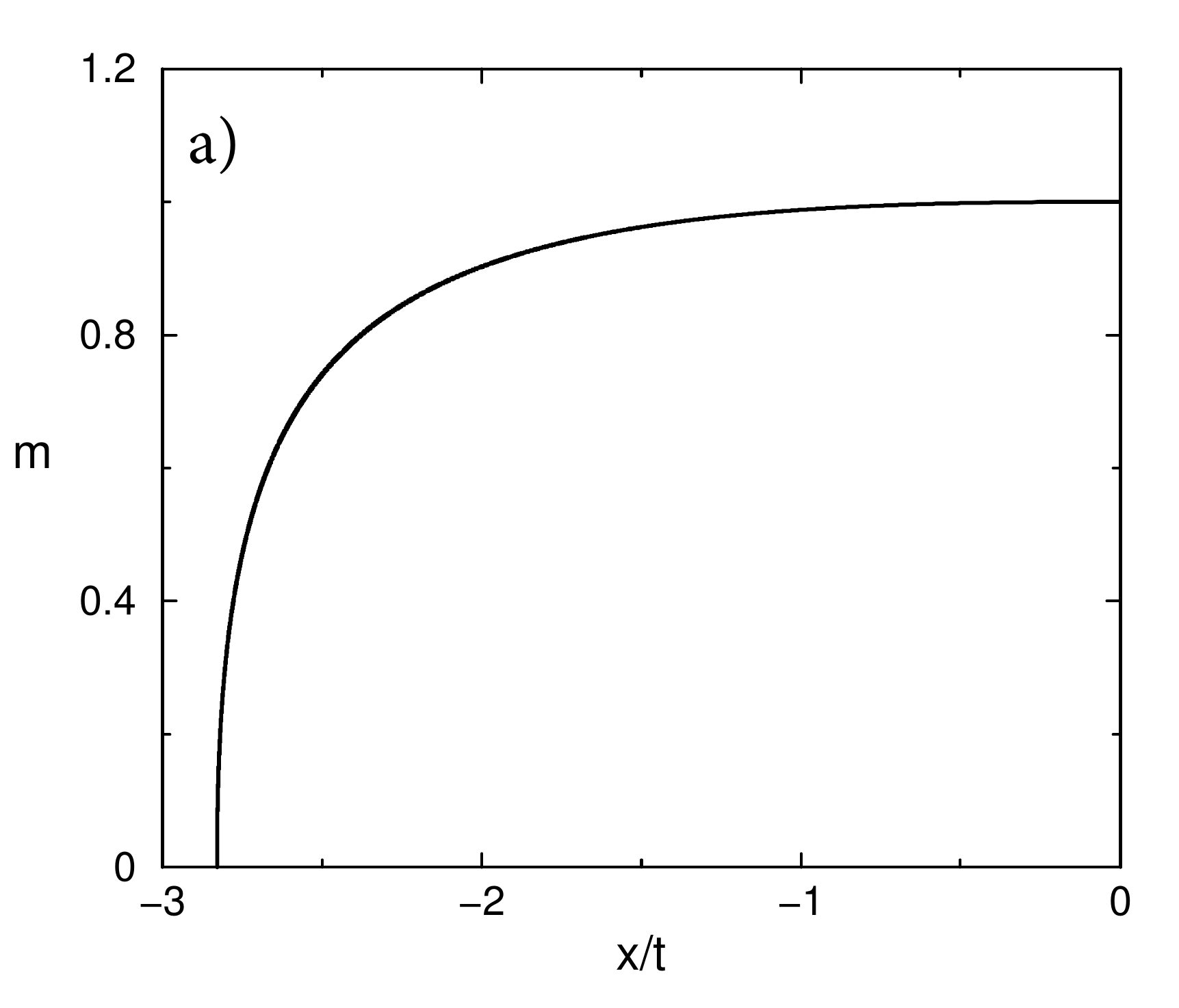} \qquad  \includegraphics[height=2in]{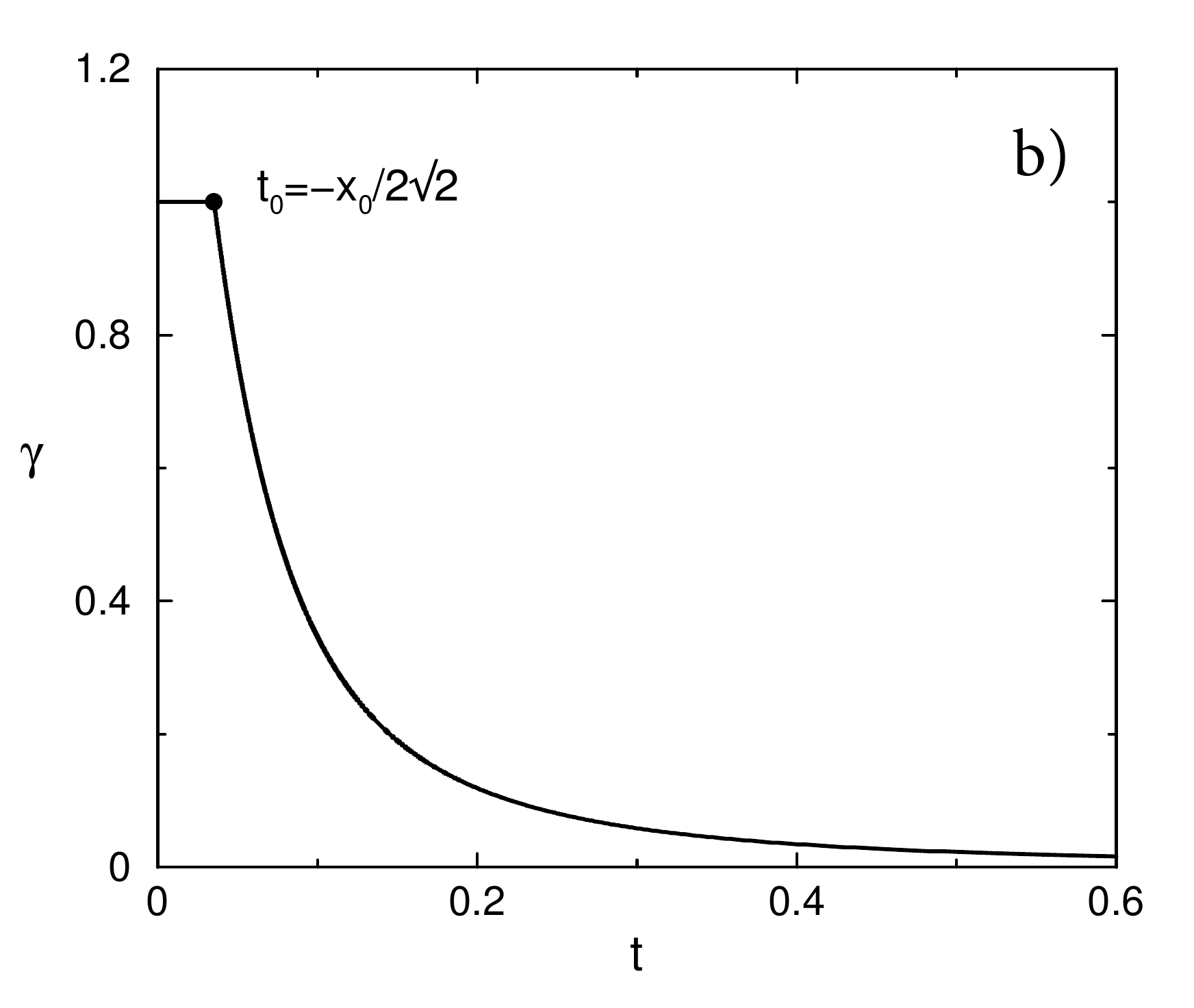}
 \caption{Stability of the dispersive dam break flow for the focusing NLS equation. a) Behaviour of the elliptic modulus $m$ as  function of $x/t$ in the modulation solution (\ref{mxt}) implying the dominance of solitons in the bulk of the wave train; b) The decay of the growth rate $\gamma = \Im V_0^{(1)}$  of the unstable nonlinear mode associated with the pair $\a_0=iq$, $\bar \a_0 =-iq$ of the spectrum branch points. The plot shows the function $\gamma(t)$  at a fixed point  $x=x_0=-0.1$ inside the wave train described by the modulation solution (\ref{simplew}).}
 \label{mgamma}
 \end{figure}

We now discuss stability of the obtained solution of the focusing dispersive dam break problem.
An important observation is that the constructed solution (\ref{cnoidal}), (\ref{simplew}) essentially describes a modulated soliton train. Indeed, the dependence of the modulus $m(x/t)$ in this solution shown in Fig. \ref{mgamma}a exhibits the values of $m$ close to unity almost over the entire oscillations zone, which means that the dispersive dam break flow in the focusing NLS is dominated by solitons with the amplitude close to $2q$.  We note that the approximation of a dispersive dam break flow in a focusing medium by a modulated train of solitary waves was successfully used in \cite{ass_smyth_2008}.
The transition from $m \approx 1$ in the bulk of the wave train to $m=0$ at the harmonic  edge occurs within a narrow (in $x/t$-coordinate) dynamic region, where the new oscillations  are generated and then quickly transform into solitons.

Despite the partial saturation of the modulational instability in the dispersive dam break flow due to the vanishing of the imaginary part of the characteristic speed $V_1$ for the solution (\ref{simplew}),  the wave train described by this solution is still subject to the instability implied by the nonzero imaginary part for the second pair of the characteristic speeds $V_0$,$\overline V_0$ associated with the constant Riemann invariants $\alpha_0=iq, \bar \a_0=-iq$.  This instability, however,  has  a weak effect on the dispersive dam break flow as such due to  the just mentioned fact that the major part of the wave train is dominated by solitons, which are modulationally stable. To quantify the effect of dispersive saturation we compute the imaginary part of  $V_0^{(1)}$ (we restore the upper index here to distinguish this speed from the value $V_0^{(0)}$ (\ref{alpha}) in the genus zero region). Using (\ref{V12}) and  (\ref{simplew}) we obtain
\begin{equation}
\label{imV1}
\gamma = \hbox{Im} [V_0^{(1)}] =  \frac{4q\mu(m)(q^2-b_1^2)}{[a_1(1-\mu(m))]^2 + [(q-b_1)+(q+b_1)\mu(m)]^2},
\end{equation}
where, we recall,  $\mu(m)=E(m)/K(m)$ (see (\ref{simplew})) and $a_1(m)$, $b_1(m)$ and  $m(x,t)$ are given by (\ref{abm}), (\ref{mxt}). The value of $\gamma$ can be viewed as a growth rate of nonlinear mode associated  with the spectral pair $\a_0$, $\bar \a_0$.

The dependence (\ref{imV1}) of $\gamma$ on $t$ for a fixed $x_0<0$ is shown in Fig. \ref{mgamma}b.  
One can see at each point $-2\sqrt{2}q t<x_0<0$ the value of $\gamma$   decreases to zero with time, thus confirming our conclusion about stability of the constructed solution.

While the initial dam break undergoes rapid dispersive saturation, the upstream  uniform plane wave state (\ref{pw}) for $x<-2\sqrt{2}qt$ remains modulationally unstable   with respect to long-wave ($k< 2q/\eps$) perturbations, and such small perturbations (a noise) are inevitably present in any physical system or numerical simulations. As a result, this instability imposes  restrictions on the `natural' lifespan of the described single-phase coherent structure of the `ideal' focusing dispersive dam break flow.  Let the typical amplitude of the noise be $\delta \ll1$. Then the linear theory prediction for  the characteristic time of the development of the fastest growing mode with $k=\sqrt{2}q/\eps$ is
\begin{equation}
\label{tm}
t_m \sim \frac{\eps}{2q^2} \ln\frac{1}{\delta},
\end{equation}
which gives an estimate for the  lifetime of the focusing dispersive dam break flow. 

In conclusion of this section we note that the focusing dispersive  dam break problem was studied experimentally in \cite{fleischer2010} in the context of diffraction from an edge in a self-focusing medium. The authors of \cite{fleischer2010} observed the expanding nonlinear oscillatory regularisation of a discontinuous intensity profile, qualitatively similar to that described by solution (\ref{cnoidal}), (\ref{simplew}).  To suppress the modulational instability of the background state  in  \cite{fleischer2010}  nonlocality  was used as suggested  by previous theoretical studies  \cite{trillo}, \cite{ass_smyth_2008}.
Solution (\ref{cnoidal}), (\ref{simplew}) was also used in \cite{abdullaev} for the modelling of the matter-wave bright soliton generation at the sharp edge  of density distribution in a BEC.

\section{Interaction of focusing dispersive dam break flows and the generation of rogue waves}

As  was pointed out in Section II, the prototypical rogue wave solutions (soitons on finite background) are associated with the degenerate genus two NLS dynamics.   This suggests that one can expect the rogue wave formation in the processes involving the interaction of two ``regular'', single-phase waves ($g=1$).   Indeed, the ``elementary'' rogue wave events  during individual soliton collisions  were observed in numerical simulations \cite{dudley2010} (see also \cite{dudley_nat2014}).  Here we consider a more general scenario where rogue waves  (not necessarily exact Akhmediev or Peregrine breathers) are formed in the interaction of two single-phase dispersive wave trains. 

We consider  the NLS equation (\ref{dh}) with $\eps \ll 1$ and initial conditions in the form of a rectangular barrier for the intensity with zero initial velocity,
\begin{equation}\label{ic1}
\rho(x,0) =\left\{
\begin{array}{ll}
q^2  &\quad \hbox{for} \quad |x| <L,\\
 0& \quad \hbox{for} \quad |x|>L;
\end{array}
\right.
\qquad  u(x,0)=0 \, .
\end{equation}
We shall refer to (\ref{dh}), (\ref{ic1}) as the NLS box problem.
The rigorous semi-classical asymptotics of the NLS box problem solution were calculated in the recent paper \cite{kenjen14}  for the initial stage of the evolution involving solutions with $g=0,1$. The analysis in \cite{kenjen14} was performed using the steepest descent for the oscillatory Riemann-Hilbert problem (RHP) \cite{DZ1} associated with the semi-classical NLS equation (\cite{KMM}, \cite{TVZ}) and, in particular, yielded the modulation solution  (\ref{cnoidal}), (\ref{simplew}). In what follows we shall use an appropriate combination of the Whitham theory and the RHP techniques  to construct a relatively simple exact modulation solution describing the dispersive dam break flow interaction ($g=2$). This solution will then be  used to predict the rogue wave formation.

\subsection {Before interaction  $(g=0,1)$}
The semi-classical evolution (\ref{nls1}), (\ref{ic1}) starts at $t=0$ with the instantaneous formation of two dispersive dam break flows  at the discontinuity points $x=\pm L$ of the initial profile.
The  corresponding modulation solution  consists of two centred fans  emanating from $x=\pm L$  and described by the  formulae (see (\ref{g1})): 
\be \label{genus1}
\begin{split}
 x + \hbox{Re}[ V_1^{(1)}(iq, -iq, \alpha_1, \bar \alpha_1)] t= \pm L \, , \\
\hbox{Im} [V_1^{(1)}(iq, -iq, \alpha_1, \bar \alpha_1)]=0 \, .
\end{split}
\ee
The upper sign solution (\ref{genus1}) is defined for $0<x<L$  and the lower sign for $-L<x<0$. Also one uses $a_1= \Re \a_1 <0$  in (\ref{genus1}) for $0<x<L$ and $a_1>0$  for $-L<x<0$  -- see (\ref{abm}). We intentionally represented solution (\ref{genus1}) in the `hodograph' form to elucidate  the natural matching of (\ref{genus1}) with the  modulation solution in the interaction region, which is of our primary interest. Explicit expressions for the modulation solution (\ref{genus1}) in terms of elliptic integrals are obtained from (\ref{simplew}) by replacing $x$ with $x \pm L$.  

Earlier we introduced the notion of a breaking curve as the line in $x$-$t$ plane separating  regions described by solutions with different genus $g$. On the first breaking curve $T_1$ separating the regions with $g=0$ and $g=1$ (see Fig.~\ref{fig6}a) one has  (see (\ref{limitsimple})):
\begin{equation}\label{T1}
T_1: \quad  m=0,    \quad \a_1=\mp q/\sqrt{2},
\end{equation}
where the minus sign applies to $0<x<L$ and the plus sign to $-L<x<0$. Now,  evaluating the limit $m \to 0$ of the solution (\ref{genus1}), we obtain the equation of the first breaking curve
\begin{equation}\label{T1}
T_1: \qquad t= \frac{L-|x|}{2 \sqrt{2}q}  \, 
\end{equation}
consisting of two symmetric parts.
\begin{center}
\begin{figure}[ht]
 \includegraphics[height=2.0in]{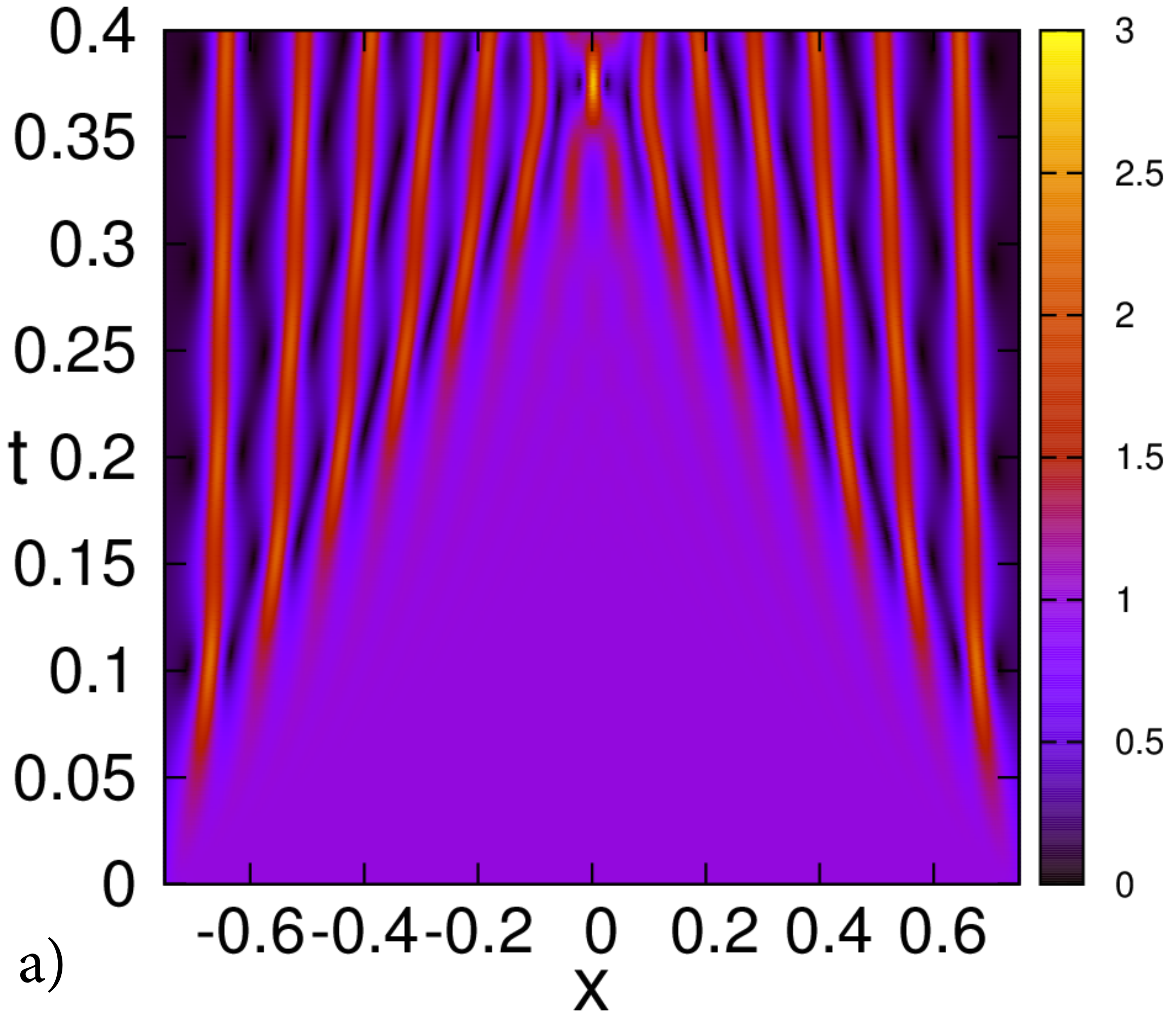}   \qquad  \includegraphics[height=2.0in]{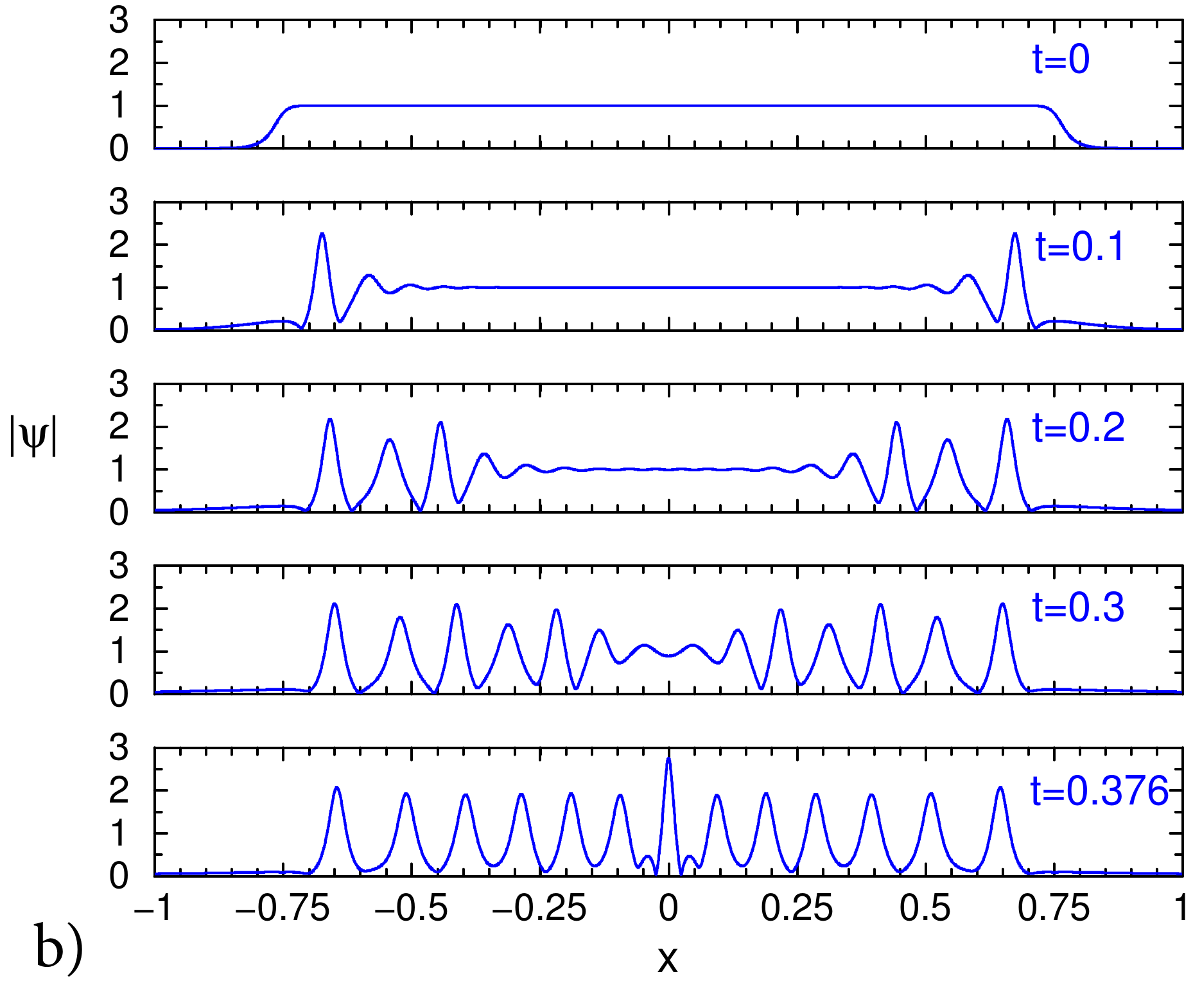}
 \caption{(Color online) Collision of two counter-propagating focusing dispersive dam break flows results in the formation of rogue waves. Numerical simulation of the NLS box problem (\ref{dh}), (\ref{ic1}) with $\eps=1/33$, $q=1$, $L=25/33$. The rogue wave is formed  at  about $t=0.376$.  a) Density plot for the amplitude $|\psi|=\sqrt{\rho}$;  \ b) Amplitude profiles for different $t$.}
 \label{fig4}
 \end{figure}
\end{center}

At $x=0$  $t=t_0=L/2\sqrt{2}q$ the counter-propagating dispersive dam break flows collide,  so the developed single-phase description (\ref{genus1}) becomes invalid for $t>t_0$. In Fig.~\ref{fig4} the results of the numerical simulations of the box problem with $L= 25/33$, $q=1$ and $\eps=1/33$ are presented for $t \le t_0$ (the smoothed initial data were used in the numerical simulations to avoid unwanted numerical effects). The numerical method used in the simulations is briefly described in Appendix B, where also the effects of smoothing the initial data are discussed. In Fig.\ref{fig4}a the density plot is presented for $|\psi(x,t)|$ while Fig.~\ref{fig4}b shows spatial profiles of $|\psi (x)|$ at different moments of time. One can see from the bottom plot that the dispersive dam break flow collision  leads to the rapid formation of a rogue wave with the maximum at $x=0$, the amplitude slightly less than 3 and the wave form typical of a breather:  a tall central peak rapidly decaying to zero and two smaller ``wings'' at both sides. 

Remarkably, at the point of collision $(x=0, t=t_0)$ the amplitudes of both single-phase modulated wavetrains 
are very small and the modulation is stable (the modulation solution yields the zero amplitude and the wavenumber $k_0=\sqrt{6}q/\eps > k_c$ at this point).  However, for $t>t_0$ the interaction between these two stable, small-amplitude tails of the counter-propagating dispersive wavetrains gives rise to the development of modulational instability resulting in the rapid formation of rogue waves.

In the next section, we shall study the development of this  process using the modulation theory for $g=2$, some results of the RHP analysis of the semi-classical NLS equation and direct numerical simulations.

\subsection{Interaction ($g \ge 2$)}

\begin{figure}[ht]
\includegraphics[height=2.1in]{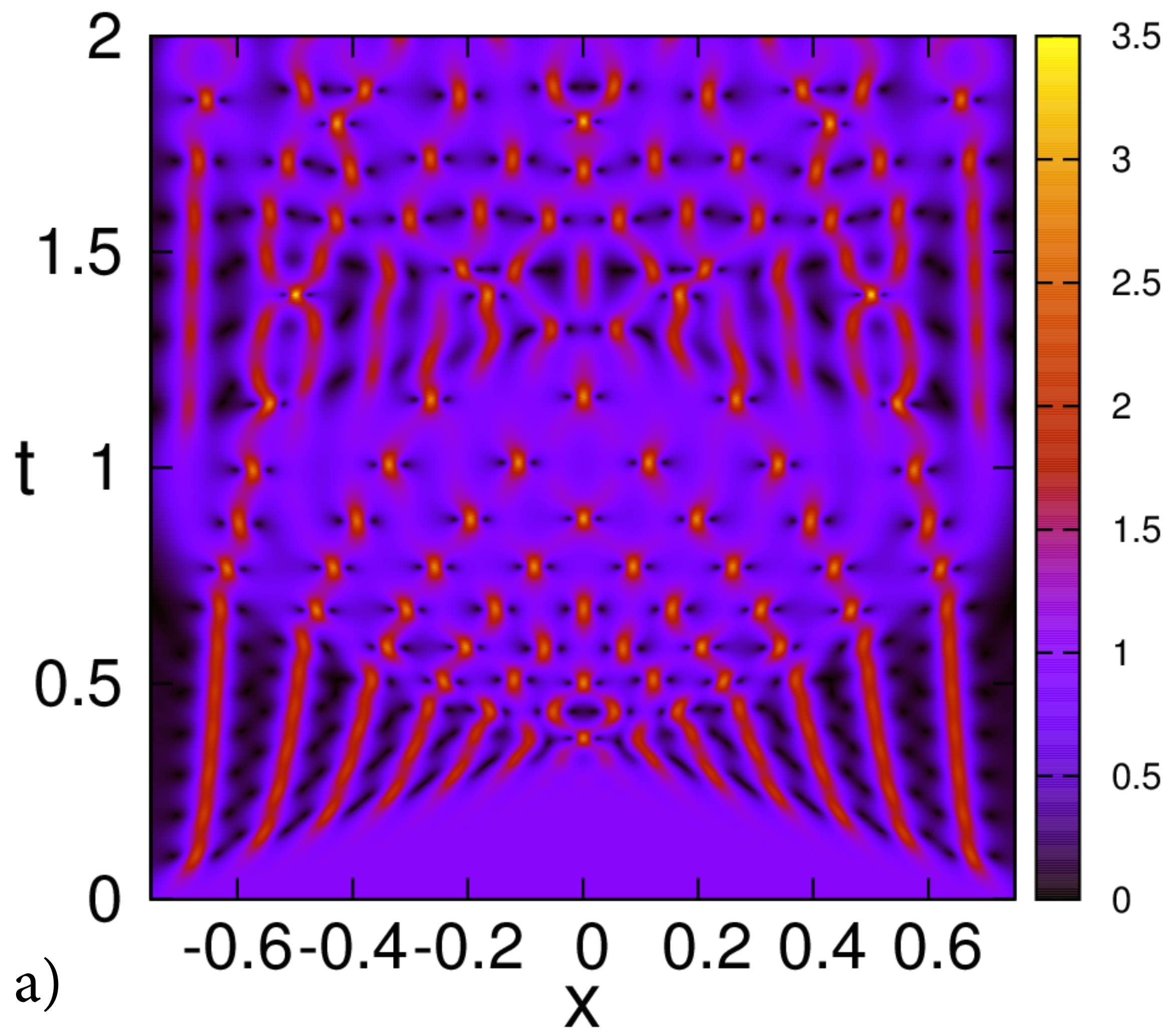} \qquad \includegraphics[height=2.1in]{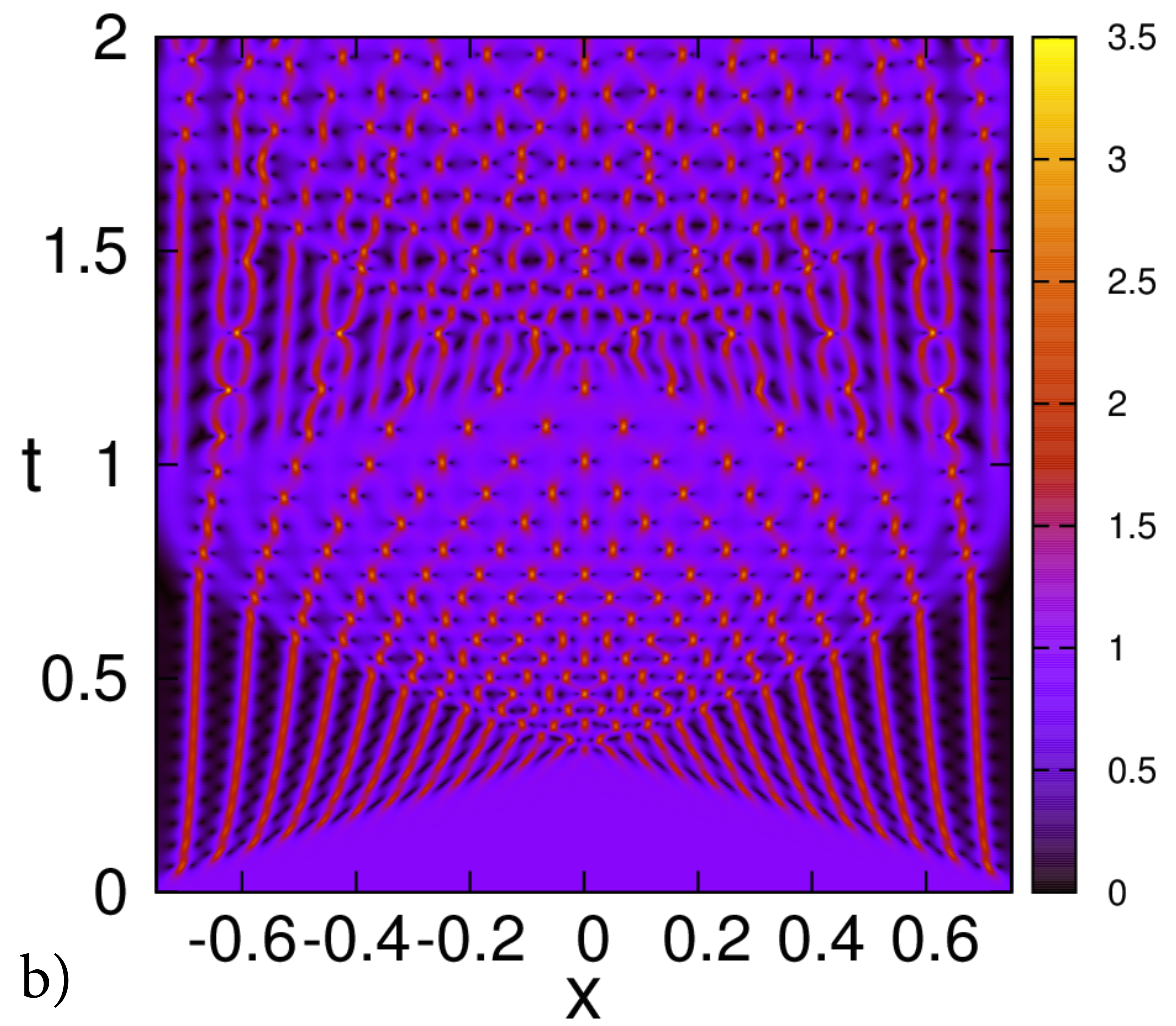}
\caption{(Color online) Density plot for $|\psi|=\sqrt{\rho}$ in the focusing NLS box problem with $q=1$, $L=25/33$. a) $\eps=1/33$;  \ b) $\eps=1/60$.}
 \label{fig5}
 \end{figure}

 We first present the results of the numerical simulations of the NLS box problem with $q=1$, $L=25/33$ and two different values of $\eps$: $1/33$ and $1/60$. The respective $x$-$t$ density plots for the amplitude $|\psi|=\sqrt{\rho}$ are presented in Fig.~\ref{fig5}a and Fig.~\ref{fig5}b. One can  see the  regions with distinctly different behaviour of the solution in both plots. Remarkably, both simulations produce very similar {\it macroscopic} patterns differing apparently only in the number of the oscillations forming the pattern.  This striking robustness of the macroscopic features of the dispersive dam break flow interaction despite the presence of modulational instability in the system,  is a strong indication of the applicability of the limiting ($\eps \to 0$), semi-classical description and of the Whitham modulation equations in particular. Indeed, as we mentioned, the existence  of the semi-classical limit was rigorously established in \cite{kenjen14} for the initial stage of the evolution involving solutions with $g=0,1$ but now we have some confidence in assuming the validity of the semi-classical description for the regions with $g>1$. 

\begin{figure}[ht]
\includegraphics[height=2.0in]{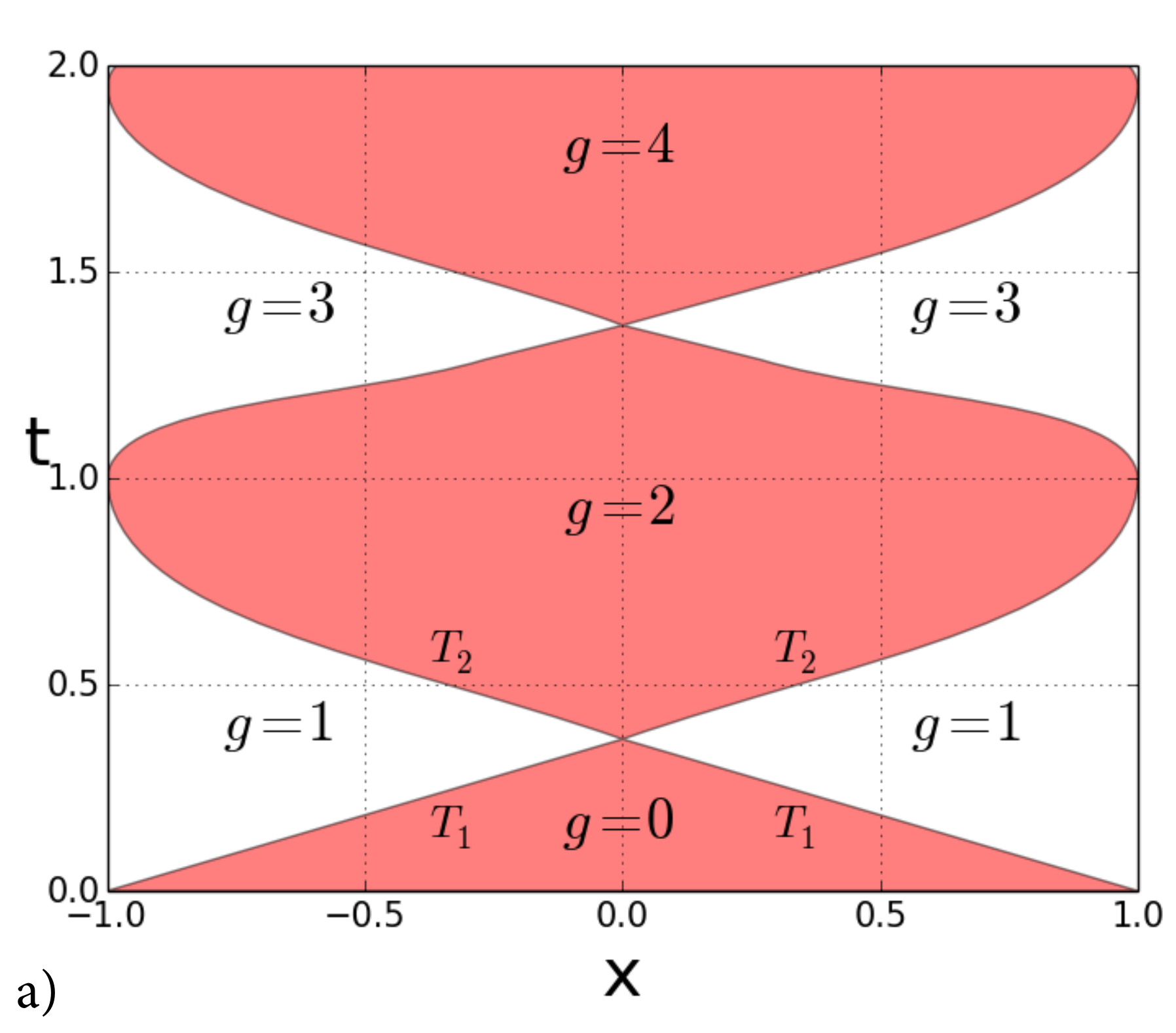} \qquad \qquad \includegraphics[height=1.6in]{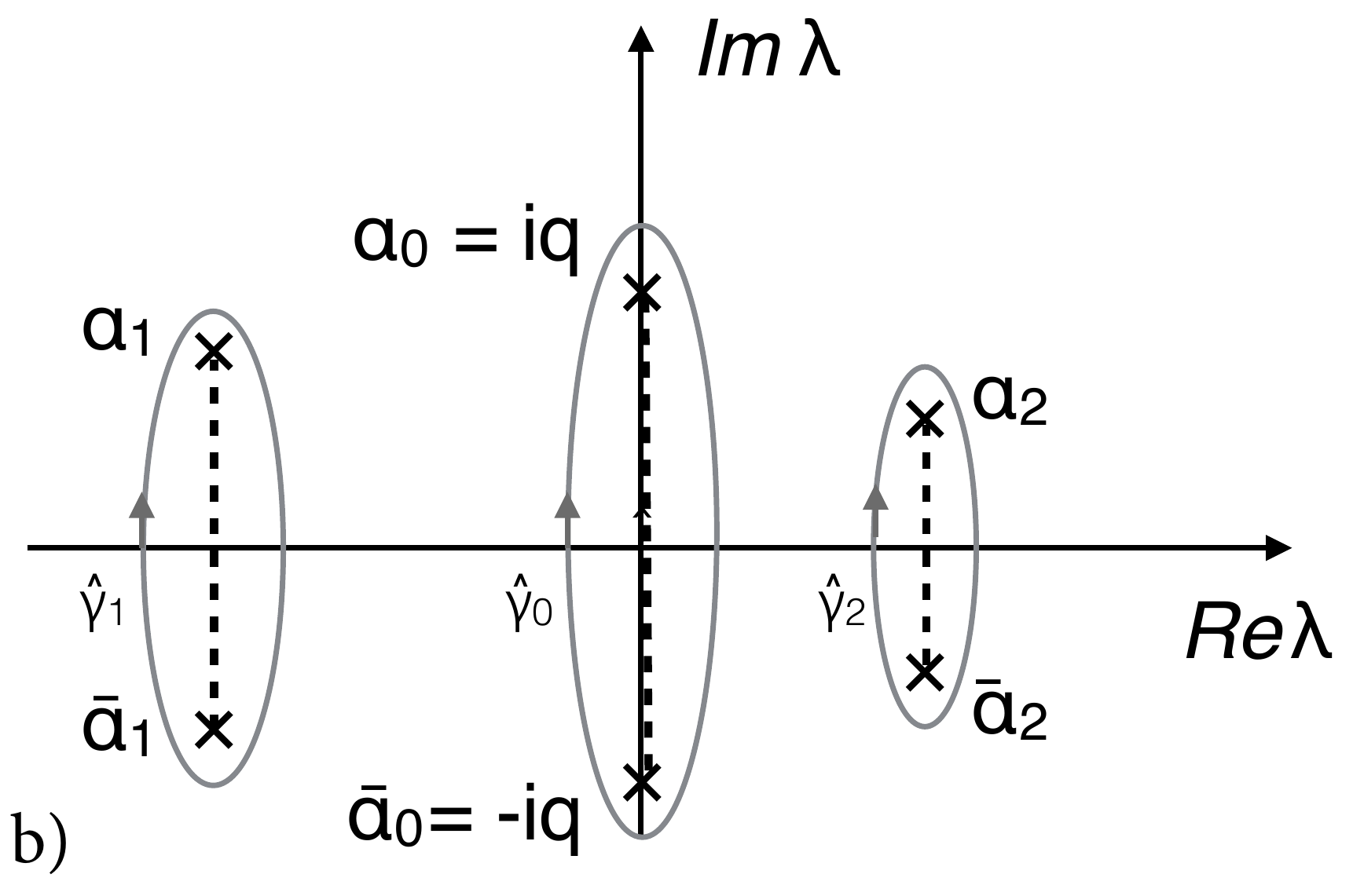} \caption{  a)  (Color online)  The structure of the $x$-$t$ plane of the NLS box problem in the semiclassical limit suggested by  the $x$-$t$ density plots in Fig.~\ref{fig5}. The diagram is normalised for $q=1$, $L=1$; b) The spectral portrait of the solution in the $g=2$ region;. }
 \label{fig6}
 \end{figure}

We have used the image filtering software (an edge-detect filter is applied to detect contours and facilitate
the sampling of points) to extract the boundaries between the regions with qualitatively different behaviours of the oscillations in the plots of Fig.~\ref{fig5}. The normalised to $q=1$, $L=1$ diagram is shown  Fig.~\ref{fig6}a  where B\'ezier curves of a few points extracted from the filtered image are used to produce the boundary lines. A minor adjustment of the image to the analytically available points (such as the collision point $(0,1/2\sqrt{2})$) was necessary to compensate for the slight time delay present in the direct simulations  in Fig.~\ref{fig5} due to the smoothing of the box edges.  

We now formulate the hypothesis about the structure of the $x$-$t$ plane in the box problem shown in Fig.~{\ref{fig6}a} .
As our numerical simulations suggest, the interaction of two single-phase dispersive dam break flows is described by the modulated two-phase ($g=2$) solution  confined to a curved rhombus-like region. The genus change pattern from $g=1$ to $g=2$ across the first breaking curve  is also consistent with the spectrum modification mechanism  shown in Fig. \ref{fig1}b as it involves the emergence of one band from a double point on the real axis of the $\la$-plane. The opposite signs of  the real parts of the evolving Riemann invariants in the two  dam break solutions (\ref{genus1}) ($g=1$) before the interaction, suggest that the spectral portait in the interaction region will have a qualitative form shown in Fig. \ref{fig6}b, i.e. will consist of two slowly evolving bands $ \gamma_1$ and $ \gamma_2$ located at opposite sides of the fixed central band  $\gamma_0$ (see also Appendix A). Indeed, in Section 4.2.1 below, we shall construct the modulation solution describing the slow evolution of this spectral portrait and show that it is consistent (i.e. provides the necessary matching) with the known ``outer'' solution (\ref{genus1}) in the $g=1$ region. 

\begin{figure}[h]
\includegraphics[height=2.2in]{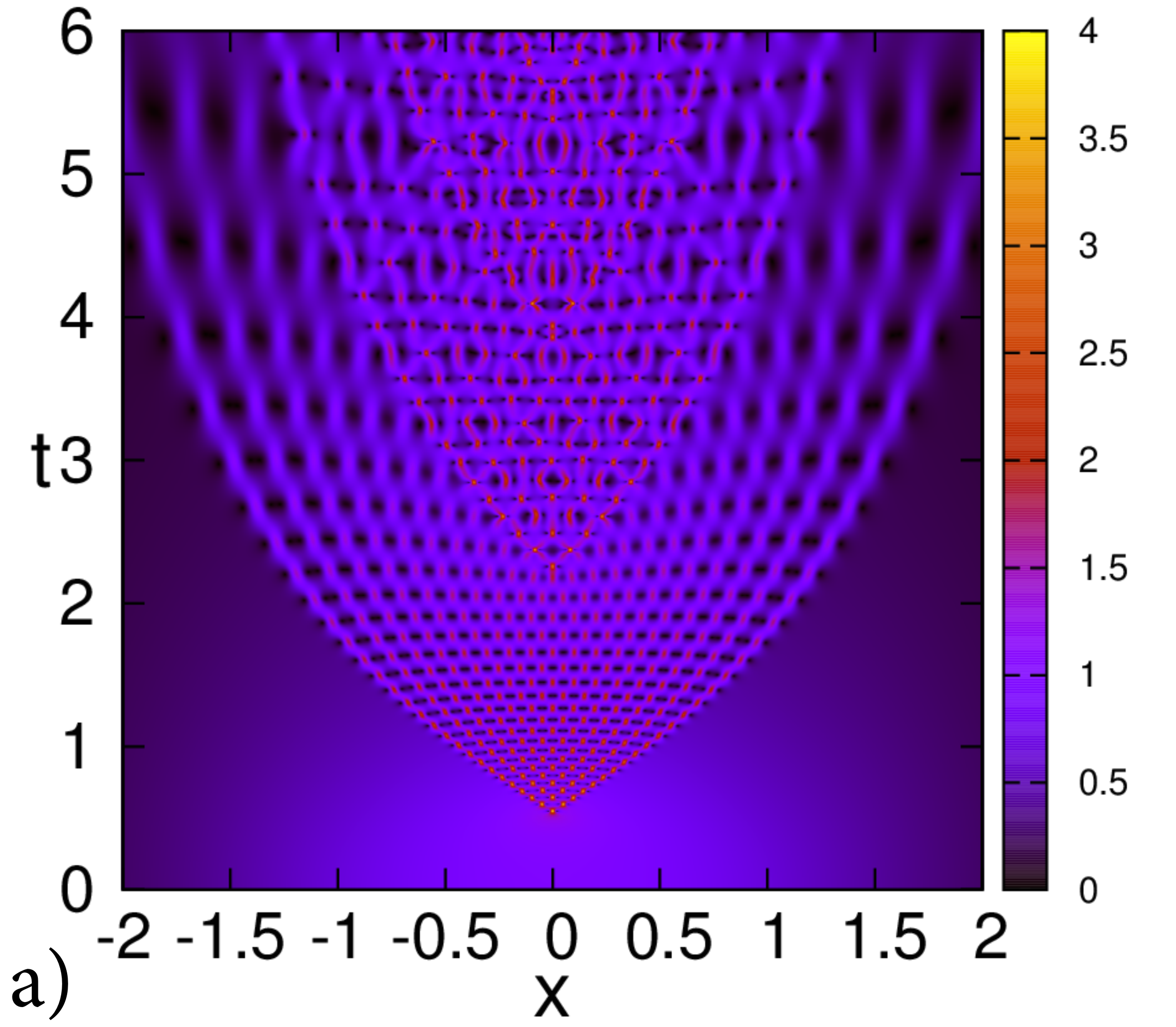} \qquad \includegraphics[height=2.2in]{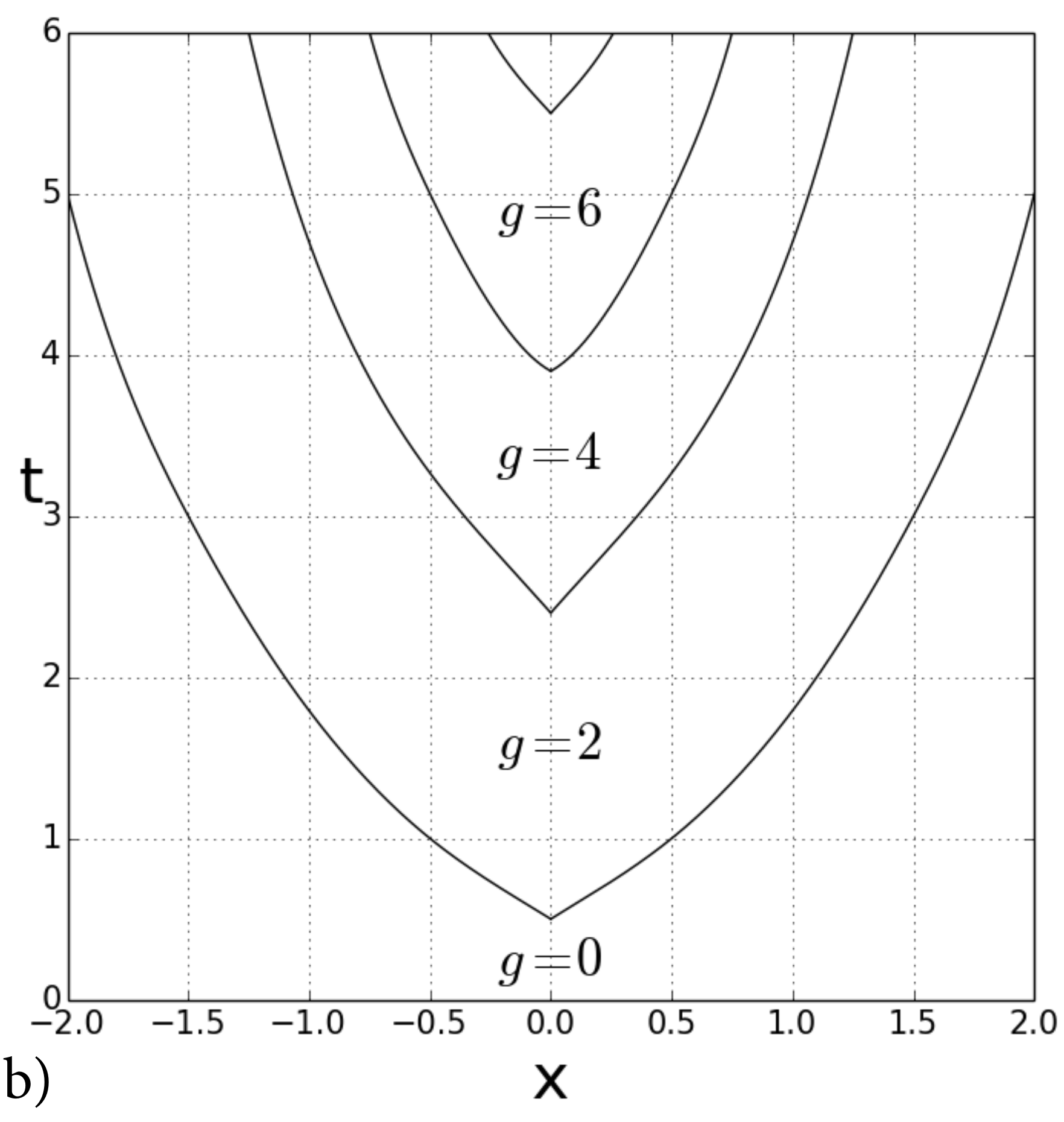}
\caption{ NLS equation (\ref{nls1}) with  $\eps=1/33$ and the analytic initial condition $\psi = \sech x$. a) (Color online)  Density plot for $|\psi(x,t)|$; b) The structure of the $x$-$t$ plane in the semi-classical limit suggested by a). }
 \label{fig7}
 \end{figure}

The subsequent evolution leads to the formation of the regions of higher genera: $g=3$, $g=4$, etc. (see Fig. \ref{fig6} b)). All the generated oscillations are confined to the original box interval $-L <x<L$. Remarkably, for any $0<|x| < L$, each crossing of the breaking curve results in the genus increment by one, while  the genus  increases by two when crossing the  breaking curves at $x=0$ . We note that the known scenario of the development of the semi-classical NLS solution for analytic, sufficiently rapidly decaying initial data (e.g. $\psi (x,0) = \sech(x)$) involves only the genus increments by two across breaking curves \cite{TVZ}, \cite{KMM}  (see Fig. \ref{fig7}). The striking difference between the two scenarios is the reflection of the two fundamentally different spectral mechanisms of the genus transformation involved. These are shown in Fig. \ref{fig1}  for the first breaking curve  but the principle remains for higher breaks as well.

Below we concentrate on the analysis of the dispersive dam break flow interaction occurring in  the  region with $g=2$.

\subsubsection{$g=2$: hodograph solution}
The generic spectral portrait of the finite-band NLS solution in the $g=2$ region of the box problem is shown in Fig.~\ref{fig6}a.
The modulation is described by three complex conjugate pairs of Riemann invariants
$\a_j, \bar \a_j$, $j=0,1,2$. Two of the Riemann invariants remain constant: $\a_0=iq$, $\bar \a_0=-iq$, so we are left with just two varying pairs: $\a_j(x,t), \bar \a_j(x,t)$, $j=1,2$. We note that, within the framework of the modulation theory,  the constancy of $\a_0, \bar \a_0$ follows from the matching, across the second breaking curve,  of the genus two modulation solution with the genus one solutiion (\ref{genus1}) for which these invariants are constant. At the same time, in the rigorous RHP analysis of \cite{kenjen14}, the branchpoints 
$\alpha_0$, $\bar \alpha_0$ for the box potential are special, as they coincide with the endpoints of the spectral interval on the imaginary axis
(this spectral interval is the locus of accumulation of the points of discrete spectrum, whose number is growing like
$O(\varepsilon^{-1})$). The spectral interval, of course, is invariant under the NLS-evolution. 
 Further, due to the symmetry $x \to -x$  of the box problem the evolving bands must always be located at the oposite sides of the imaginary axis $\hbox{Re} [\la] = 0$, hence  at $x=0$ one has $\a_1 = - \bar \a_2$.  It is worth mentioning  that the opposite signs for $\hbox{Re}[\a_1]$ and $\hbox{Re}[\a_2]$  most readily follow from the modulation theory  as  they are requuired by the matching, along the second breaking curve, of the modulation solution in the genus two region, with two genus one modulation solutions  for the dispersive dam break flows propagating in opposite directions (towards each other). By construction, the varying Riemann invariants for these flows have real parts of opposite signs -- see  Section 4.1. The same result follows from the rigorous analysis of \cite{kenjen14}. The solution for the moving invariants $\a_j, \bar \a_j$, $j=1,2$ can be found via the Tsarev generalized hodograph transform \cite{tsarev85}. This method was originally developed for hyperbolic systems of hydrodynamic type but  is equally applicable to elliptic systems. The Tsarev result in the application to our present problem can be formulated as follows. Any  local non-constant solution of the modulation system (\ref{whitham-g}) in the genus two region is given in an implicit form by the system of two algebraic equations with complex coefficients
\begin{equation}
\label{w0}
x+V_j(\bos {\alpha}, \bos{\bar \alpha})t = w_j(\bos{\alpha}, \bos{\bar \alpha}) , \quad  x+\overline V_j(\bos{\alpha}, \bos{\bar \alpha})t = \bar w_j(\bos{\alpha},  \bos{\bar \alpha}),  \qquad j= 1,2,
\end{equation}
where $\bos{\a} =(iq,  \a_1,  \a_2)$; the characteristic speeds $V_j (\bos{\alpha}, \bos{\bar \alpha}) \equiv V^{(2)}_j (\bos{\alpha}, \bos{\bar \alpha})$ are given by (\ref{Vjg}), (\ref{kj}) (or, equivalently, by (\ref{Vjexp})). The four unknown functions  $w_j$, $\bar w_j$, $j=1,2$ satisfy the system of four {\it linear} partial differential equations
\be\label{tsarev}
\frac{\partial_{\a_j}w_k}{w_k-w_j}=\frac{\partial_{\a_j}V_k}{V_k-V_j}  \quad \hbox{and \ c.c.};  \quad j, k=1,2, \quad k\neq j \, ,
\ee
where $\partial_{\a_j} \equiv \frac{\partial}{\partial \a_j}$. 

For (\ref{w0}), (\ref{tsarev}) to describe the interaction of two dispersive dam break flows in the box problem, equations (\ref{tsarev}) must be supplied with  appropriate boundary conditions.  These conditions follow from the requirement of continuous matching,  on the second breaking curve $t=T_2(x)$, of the hodograph solution (\ref{w0}) with the known solution  (\ref{genus1}) (in which $\a_1$ should be replaced by $\a_2$ for $0<x<L$ in the ``plus'' branch of the solution). Similar to the first breaking curve $T_1$, the curve $T_2$ is a free boundary, on which two of the Riemann invariants merge (cf. Fig.\ref{fig1}b showing the prototypical spectrum modification across a breaking curve in the box problem). It can be shown that the corresponding merged velocities are real (see \cite{el_tovbis_2016}) so the boundary $T_2$ represents a symmetric curve consisting of two real characteristics that carry over the constant values of the merged Riemann invariants brought by  the two branches of $T_1$ (see \eqref{T1})
\be \label{T2}
T_2:  \qquad \a_1=\bar \a_1 = - q/\sqrt{2} \ \  \hbox{for} \ \  -L<x<0  \quad  \hbox{and}   \quad \a_2= \bar \a_2 =  q/\sqrt{2} \ \ \hbox{for} \ \  0<x<L \, .
\ee  
%\red{(Note that one has to replace $\a_1 \to \a_2$ for the ``plus'' branch of solution (\ref{abm}), (\ref{mxt}) for $0<x<L$ to explicitly distinuish between the two ``counter-propagating'' solutions in the $g=1$ region participating in the matching conditions on $T_2$).}
The matching regularisation procedure for the  the focusing NLS is in many respects analogous to that in the defocusing NLS theory (see  \cite{ek95}).  We won't describe it in much  detail but just mention that, having found the solution $w_j$, $\bar w_j$ of the  Tsarev equations, one then needs  to verify that the resulting hodograph equations (\ref{w0}) are invertible, i.e. that they specify functions $\bos \a(x,t)$, $\bos {\bar \a} (x,t)$ in a certain region of $x$-$t$ plane.  See \cite{grava} for the `hyperbolic' counterpart of this construction arising in the description of the DSW interaction in the KdV modulation theory. 

We by-pass the outlined above direct matching regularisation  procedure  by taking advantage of  the available mathematical results from the RHP analysis \cite{TV2010} of the semi-classical focusing NLS,  and applying them to the genus two region in the box problem.  
More specifically, we shall express Tsarev's $w_j$'s in terms of the ``modulation phase shift'' functions $\Upsilon_j(\boldsymbol{\a, \bar \a})$ introduced in Section II. These phase functions depend in a simple way on the the semi-classical scattering data for the box potential  \cite{kenjen14}.  We shall  then verify that the functions $w_j$ generated by the phases $\Upsilon_j$: i) satisfy equations (\ref{tsarev}); and ii) provide, via the hodograph formulae (\ref{w0}), the required matching for $\a_j(x,t)$ on the breaking curve.  Below we present the  formal derivation of the modulation solution along these lines. The outline of the self-contained rigorous RHP construction underlying this derivation and explaining the connection between the modulation solution and the semi-classical RHP  can be found  in Appendix A.

We start with recalling that, for the solution $\psi (x,t; \eps)$ of the semi-classical NLS to have an  asymptotic representation in  the form of the modulated finite-band potential (\ref{fingap}),  (\ref{whitham-g}) depending on $g$  phases of the form $\eps^{-1}\eta_j=\eps^{-1}(k_jx+\omega_j t + \eta_j^0)$ (i.e. linear functions of $x$ and $t$) one must require that the ``initial phases'' $\eta_j^0$ are functions of $\bos \a, \bos{\bar \a}$  so that the general kinematic conditions  (\ref{kom_local}) are satisfied.  To this end we introduce  $\eta_j^0= -\Upsilon_j(\boldsymbol {\a, \bar \a})$ and use the first condition (\ref{kom_local}), 
\be
\frac{\partial (k_j x+ \omega_j t - \Upsilon_j)}{\partial x} = k_j , \quad j=1, 2,
 \ee
to obtain
 \be \label{TsaUps}
\frac{\part k_j}{\part \a_m}  x+\frac{\part \omega_j}{\part \a_m} t = \frac{\part \Upsilon_j}{\part \a_m} \, , \quad \hbox{and c.c.} ,\quad j,m=1,2,
 \ee
provided $\partial_x \a_j \ne 0$, $\partial_x \bar \a_j \ne 0$, $j=1,2$.  Here $k_j({\bos \a}, {\bos \bar \a})$ and $\omega_j(\bos{\a}, \bos{\bar \a})$ are defined by (\ref{kj}), and $\Upsilon_j$'s are yet to be found.  The second condition (\ref{kom_local}) leads to the same set of equations (\ref{TsaUps}). As we shall see, only half of the equations (\ref{TsaUps}) are independent, so it is sufficient to consider either $j=1$ or $j=2$.  We also note that equations (\ref{TsaUps}) admit a compact and elegant representation in the form of the stationary phase conditions:
\be \label{stat_phase}
\frac{\partial \eta_j }{\partial  \a_m} =0, \qquad \frac{\partial \eta_j }{\partial \bar \a_m} =0, \quad j,m=1,2.
\ee
For given functions $\Upsilon_j(\bos{\a}, \bos{\bar \a})$ equations (\ref{TsaUps}) fully define the modulations $\bos{\a}(x,t), \bos{\bar \a}(x,t)$ (assuming invertibility of (\ref{TsaUps}), which is not guaranteed {\it a-priori}). 
Comparing equations (\ref{TsaUps}) with  the hodograph solution (\ref{w0})  and using the representation (\ref{Vjg}) for the characteristic speeds $V_j(\bos \a, \bos{\bar \a})$ in (\ref{w0}) one readily makes the identification
\be \label{wUps}
w_m= \frac{\partial_{\a_m} \Upsilon_j}{\partial_{\a_m} k_j} \quad \hbox{and c.c.}, \quad j,m=1,2. 
\ee
Now we observe  that formula (\ref{wUps})  must yield the same function $w_m(\bos \a, \bos{\bar \a})$ for both values of $j$.  This is a consequence of the consistency of the genus two Whitham modulation system with two ``extra'' (wave number) conservation laws (\ref{wncl})  (the same argument  was used to establish the   ``nonlinear group velocity'' representation  (\ref{Vjg}) for the characteristic speeds of the Whitham modulation system). Thus, it is sufficient to consider only half of the equations (\ref{TsaUps}).

Within the RHP approach  the phases $\Upsilon_j$  are determined in terms of  $g+1$ functions $f_k(\z)$, $k=0, 1, \dots, g$ containing the information about the scattering data for the initial potential  (see Appendix A and \cite{el_tovbis_2016}). For $g=2$ we have
\be\label{Upsilon}
\Upsilon_j=\frac{1}{4\pi i}\sum_{k=0}^2\oint_{\gt_k}\frac{f_k(\z)p_j(\z)d\z}{\Rscr(\z)}, \quad j=1,2,
\ee 
where
\be \label{pj}
p_j(\l)=\k_{j,1}\l+\k_{j,2},
\ee
$\Rscr(\l) \equiv \Rscr_2(\l, \bos{\a}, \bos{\bar \a})$ is given by (\ref{rsurf}), and the coefficients $\k_{j,1}$, $\k_{j,2}$ are determined by conditions (\ref{holonorm}) with $g=2$.  
We have verified that functions $w_j$ defined by (\ref{w0}), (\ref{wUps}), (\ref{Upsilon})  satisfy Tsarev's equations (\ref{tsarev}) for arbitrary $f_k(\z)$, thus  defining the general {\it local} solution to the genus two Whitham equations.   We mention in passing that the quantities $d \Omega_j=p_j(\l) d \l /\Rscr(\l, \bos{\a}, \bos{\bar \a})$ represent the normalised holomorphic differentials playing important role in the construction of finite-band solutions of the NLS equation (see e.g. \cite{tracy_chen}). These differentials also serve as the generating functions for  solutions of the Whitham equations  associated with KdV hierarchy \cite{el96}, \cite{ekv2001}. One can see from (\ref{Upsilon}), (\ref{wUps}) that they play essentially the same role in the NLS  modulation theory.

For the box potential (\ref{ic1}) we have $\a_0=iq$ so $\Rscr_g (\l)=R(\l)\nu(\l)$, where
\be\label{Rnu}
R(\l)=\sqrt{(\la - \a_1)(\la - \bar \a_1) \dots (\la - \a_{g})(\la - \bar \a_{g})}, \quad  \nu(\l)=\sqrt{\l^2+q^2}.
\ee
The functions $f_k(\l)$ in (\ref{Upsilon}) can be  inferred from the results of \cite{kenjen14}:
\be \label{fk}
 f_0(\l) = f_1(\l)=-2L\l, \quad f_2(\l) =-2L\l+ 4L \nu(\l) \, .
\ee
Using (\ref{wUps}), (\ref{w0}) one can verify that the necessary matching conditions with the genus one solution (\ref{genus1}) on the second breaking curve are satisfied. In what follows we shall only need to explicitly verify these conditions at  the point of the dispersive dam break flow collision $x=0, t=L/2\sqrt{2}q$, which is the common point for $T_1$ and $T_2$. At this point one has $\a_1= - q / \sqrt{2}$ and $\a_2 = q/\sqrt{2}$ (see (\ref{T2})).

\subsubsection{Rogue wave formation}
\begin{figure}[ht]
\centerline{\includegraphics[height=2.5in]{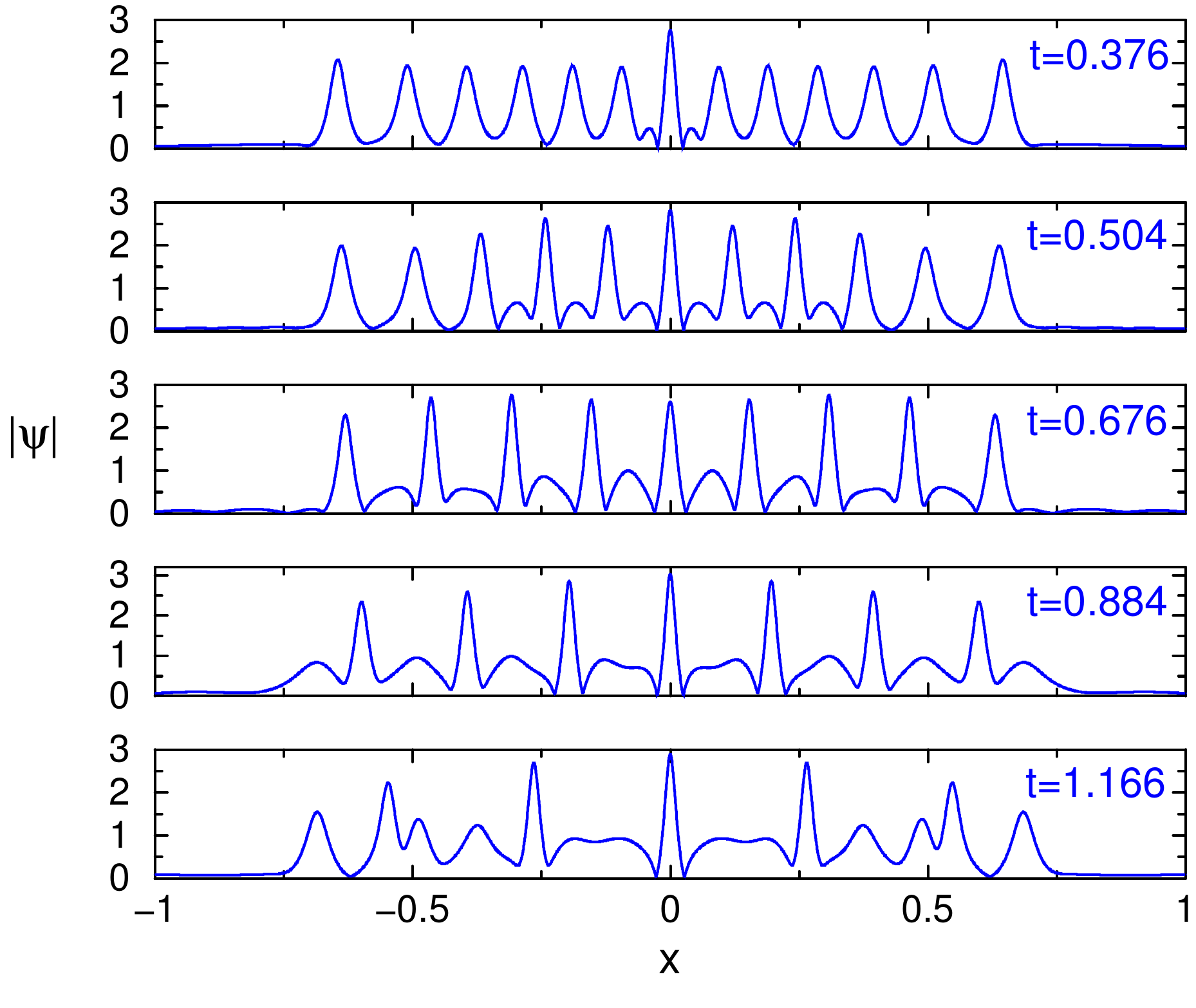} }
\caption{(Color online) Emergence and development of the modulated breather lattice ($g=2$) due to the interaction of dispersive dam break flows ($g=1$).  Numerical solution $|\psi(x,t)|$ of the  NLS equation (\ref{nls1}) with $\eps=1/33$ and the box initial conditions (\ref{ic1}) with $q=1$, $L=25/33$. See Fig.~\ref{fig5}a for the corresponding density plot.}
\label{fig8} \end{figure}

We shall call the modulated quasi-periodic  wave in the dam break flow interaction ($g=2$) region the {\it modulated  breather lattice},  by analogy with the term `modulated soliton lattice' used for the slowly varying finite-band solutions of the KdV equation \cite{dn89}.
We first present numerical solution for the amplitude $|\psi(x,t)|$ in the  interaction region. The spatial profiles shown in Fig. \ref{fig8} represent the snapshots of the  solution in Fig. \ref{fig5}a  taken at the times corresponding to the temporal maxima of the breather lattice.

We shall use the  modulation solution obtained in the previous subsection to analyse the dam break flow interaction dynamics. For that, we shall look at the temporal  behaviour of the modulations at $x=0$, which, due to the symmetry $x \to -x$ of the initial data, allows for  a significant simplification of the analytic expressions.  As we shall see, the modulation solution at $x=0$  provides  a major insight into the interaction dynamics in the entire genus two region.  From (\ref{TsaUps}) we obtain:
\be \label{modx0}
x=0: \qquad \frac{\part \omega_j}{\part \a_m} t =  \frac{\part \Upsilon_j}{\part \a_m} \, , \quad \hbox{and c.c.} ,\quad m,j=1,2.
\ee

As we have already mentioned, in view of the  symmetry $x \to -x$ of the box problem the solution at $x=0$ must also exhibit the spectral ``portrait'' symmetric with respect to the imaginary axis $\hbox{Re} [ \la] = 0$ so that we have $\a_2(0,t) = - \bar \a_1(0,t)$,  and the coefficients $\k_{i,j}$  entering the expressions for $\omega_{1,2}$ and $\Upsilon_{1,2}$ (see (\ref{kj}), (\ref{holonorm}) and  (\ref{Upsilon}), (\ref{pj})) can be evaluated as (see Appendix A) 

 \begin{equation}\label{kappa12}
x=0: \quad   \k_{1,1} = - \k_{2,1}= - \frac{1}{4 i}  \left( \int \limits_{q}^{\infty} \frac{z dz}{Q(z)\mu(z)} \, ,   \right)^{-1}  \, , 
\quad  \k_{1,2}=  \k_{2,2} = - \frac{1}{2 i}  \left( \int \limits_{-q}^{q} \frac{d z}{Q(z)\mu(z)}    \right)^{-1} ,  \end{equation}
where
\be \label{Qmu}
Q(z) = \sqrt{[(z-b)^2 +a^2][(z+b)^2 +a^2]}, \qquad \mu(z) = \sqrt{q^2  - z^2} \, .
\ee
As follows from our previous analysis, only half of the equations in the system (\ref{modx0}) are independent  so it is sufficient to consider only $j=1$ or $j=2$ which leaves us  with two c.c. equations. The symmetry $\a_2 = - \bar \a_1$ at $x=0$ reduces the number of independent  equations to just one.  As a result, the hodograph modulation solution (\ref{modx0}) can be represented in the form (see Appendix A for the outline of the calculation)

\be\label{abx0}
\frac{L}{2\pi}\int \limits_{-q}^{q} \frac{dz}{Q(z)\mu(z)} \int \limits^{\infty}_{-\infty} \frac{(z-b) - ia}{|z+i\a|^2}\frac{dz}{Q(z)}  
= \int \limits^q_{-q} \frac{(z-b) - ia}{|z+i\a|^2 } \frac{d z}{Q(z) \mu(z)}
\left( -\frac{t}{2} + \frac{L}{2\pi} \int \limits^{\infty}_{-\infty} \frac{dz}{Q(z)}\right) ,
\ee
where
\be \label{abt}
\a(t) = a(t)+ib(t) =\a_2(0,t)  .
\ee
It is not difficult to verify that at the collision point $t=L/2\sqrt{2}q$ one has $a=q/\sqrt{2}$, $b=0$ as required by the matching conditions (\ref{T1}) and (\ref{T2}). Thus, the obtained solution (\ref{abx0}) indeed satisfies the matching conditions at the breaking curve $T_1$.
\begin{figure}[ht]
 \includegraphics[height=2.in]{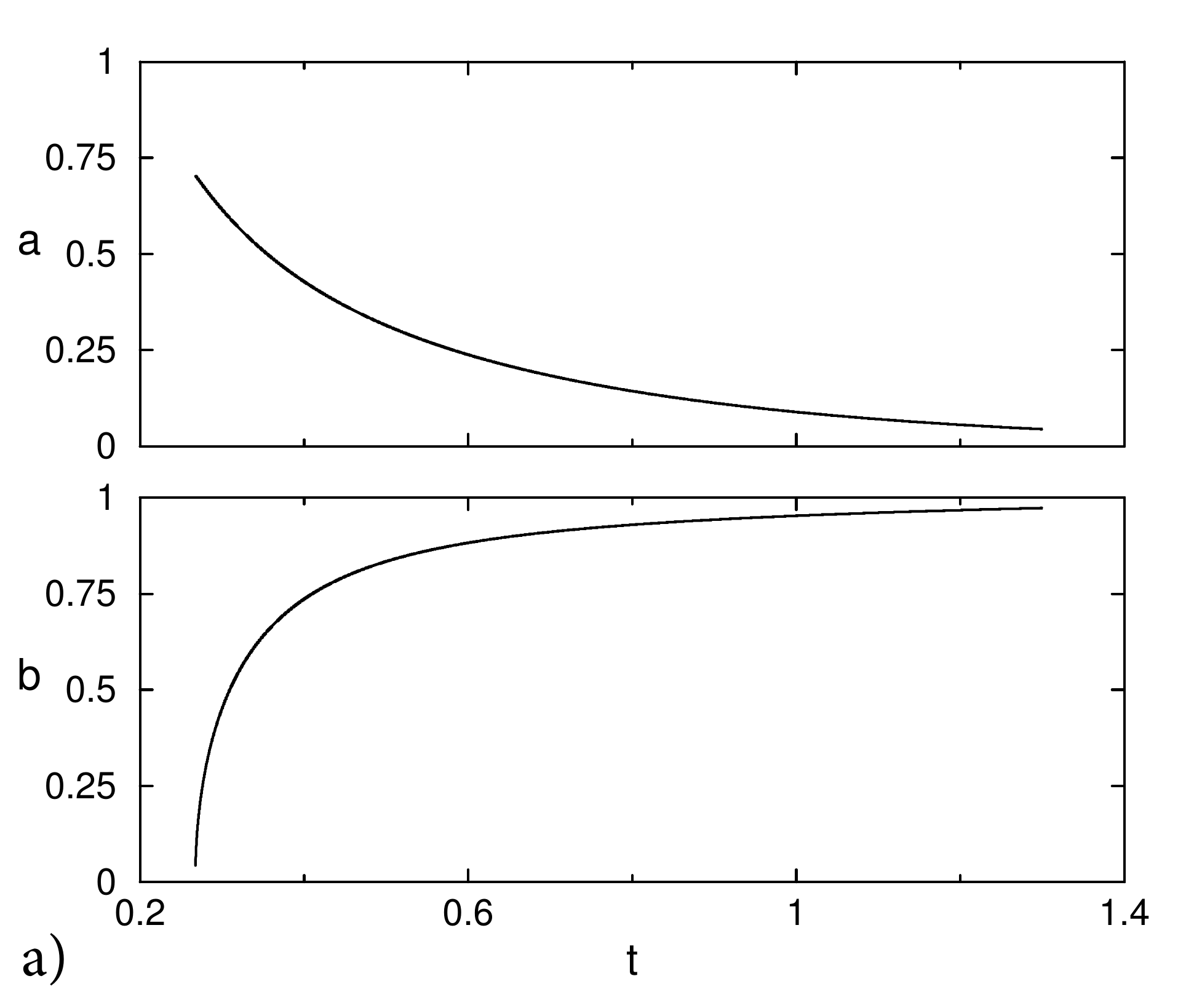}  \quad  \quad \includegraphics[height=2.in]{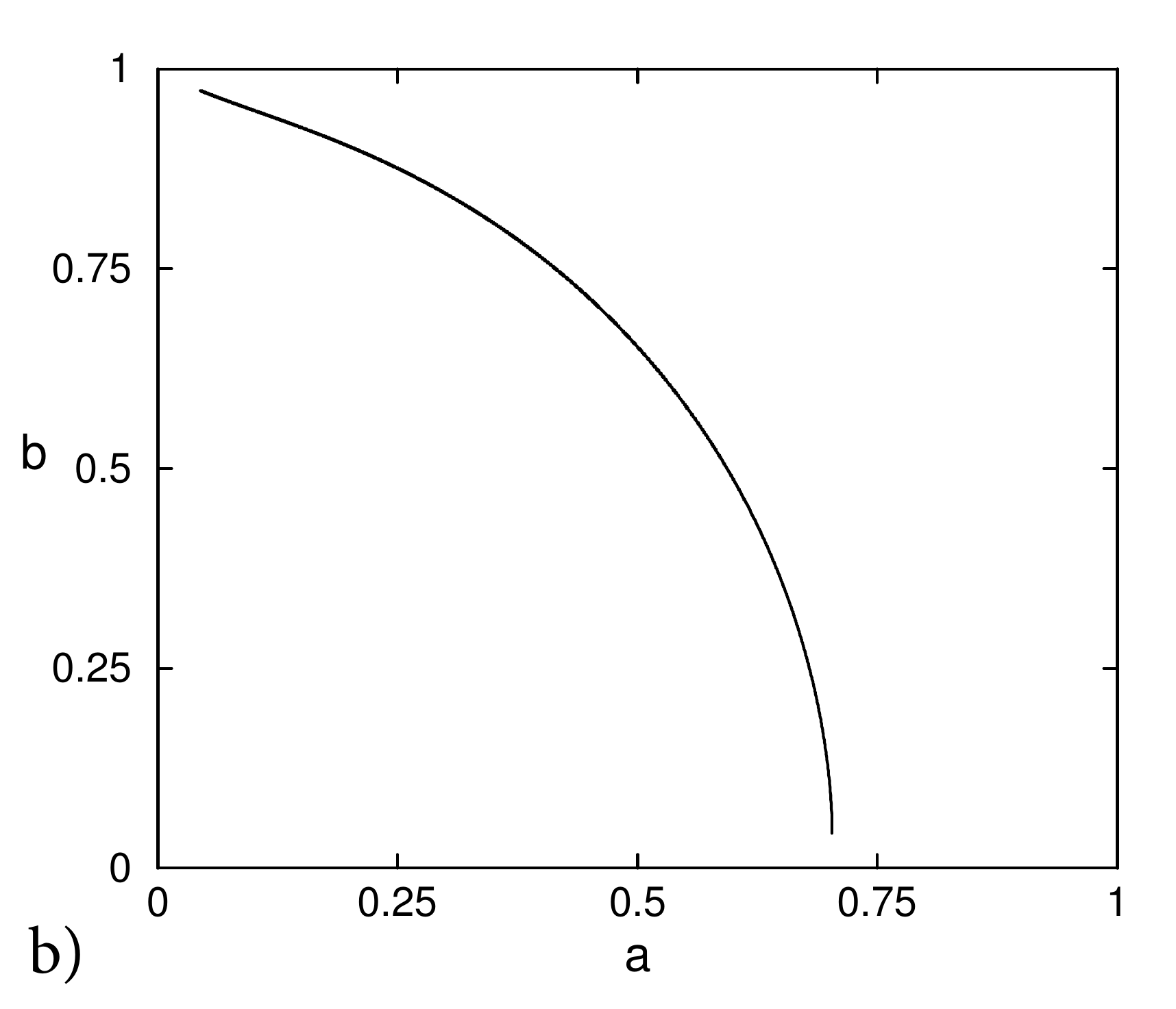}
 \caption{Evolution of the spectral branch point $\alpha_2 = - \bar \a_1= a+ ib $ at $x=0$ in the genus two interaction region -- formula (\ref{abx0}) with $q=1$, $L=25/33$, as in Figs. \ref{fig5}, \ref{fig8}. The evolution starts on the real axis, $b=0$, $a=1/\sqrt{2}$ at $t=t_0=L/2\sqrt{2} \approx 0.27$.  One can see that $b$ rapidly grows with time while $a$ decreases, so both  branch points $\alpha_2$ and $\a_1$ approach the imaginary axis close to $b =1$
 at the end of the genus two region. a) Dependencies $a(t)$ and $b(t)$; (b) Trajectory of $\a_2=-\bar \a_1$ in the complex plane.}
 \label{fig9}
 \end{figure}
 
Separating real and imaginary parts in (\ref{abx0}) we obtain a system of two equations for  $a(t)$ and $b(t)$. We then solve the obtained system numerically, using the Broyden method \cite{Broyden}.  This  method  is a generalization of the secant method,  also known as a quasi-Newton method, to systems of nonlinear equations, and involves the Jacobian matrix of the system.
 Instead of the analytical evaluation
of derivatives  in (\ref{abx0}), we constructed an approximation to the Jacobian matrix, 
which was updated at each iteration. The initial Jacobian can be set as the identity matrix or a
finite difference approximation, like in the first-order forward difference method. The iterations stop when some tolerance value, e.g. $10^{-8}$, is reached.  The resulting plots $a(t)$, $b(t)$ and $b(a)$ are presented in Fig.~\ref{fig9}. One can see that  the imaginary part of $\a$ rapidly grows, while the real one decreases.

 \begin{figure}[ht]
\includegraphics[height=2.in]{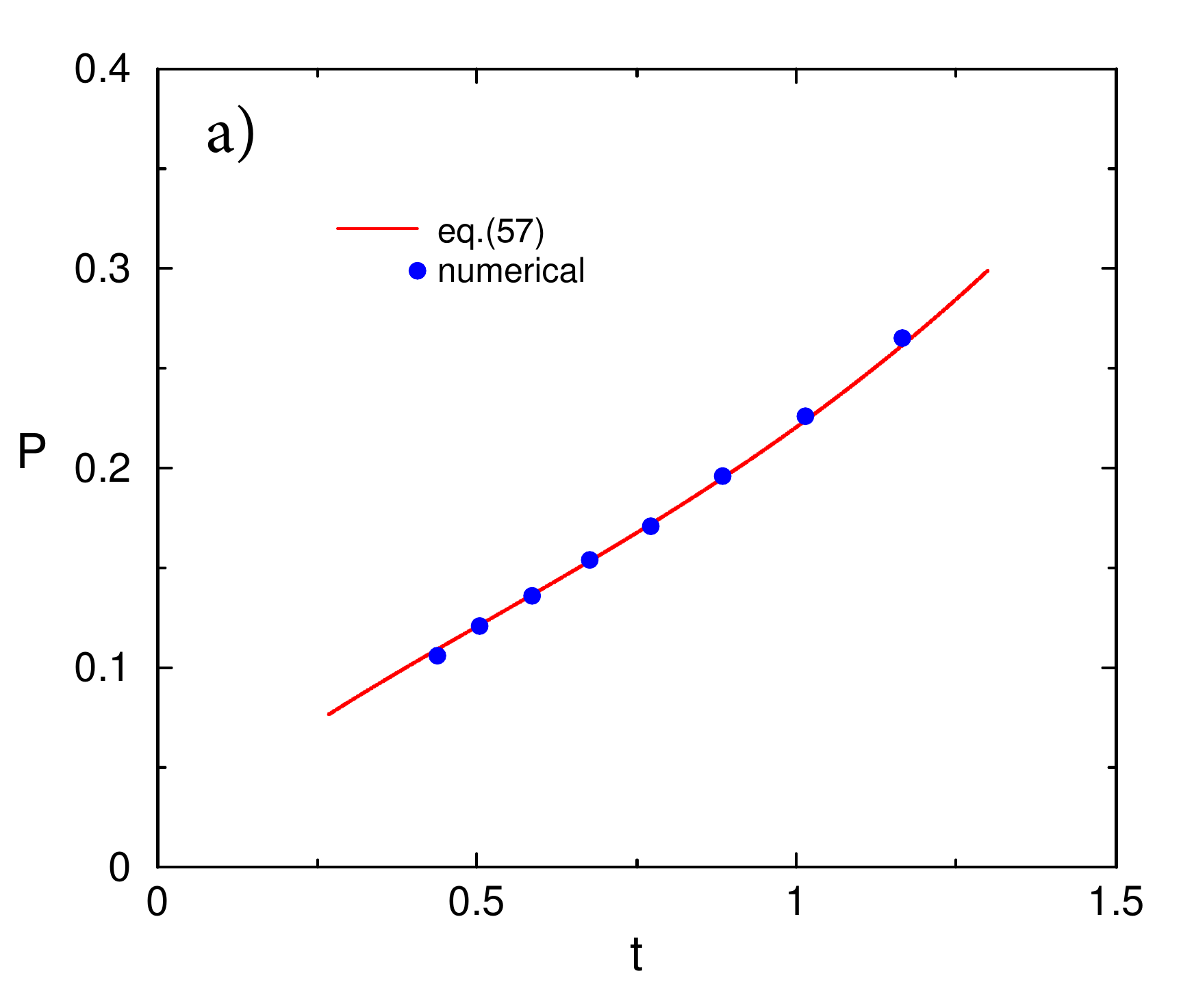} \qquad  \includegraphics[height=2.in]{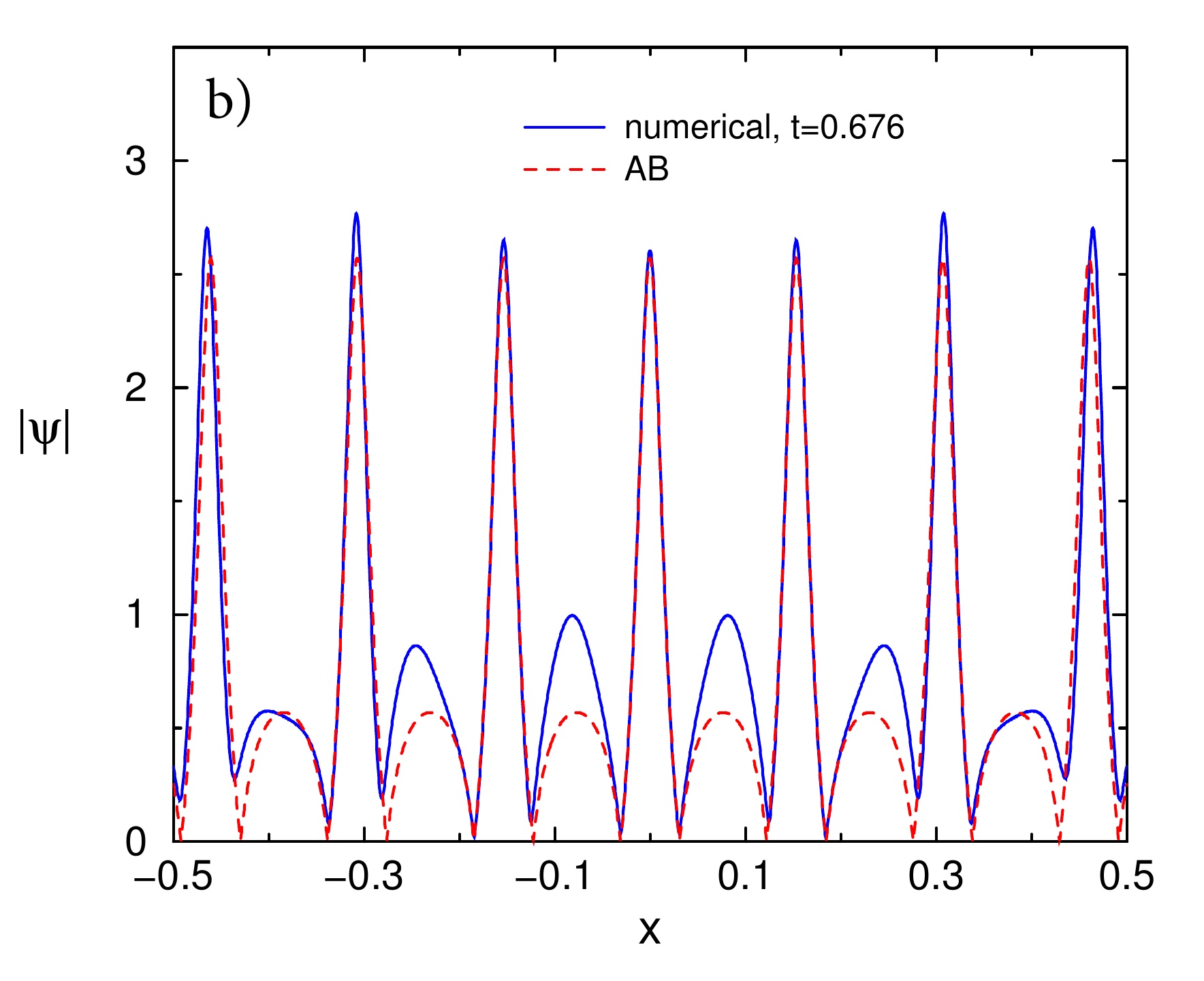}
\caption{(Color online) a) Dependence of the spatial period (the wavelength) on time in the genus two interaction region at $x=0$. Solid line -- formula (\ref{periodx0}); dots -- the average distance between the peaks of the breather lattice adjacent to the central peak at $x=0$ in the numerical solution of the NLS box problem ($q=1,$ $L=25/33$, $\eps=1/33$), at different moments of time; b) Comparison of the numerical solution for the amplitude $|\psi|$ in the box problem at $t=0.676$ (solid line) with the (shifted) amplitude profile of Akhmediev breather  (\ref{AB}) with the spatial period $2\pi \eps/p=P(0.676)=0.154$ found from (\ref{periodx0}) (dashed line). }
\label{fig10} \end{figure}
 \begin{figure}[ht]
  \includegraphics[height=2.in]{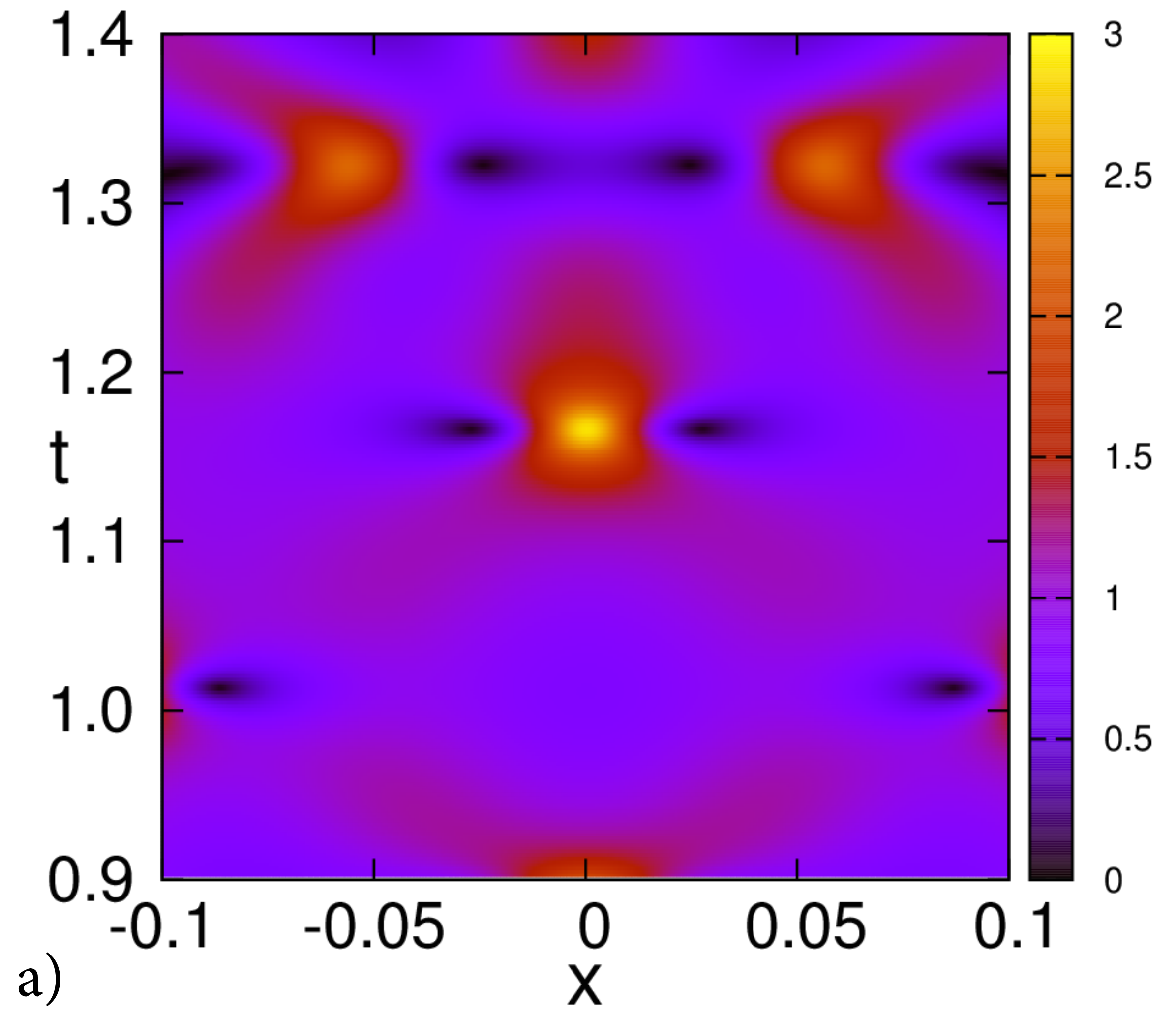} \qquad \includegraphics[height=2.in]{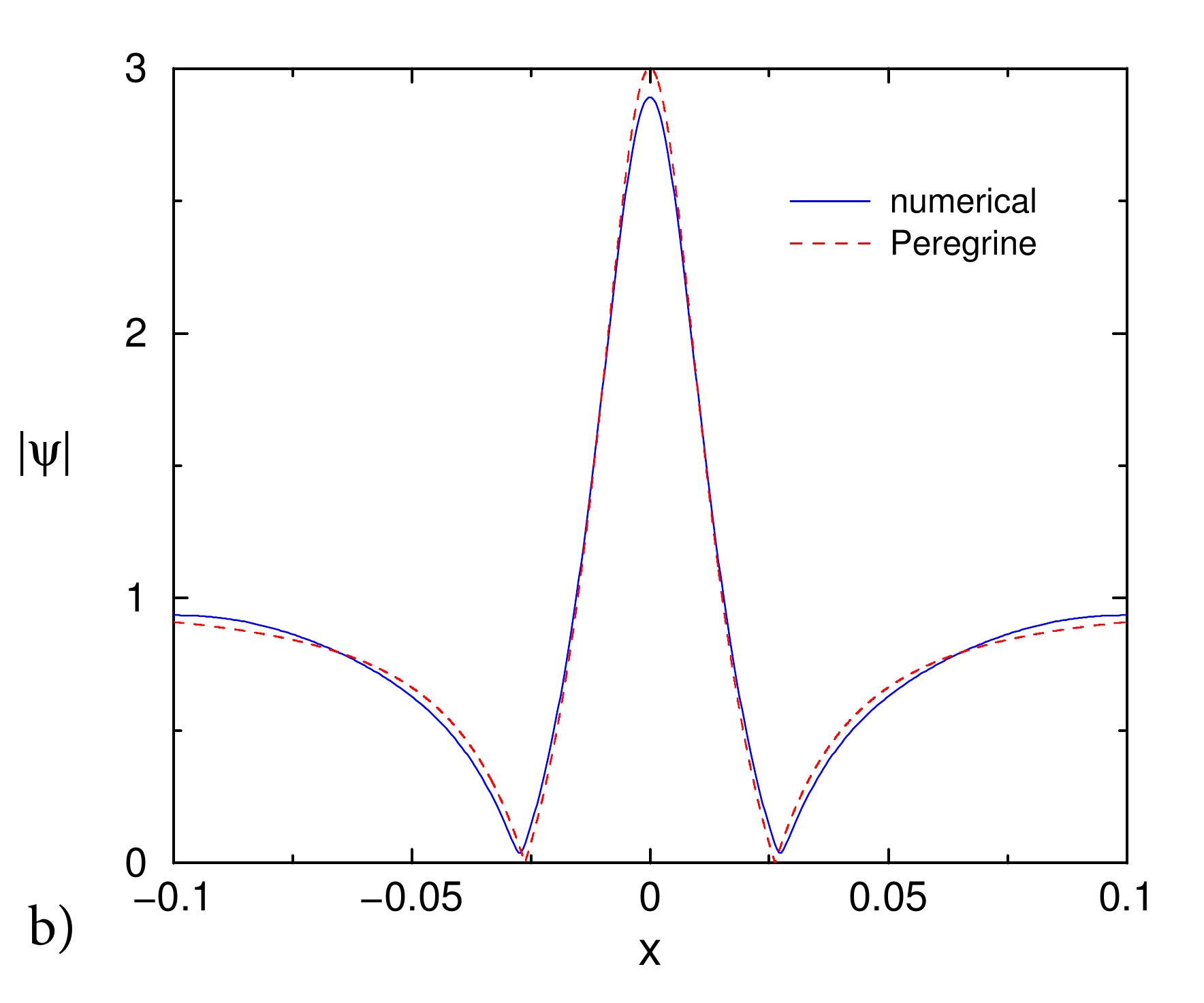}
 \caption{(Color online) a) Density plot for  $|\psi|$ in the NLS box problem with $q=1$, $L=25/33$, $\eps=1/33$ -- the zoom in on the region of the rogue wave formation at $t=1.166$; b) The rogue wave profile:   solid line -- the numerical solution of the box problem zoomed in near  $x=0$, $t=1.166$;  dashed line -- the Peregrine breather (\ref{per}) for $q=1$. }
 \label{fig11}
 \end{figure}
As follows from  (\ref{kj}), (\ref{kappa12}),   the $\eps$-normalised wavenumbers and frequencies are calculated at $x=0$ as $k_1=-k_2 =-4\pi i \k_{1,1}$ and  $\omega_1=\omega_2=-4\pi i \k_{1,2}$ respectively. Thus, in the vicinity of $x=0$ the wave can be viewed as periodic with the spatial $P$ and temporal (`breathing') $T$ periods  slowly depending on time
 \be \label{periodx0}
P(t)=  \frac{2 \pi \eps}{k_1}=
 2 \eps \int \limits_{q}^{\infty} \frac{ z d z}{Q(z)\mu(z)} , \qquad    T(t)= \frac{2\pi \eps}{\omega_1}= \eps \int \limits_{-q}^{q} \frac{ d z}{Q(z) \mu(z)}\, .
\ee
Note that $P(t)$ and $T(t)$ have the meaning of the {\it local}  (i.e. defined at a point $(0,t)$) periods  of the slowly modulated finite-band solution --  see the definition (\ref{kom_local}) of the local wavenumbers and local frequencies. Thus, the solution at $x=0$ considered  for any fixed time $t$ within the genus two region has a {\it single} local spatial period and thus, can be approximated by an appropriate periodic NLS solution in some vicinity of $x=0$.   The natural candidate for such an approximation  is the Akhmediev breather (AB)  (\ref{AB}), which is a limiting wave form of the two-gap NLS solution and has a single spatial period $2 \pi \eps /p$ (we note that AB (\ref{AB}) is also a single-parameter solution so its period also defines the amplitude). To this end we use the dependence $P(t)$ as the period for the AB and compare the 
profiles of intensity in the quasiperiodic breather lattice observed in the numerical solution (see Figs.~ \ref{fig5}, \ref{fig8}) with the spatial profile of the AB (\ref{AB}) with the  period $2\pi \eps /p=P(t)$ and appropriately chosen phase. Such a comparison for $t=0.676$ ($P=0.154$) is shown in Fig. \ref{fig10}b. One can see excellent agreement for the amplitudes, positions and detailed profiles of the main peaks (obviously some fitting of the AB phase was necessary). Remarkably, the agreement  within the genus two region remains very good (and even improves with respect to the lower maxima) away from $x=0$. Thus, the amplitude profile of the breather lattice in the bulk of the genus two region can be approximated by the modulus \blue{$|\psi|$} of  \blue{the} modulated ``time-periodic AB'' with the slowly varying spatial and temporal periods given by (\ref{periodx0}). 
\begin{figure}[ht]
\centerline{  \includegraphics[height=2.5in]{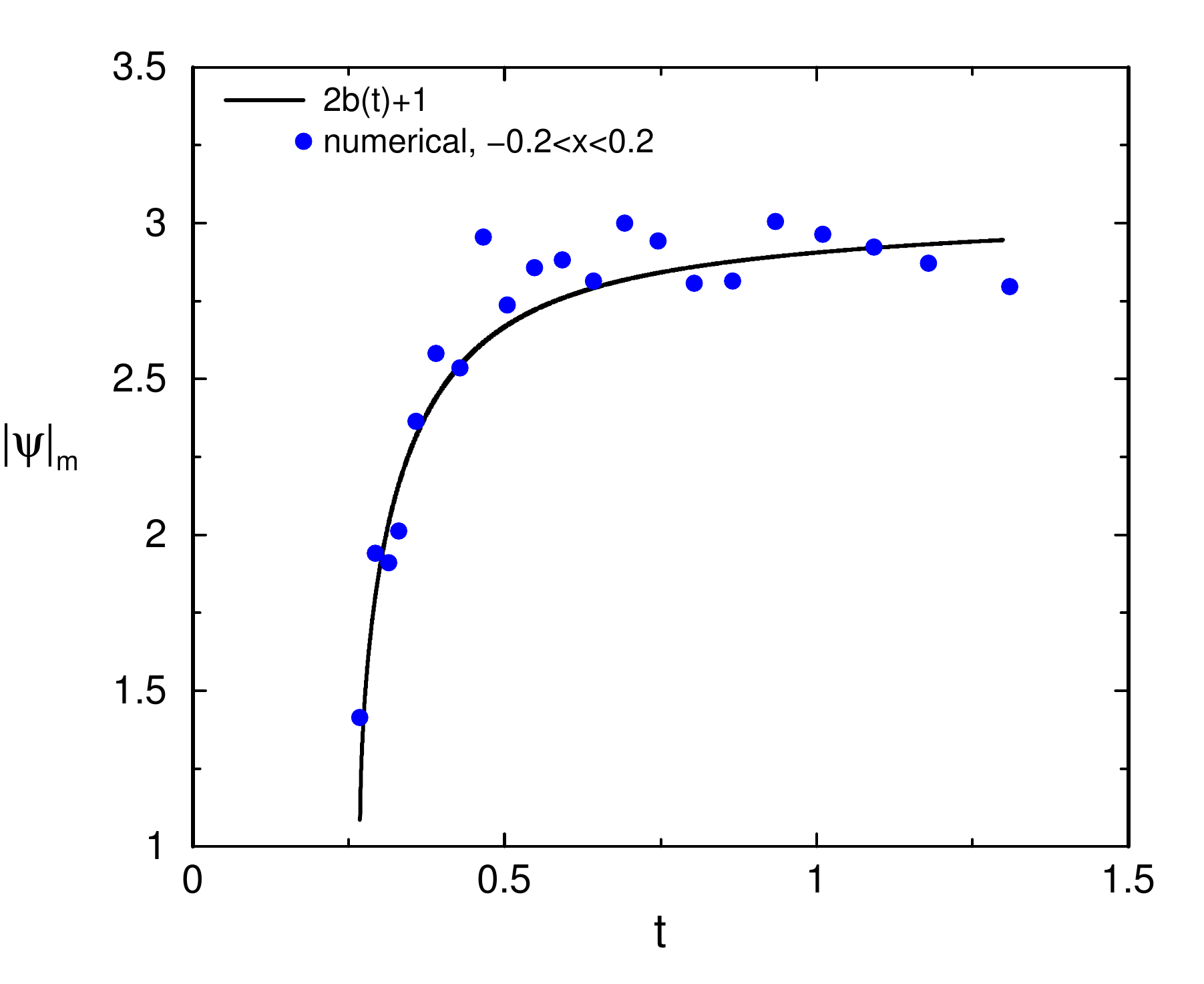}}
 \caption{(Color online) Maximum amplitude $|\psi|_m$ as function of time in the genus two interaction region. Solid line: the estimate $2b(t)+1$ for $|\psi|_m$ constructed from the modulation solution (\ref{abx0}) at $x=0$; Dots: the values of $|\psi|_m$ extracted from the numerical solution of the NLS box problem with $q=1$, $L=25/33$, $\eps=1/60$ (see Fig. \ref{fig5}b) within the strip $-0.2< x < 0.2$.}
 \label{fig12}
 \end{figure}
   
As one can see, the spatial period  $P$ increases with time which is indicative of the tendency of the approximating AB  breather towards the Peregrine soliton. Indeed, from the trajectory of the branch point $\alpha$  in the complex plane shown in  Fig. \ref{fig9}b it is seen that near the upper boundary of the genus two region, ($t \approx 1.25$ -- see Fig. \ref{fig5}), the imaginary part $b$ of the moving branch point approaches the value $q=1$ whereas $a$ approaches zero. Since $\a = \a_2 = - \bar \a_1$  one can conclude that the spectral portrait of the solution approaches that of the Peregrine soliton (two double points placed at the endpoints $\pm iq$ of the basic branchcut). Indeed, the comparison of the wave form of the large oscillation observed in the numerical simulations at $t=1.16$, $x=0$ with the plot of the amplitude $|\psi|$ for the Peregrine soliton (\ref{per}) shows excellent agreement  -- see Fig. \ref{fig11}b (we have also checked the agreement for the phase but do not show it on the plot).

Finally, we study the behaviour of the maximum wave height as a function of time  within the genus two region. The maximum wave hight in a (non-modulated)  finite-band NLS solution with $ 0 \le g \le 2$ can be found from the simple formula $|\psi|_m=\sum b_j$, where $b_j = \hbox{Im}[ \a_j]$ \cite{wright}. This formula is obvious for $g=0,1$ (see (\ref{cnoidal})). It can also be readily obtained for $g=2$ in the particular case when $b_1=b_2$ (see e.g. \cite{osborne_book}). To this end we plot the function $1+2b(t)$ using the solution $b(t)$ of the modulation equation (\ref{abx0}) and compare the result with the values of $|\psi|_m$ extracted from the numerical solution of the box problem for the NLS equation (\ref{nls1}) with very small  $\eps=1/60$ (see Fig.~\ref{fig5}b for the corresponding amplitude density plot)  in the strip $-0.2< x < 0.2$ containing  at least three peaks in an  $\eps$-neighbourhood of most points $(0,t)$ within the $g=2$ region. The comparison is presented in Fig. \ref{fig12}. One can see that the curve  $1+2b(t)$ provides a very good approximation for the dependence of $|\psi|_m(t)$, which exhibits  rapid growth within the two-phase interaction region, further supporting the proposed mechanism of the rogue wave formation.  

\subsubsection{Long-time behaviour}
Of particular interest is the long-time behaviour of the solution to the semi-classical NLS box problem.  Here we only present a hypothesis about the asymptotic structure of the solution based on the results of  the numerical simulations,  leaving a detailed analytical study to  future work. 

The small-dispersion NLS evolution in the box problem (\ref{nls1}), (\ref{ic1}) with zero initial `velocity' $u$  has two key macroscopic features: (i) the oscillations for all times are confined to the spatial domain $-L < x < L$; and  (ii) the solution genus (the number of nonlinear oscillatory modes) grows with time.  Both features are illustrated in the diagram in Fig.~\ref{fig6}a. Our numerical simulations 
suggest that  the pattern of the  $x$-$t$ plane  splitting into the regions of different genera shown in Fig.~\ref{fig6}a persists as time increases (we ran the computations up to about $t=6$ for the box problem with $\eps = 1/33$, $q=1$, $L=25/33$). Motivated by this observation, we put forward a plausible hypothesis that for $ t \gg 1$  the solution genus  $g \sim t$, as long as $t \ll \eps^{-1}$. Then a pertinent question arises: what is the long-time asymptotic distribution of the spectral bands in the complex plane?  In the KdV theory the consideration of the thermodynamic type infinite-genus  limit of finite-band potentials \cite{ekmv01} and of the associated Whitham equations \cite{el03} has lead to  the kinetic description of a soliton gas \cite{ek05}, \cite{ekpz2011}. The focusing NLS counterpart of this theory would include the `breather gas' description which is yet to be developed. The long-time asymptotics of the NLS box problem, thus, could provide insight into the properties of  strong integrable NLS turbulence, which has recently become  the subject of an active research (see  \cite{agaf_zakh2014} and  \cite{suret2014} for the recent numerical and experimental results respectively). 
\begin{figure}[ht]
\includegraphics[height=2.in]{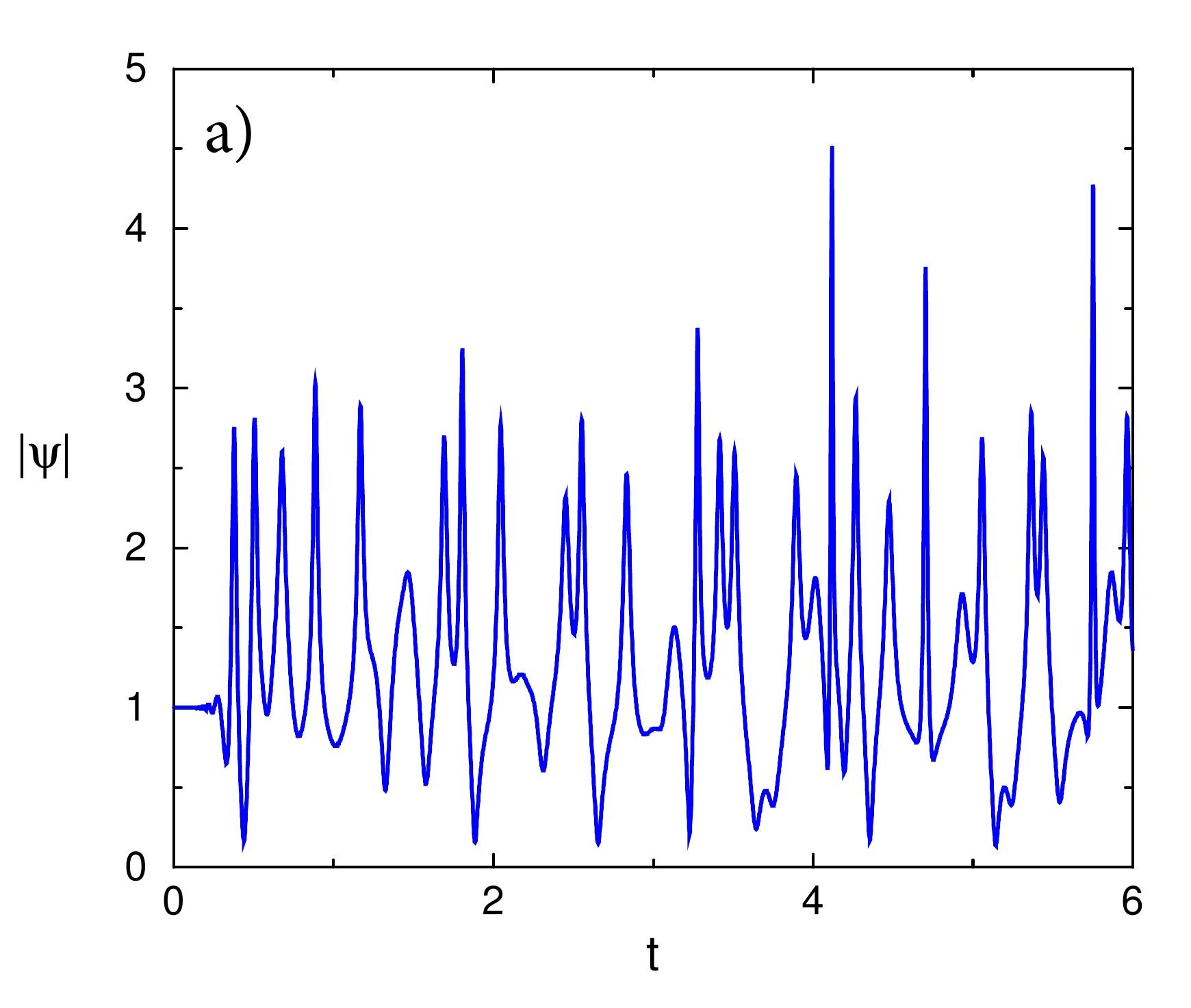} \qquad \includegraphics[height=2.in]{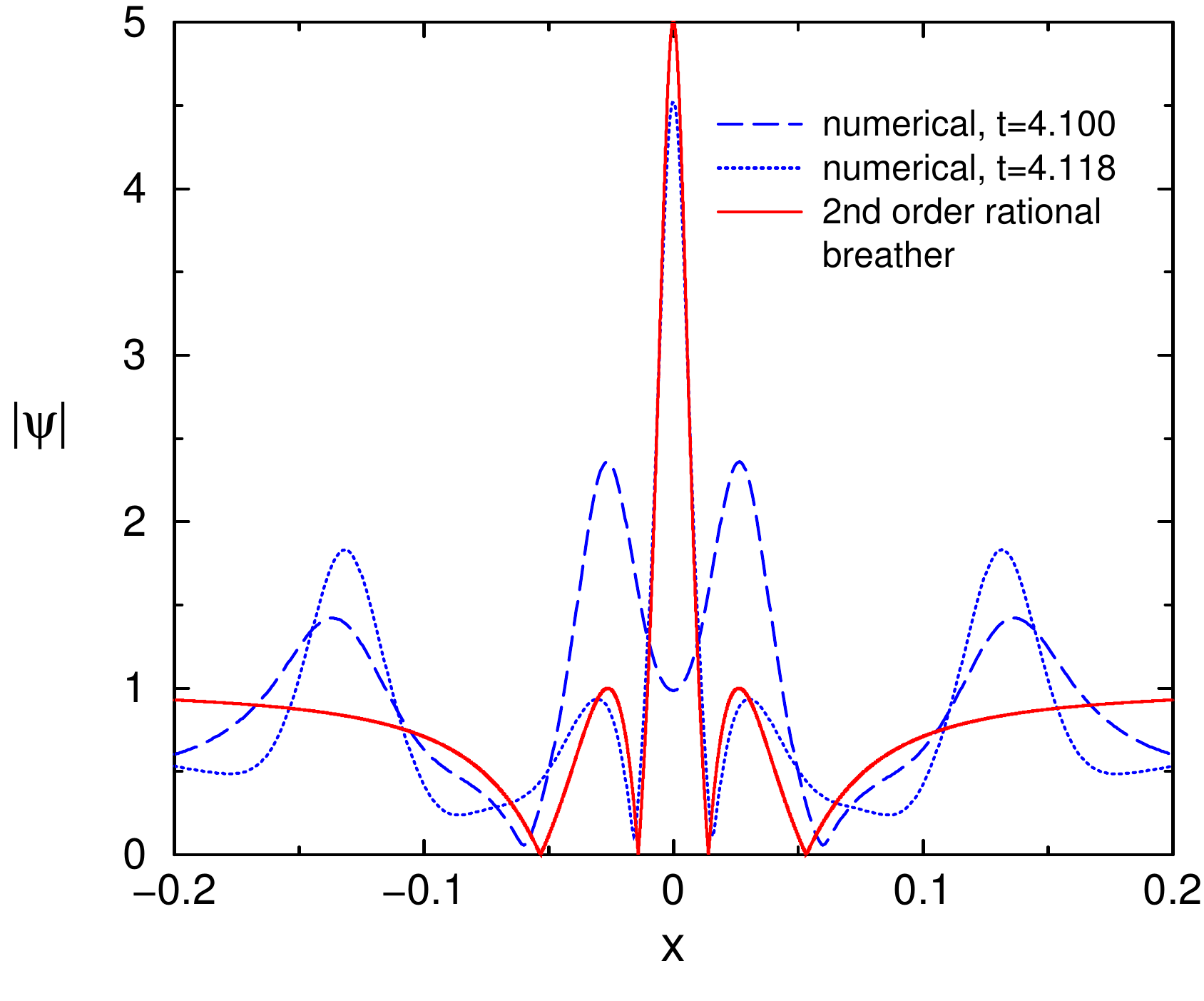} 
\caption{(Color online) Formation of higher order rogue  waves in the long-time evolution in the NLS box problem  with $q=1$, $L=25/33$, $\eps=1/33$. a) Dependence of the wave amplitude $|\psi|$ on $t$ at $x=0$. One can see the presence of the rogue wave with $|\psi|_m \approx 4.5$ at $t =4.118$; b) Profile of the higher-order rogue wave: numerical solution of the box problem at $t=4.100; 4.118$ (dashed and dotted-dashed lines respectively) near $x=0$ and second-order rational breather solution (\ref{2nd_order}) (solid line).}
\label{fig13} \end{figure}

In the higher-genus ($g>2$) regions generated in the course of the evolution in the NLS box problem, there arises a possibility of the formation of higher-order rogue waves with the maximum height significantly exceeding that of the `regular' rogue waves.  Probably the simplest example of such a ``super rogue wave'' is the second-order rational breather solution of the NLS equation, which has the form \cite{akhmediev_taki_09}, \cite{akhmediev_PRE09}
\be \label{2nd_order}
\psi=q\left[ 1- 4 \frac{G+iH}{D}  \right]e^{iq^2t/\eps},
\ee
where $G$, $H$ and $D$ are given by:
\be \label{5breather}
\begin{split}
G=-\frac{3}{16}+\frac{3}{2}X^2 +X^4 + \frac{9}{2}T^2+6V^2T^2 + 5X^4 \,  ,  \\ \quad H=T \left(-\frac{15}{8}
-3X^2 +2X^2 + X^4 +4X^2T^2 + 2T^4 \right), \\
D=\frac{3}{64} + \frac{9X^2}{16}+\frac{X^4}{4} + \frac{X^6}{3} +\frac{33}{16}T^2 - \frac{3}{2}X^2 T^2 + X^4T^2 + \frac{9}{4}T^4 + X^2T^4 + \frac{T^6}{3} \, .
\end{split}
\ee
Here $X=x/\eps$, $T=t/\eps$. The maximum hight of the breather  (\ref{5breather}) is $5q$. 

Our numerical simulations show that the higher order rogue wave indeed appear in the NLS box problem. In Fig.~\ref{fig13}a the plot of the numerical solution for the  amplitude $|\psi(0,t)|$ is presented for the box problem with $\eps=1/33$, $q=1$ and $L=25/33$ in the interval $0 < t <6$. One can see a very high peak with $|\psi|_m \approx 4.5$ at about $t = 4.1$. The comparison of the wave profile at $t=4.118$ with the second-order rational breather profile (\ref{5breather})  is shown in Fig.~\ref{fig13}b and demonstrates good agreement. This `super rogue wave' can be interpreted as a result of the collision of two lower-order breathers shown in dashed line corresponding to the solution at $t=4.100$, just prior the formation of the higher order rogue wave. A more accurate interpretation of this effect is the interaction of nonlinear modes within the modulated $g$-phase ($g$-band) solution.  In this regard it is worth noting that,  while the local profile of this solution around $x=0$ is reasonably close to the rational breather, it is not an exact solitary wave (see also  the comparisons with Akhmediev and Peregrine breathers in the $g=2$ region).   Importantly, the emergence of a single large amplitude oscillation within a multiphase solution (\ref{fingap}) at a particular $x,t$-point requires that  a certain precise relationship between the phases $\eps^{-1}\eta_j$ is satisfied, and so is sensitive to changes of $\eps$.  This sensitivity is expected to increase with growth of $g$.  Thus, the exact prediction of the emergence of higher-order rogue waves in the regions with sufficiently large $g$ is impractical and should be replaced with a statistical description within the general integrable turbulence theory even though the original formulation of the semi-classical NLS box problem is purely deterministic. This is in line with the proposition made in the beginning of this section that the long-time asymptotic behaviour of the semiclassical NLS  should be generally described in statistical terms.  In this connection we  note that the statistical description of the long-time asymptotic solution of the small-dispersion KdV equation with deterministic initial conditions defined on the entire $x$-axis was considered in \cite{gurzybel99}, \cite{gurzybmaz00}.

\section{Effects of perturbations on the semi-classical evolution}
In Sections II - IV we have described an analytically tractable scenario of the rogue wave generation in the framework of the semi-classical focusing NLS equation with the inital data in the form of a real-valued rectangular potential (the `box').    In practice (physical or numerical experiment) this  idealised  scenario may be affected by at least two factors: (i) the presence of a noise (physical or numerical); and (ii) higher-order physical effects (e.g. Raman scattering,  saturable nonlinearity etc). The first factor is inevitably present in any physical system and, due to modulational instability, will impose natural restrictions on the admissible values of $q$ and $L$ characterising the initial box potential. The second factor generally destroys integrability of the NLS equation and thus, can affect the very existence of the multi-phase solutions and the corresponding semi-classical limits. 
While, obviously, the quantitative effect of both factors depends on their magnitudes, it is important to understand what qualitative changes may occur in the system due to their presence.   While the detailed study of this important issue is beyond the scope of this paper, we present below some estimates and numerical simulations illustrating the effects of the noise and non-integrable perturbations on the solution of the NLS box problem.

\begin{figure}[ht]
 \includegraphics[height=2.in]{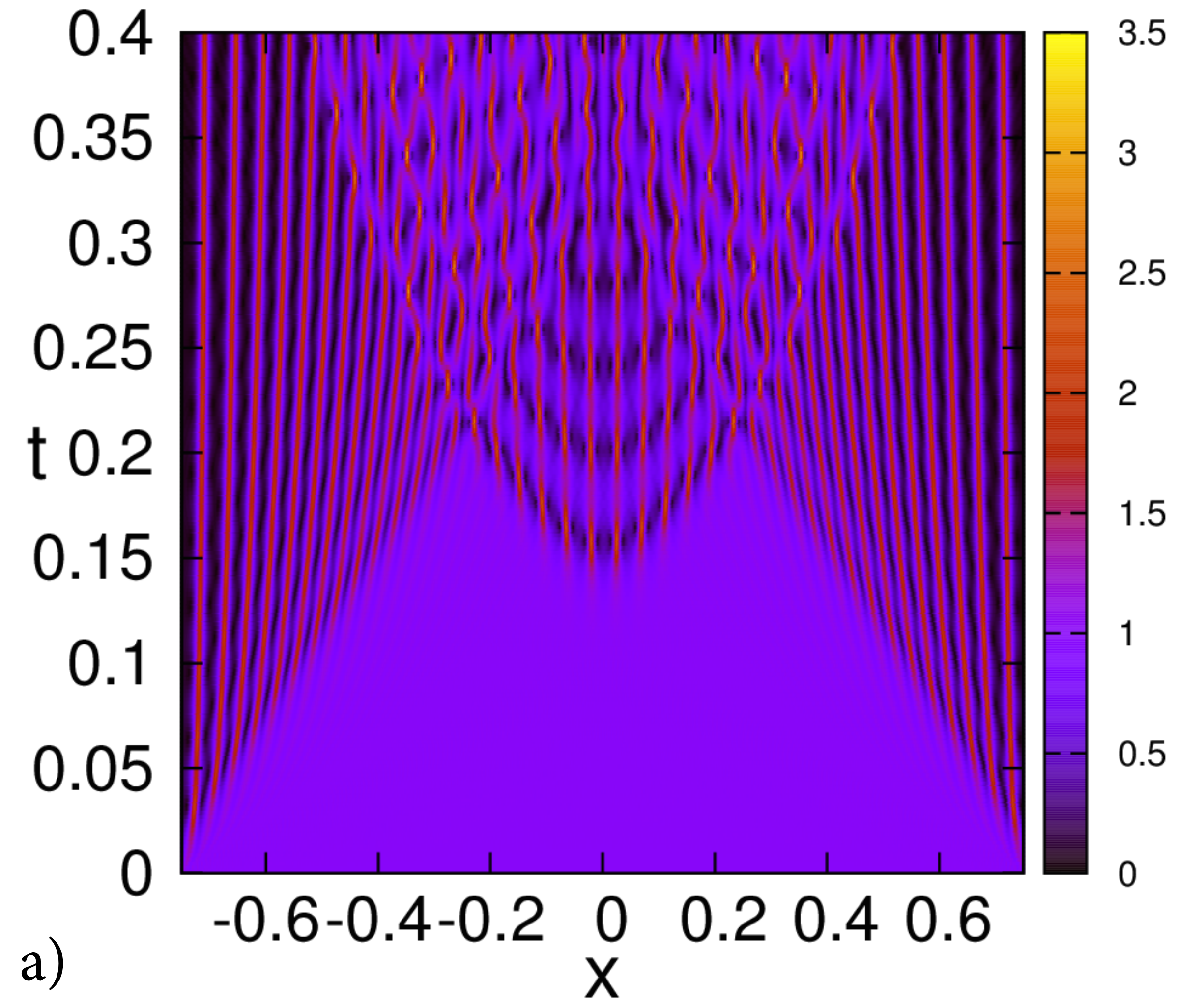}   \qquad  \includegraphics[height=2.in]{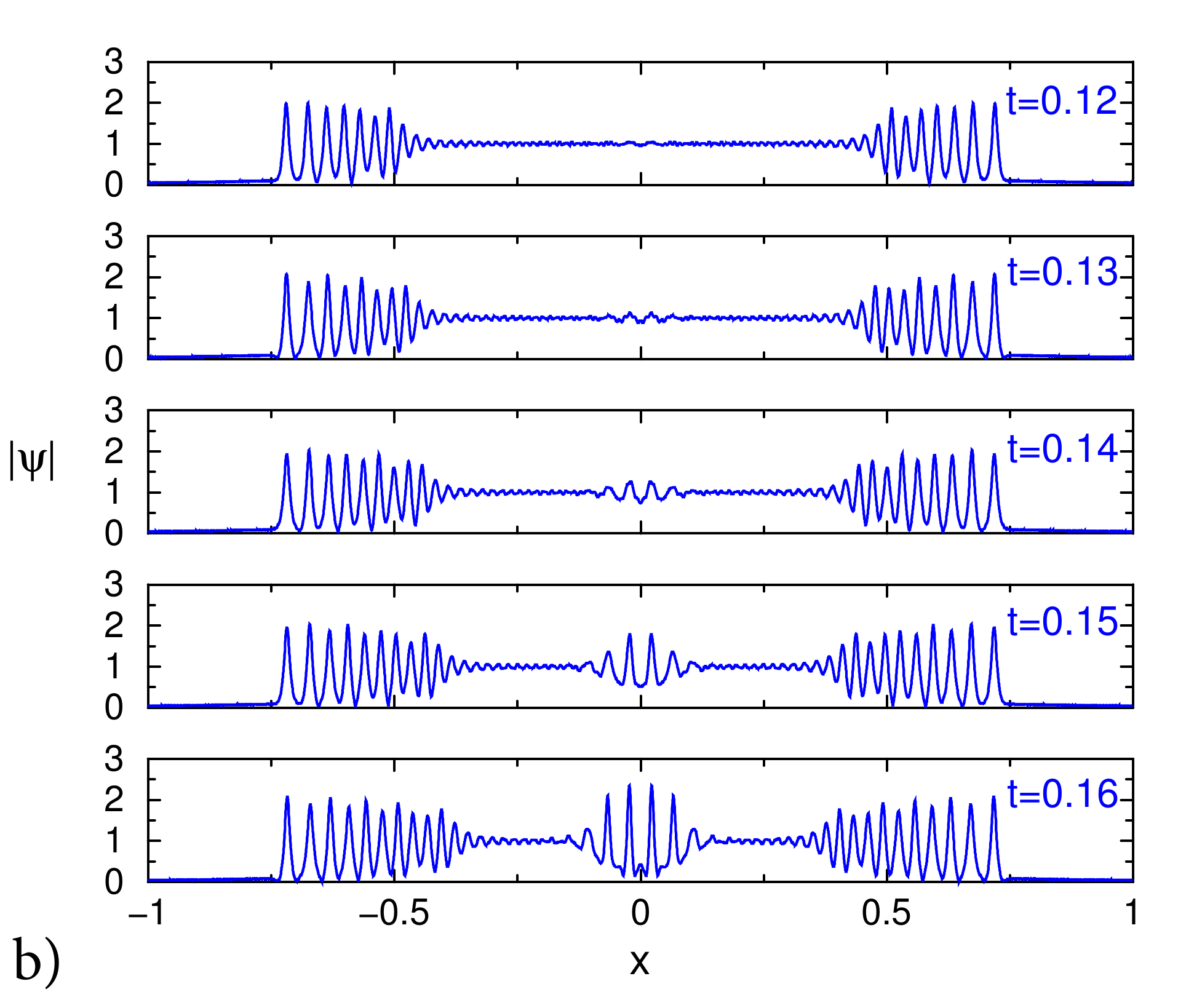}
 \caption{(Color online) ``Growing the beard'':  the effect of the noise on the semi-classical box evolution. Numerical simulation of the NLS box problem with $q=1$, $L=25/33$, $\eps=0.01$. a) density plot for $|\psi|$;
 b) amplitude profiles for different $t$.}
 \label{fig14}
 \end{figure}

The effect of the external noise on the  dispersive dam break flow evolution was briefly discussed at the end of  Section III (see formula (\ref{tm}) for the estimate of the dispersive dam break flow lifetime
due to the development of modulational instability of the external condensate (plane wave)).
In the box problem involving the generation of two dispersive dam break flows, the presence of the noise will impose some restrictions on the admissible initial box parameters for which the semi-classical NLS description is valid in ``practical terms''. E.g. if the box is too wide or too tall, the noise perturbations of the condensate in the central part of the box  will have enough time to develop before the harmonic edges of the counter-propagating dispersive dam break flows will meet at $x=0$ and saturate the instability. In Fig. \ref{fig14} the amplitude density plot is presented for the NLS evolution of the initial box with our standard parameters $q=1$, $L=25/33$ but with  $\eps=0.01$ in the NLS equation (\ref{nls1}), which is significantly smaller than our usual value $\eps=1/33$. One can see an extra oscillatory structure forming in the central part of the box, inside the region $g=0$.  Apparently, a related phenomenon was observed  in the numerical simulations in \cite{Clarke-Miller} where it was aptly named ``the beard''.  In \cite{Clarke-Miller}  the beard phenomenon was ascribed to non-analyticity of the initial data. Our numerical experiments with the box initial conditions suggest that the origin of the beard phenomenon lies in the development of the modulational instability due to the presence of a numerical noise. 
%\red{Indeed, the wave pattern of the ``beard'' is consistent with the theoretical predictions of \cite{biondini_prl_2016} establishing the universal nature of the nonlinear stage of modulational instability.}

We use the estimate (\ref{tm}) for the lifetime $t_m$ of the dispersive dam break  flow to obtain  an estimate for the ``critical'' parameters $L_c$, $q_c$ of the initial box with respect to the beard formation phenomenon. 
For that, we use the balance relation $t_m=t_0$, where $t_0=L/2\sqrt{2}q$ is the time at which  the collision of two dispersive dam break flow occurs at $x=0$ (see (\ref{T1})) so that the modulational instability of the plane wave in the genus zero region is saturated by the formation of breather lattices. As a result we obtain the critical value of the initial ``mass'' $A=qL$: 
\be\label{critbox}
A_c=(q L)_c=\sqrt{2} \eps \ln\frac{1}{\delta} \, ,
\ee
where $\delta$ is the typical amplitude of the noise. The boxes with $A>A_c$ will ``grow'' the beard.

The higher order physical effects are described by additional/modified terms in the NLS equation. In most cases these modifications lead to the loss of integrability and hence, the IST and  the semi-classical analysis via the Riemann-Hilbert steepest descent approach are no longer available. Still, periodic solutions and the Whitham equations can be derived, so the formal analytical description of dispersive dam breaks flows is possible. It is interesting to see whether the interaction pattern for  counter-propagating dam break flows will persist  despite non-integrability of the problem. 

\begin{figure}[ht]
 \includegraphics[height=2.in]{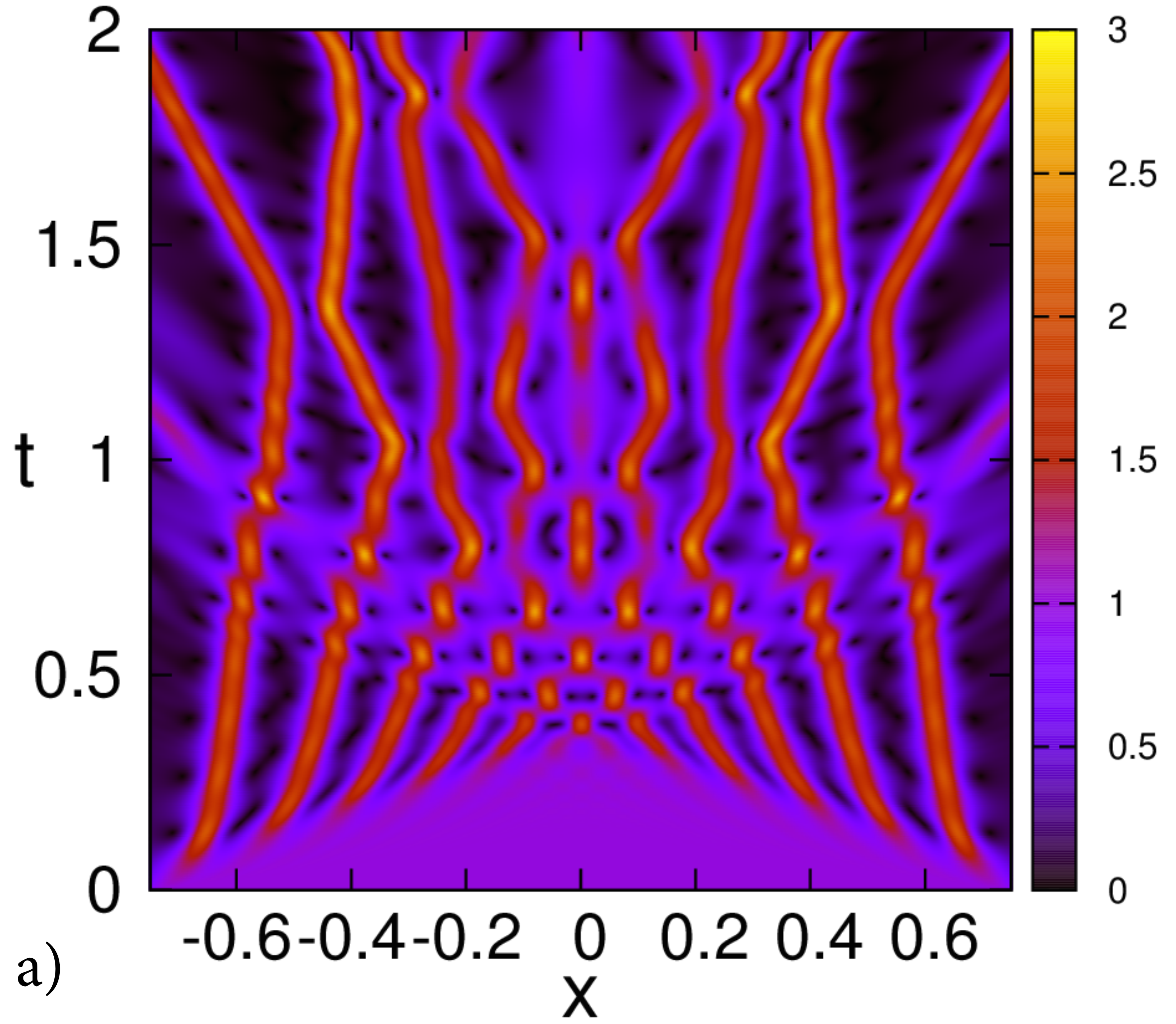} \qquad \includegraphics[height=2.in]{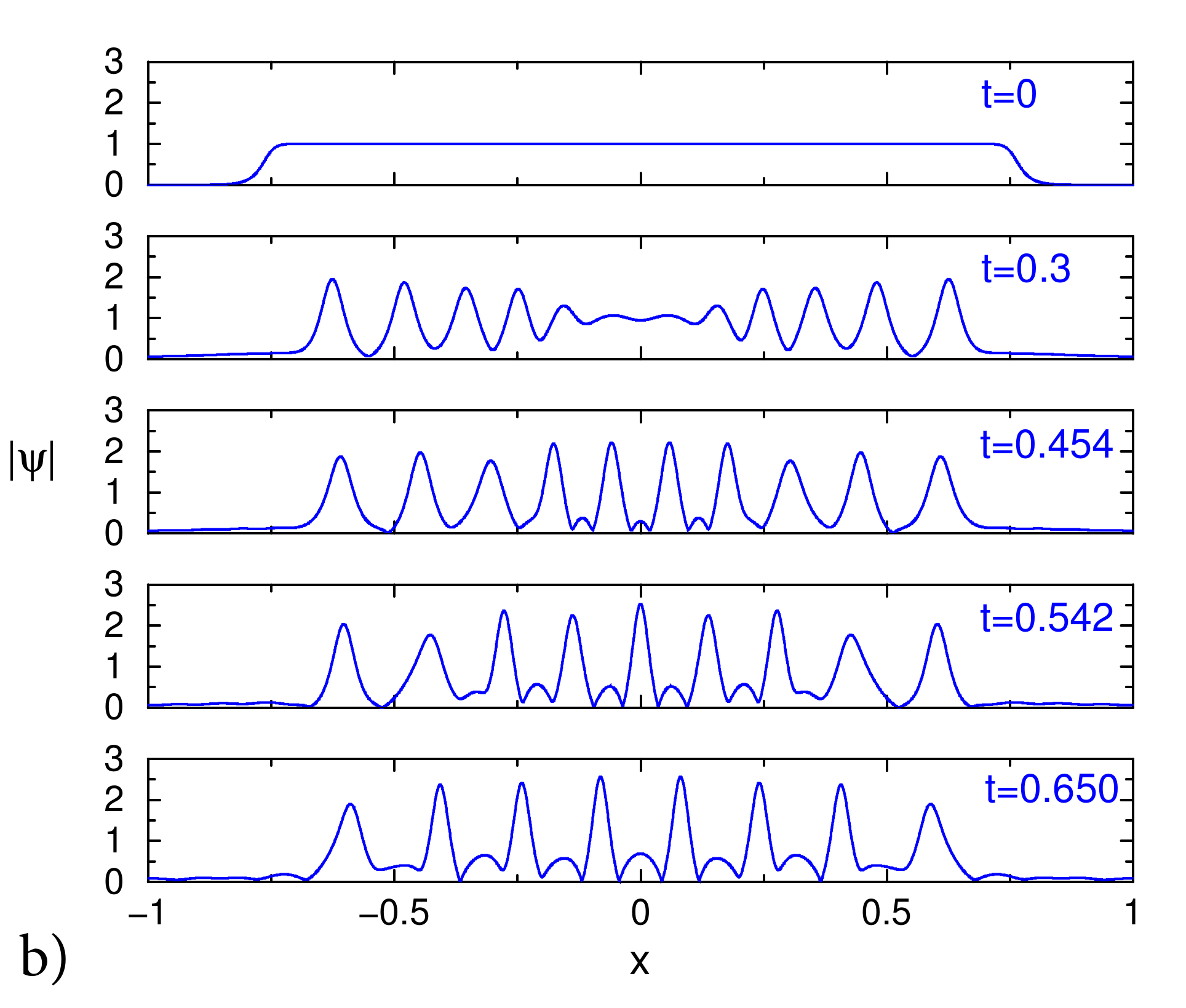} 
 \caption{(Color online) Effect of nonlinear saturation  on the formation of a breather lattice in the small-dispersion box problem (cf. Figs. \ref{fig5}, \ref{fig8}). Numerical solution of the box problem ($q=1$, $L=25/33$) for the saturable NLS equation (\ref{nls_saturable}) with $\gamma=0.1$, $\varepsilon=1/30$. a) Density plot for $|\psi(x,t)|$;  b) Spatial profiles of the amplitude $|\psi(x)|$ at different moments of time.}
 \label{fig15}
 \end{figure}

To be specific, we consider the NLS equation with saturable nonlinearity, which arises in the description of light propagation in some media with highly nonlinear optical properties  (see e.g. \cite{gatz})
\begin{equation}\label{nls_saturable}
i  \eps \psi_{t} +  \frac12 \eps^2 \psi_{xx} + \frac{|\psi|^2}{1+ \gamma |\psi|^2} \psi = 0 \, ,
\end{equation}
where $\gamma$ is the coefficient characterising the strength of the saturation effect. For $\gamma \ll 1$ equation (\ref{nls_saturable}) is a perturbed NLS equation (\ref{nls1}).

In Fig.\ref{fig15} the results of the numerical simulations of the box problem with $q=1$, $L=25/33$ for the saturable NLS (\ref{nls_saturable}) with $\gamma = 0.1$, $\eps=1/30$ are presented. We have chosen a relatively large value of the parameter $\gamma$ to elucidate the qualitative effects of saturable nonlinearity. 
One can draw some immediate conclusions from the simulations shown in Fig.~\ref{fig15}:  (i) the initial evolution for the saturable case is qualitatively similar to that in the pure, cubic NLS case:
one can clearly see the formation of two dispersive dam break flows (cf. Fig.~\ref{fig5});  (ii) quite remarkably, the qualitative agreement with the cubic case stretches beyond the evolution of the modulated periodic solutions. Indeed, one can see that  the interaction of two dispersive dam break flows leads to the formation of the two-phase region dominated by the breather lattices {\it despite non-integrability} of the saturable NLS (\ref{nls_saturable}) (we note that a similar persistence of the two-phase pattern despite non-integrability of the governing equation was observed in the numerical simulations of the DSW interaction for the defocusing saturable NLS  \cite{khamis} and in the analysis of DSWs in viscous fluid conduits \cite{conduit}). One should also note that the amplitudes of the breathers generated in the saturable NLS are noticeably smaller than in the cubic nonlinearity case (cf. Fig. \ref{fig15}b and Fig.\ref{fig8}); (iii) the evolution beyond the two-phase region is qualitatively and quantitatively very different compared to the integrable case.

  \section{Conclusions}
  
In this work, we have proposed a novel mechanism of the rogue wave formation  described in the framework of the focusing nonlinear Schr\"odinger (NLS) equation  with small dispersion parameter.   The key role in our construction is played by the dispersive focusing dam break flow --- a DSW-like nonlinear wave train regularising an initial sharp transition between the uniform plane wave  and the zero-intensity, vacuum state.  We have considered the NLS evolution of a square profile (a ``box'') giving rise to  two such counter-propagating dispersive dam break flows,  whose interaction has been shown to result in the emergence of a modulated two-phase large-amplitude breather lattice closely approximated by a sequence of Akhmediev and Peregrine breathers within certain time-space domain.  We have used a combination of  the nonlinear modulation (Whitham) theory and  elements of the  steepest descent method for the  Riemann-Hilbert problem associated with the semi-classical NLS equation, to construct the exact modulation solution describing the two-phase interaction in the box problem, and predict the parameters of the emerging rogue waves.   Our semi-classical analytical results are shown to be in excellent agreement  with direct numerical simulations of the small-dispersion NLS box problem.  We also show that the proposed  rogue wave generation mechanism  is  different, both physically and mathematically,  from the generation of the Peregrine breathers during the regularisation of the generic gradient catastrophe in the semi-classical NLS equation with analytic, bell-shaped initial data considered in  \cite{BT1, BT2, grim_tovbis, hoefer_tovbis}.

Our numerical simulations of the NLS box problem for longer times suggest that the evolution beyond the two-phase  interaction dynamics leads to the generation of the regions filled with  quasi-periodic waves with the number of interacting nonlinear modes (phases) increasing with time at each point within the spatial location of the initial box potential.  We put forward a hypothesis that for $1 \ll t \ll \eps^{-1} $ the number of oscillating phases (the genus of the solution) grows as $g \sim t$. It is argued that such a complex multi-phase wave structure would require a statistical description similar to that constructed in \cite{ekmv01, el03, ek05} for the KdV soliton gas/soliton turbulence.  Such a description is yet to be developed and could provide an important insight into  properties of the NLS integrable turbulence.

Finally, we numerically considered  effects of small perturbations on the qualitative structure of the small-dispersion NLS box problem solution.  
We first looked at the effect of  noise, inevitably present in any physically (and numerically) realistic setting. This effect  becomes essential for small values of the dispersive parameter $\eps$ and is manifested in  the generation of an additional oscillatory structure -- the `beard' -- in the central part of the box. This structure is not captured by the semi-classical NLS box problem solution. Secondly, we performed numerical simulation of the box problem for the small-dispersion NLS equation with saturable nonlinearity, a perturbed version of the cubic NLS equation.  Although the weakly saturated nonlinearity introduces a non-integrable perturbation, our simulations show that, remarkably,  this does not destroy the qualitative structure of the breather lattice even for relatively large values of the saturation parameter, although the amplitude of the breathers  is noticeably smaller than in the unperturbed, cubic,  case.
 
The proposed mechanism of the rogue wave formation can be realised in fibre optics experiments. The  obtained analytical solutions for the interaction of dispersive dam break flows can also find applications  in oceanography and BEC dynamics (see the physical estimates in \cite{grim_tovbis} and \cite{hoefer_tovbis} showing the relevance  of the small-dispersion focusing NLS to these two areas).

The general conclusion to be drawn from this work is that the semi-classical NLS equation provides a powerful analytical framework for the description of physically important effects related to the rogue wave formation and transition to integrable turbulence.

\section*{Acknowledgments}
This work was initiated owing to partial financial support from the London Mathematical Society (grant number 41435 -- G.A.E.  and A.T.). G.A.E. is grateful to A.M. Kamchanov for numerous stimulating discussions, and also thanks the Department of Mathematics at the University of Central Florida
for hospitality. E.G.K. thanks A. Gammal for useful discussions. E.G.K. is also grateful to CNPq and FAPESP (Brazil) for financial support.

\appendix

\section{ Elements of the Riemann-Hilbert problem analysis for the semiclassical focusing NLS equation with the box-like initial data.}

\subsection{$\ggg$-function}
The 1D  NLS equation   with cubic nonlinearity considered in this paper is an integrable equation \cite{ZS},
which can be solved through the inverse scattering method. It was observed in \cite{Shabat}  that the inverse scattering 
transform (IST), which is in the core of the method, can be written as a (multiplicative) matrix Riemann-Hilbert Problem (RHP).
In the study of the semiclassical (zero dispersion) limit of the NLS we are naturally  interested in the corresponding $\eps \ra 0$
limit of the  IST.  The recently developed nonlinear steepest descent method of Deift and Zhou \cite{DZ1}
is a very effective tool for small or large parameter asymptotics of matrix RHPs. The nonlinear steepest descent asymptotics
of the semiclassical NLS eq. \eqref{nls1} was first obtained in \cite{KMM} (pure soliton case) and \cite{TVZ} (solitons
and radiation). The central object of this analysis is the so-called $\ggg$-function $\ggg(\l)=\ggg(\l ;x,t)$, which is an analytic
(and Schwarz-symmetrical) function on the Riemann surface $\Gamma_g(x,t)$, whose 
branch points $\a_j=\a_j(x,t)$ are determined by the Whitham equations. 
The framework of this paper does not provide enough room to properly describe $\ggg$, we only mention that
$\ggg(\l;x,t)$ satisfies the following jump conditions
\be\label{jumpg}
\ggg_+(\l)+\ggg_-(\l)=\th_k(\l)+\eta_k  , \qquad k=0,1,\dots,g \, ,
\ee
where $\ggg_{\pm}$ are the values of the $\ggg$-function at the opposite  sides of  the oriented branchcut $\g_k$ between 
$\bar \a_k$ and $\a_k$;   $\eta_k$ are some real constants (in $\la$), and $\eta_0=0$. 
Here 
\be\label{f}
\th_k(\l;x,t)=f_k(\l)+2t\l^2+2x\l,
\ee
where the functions $f_k(z)$ contain the information about scattering data for a particular initial condition (potential).
For the box potential \cite{kenjen14} $f_0, f_1$ and $f_2$ are given by (\ref{fk}).
%\be \label{fkapp}
% f_0(\l) = f_1(\l)=-2L\l, \quad f_2(\l) =-2L\l+ 4L \nu(\l) \, .
%\ee

%{\bf The branchcuts $\g_k$  are also known as spectral bands ???}.

In general, $\ggg$ may have additional constant jumps along some gaps  connecting the neighboring bands,
but in the case of the box potential there are no such  jumps. Thus, 
the function $\ggg(\l)$ is analytic everywhere except the branchcuts $\g_k$. In particular, it is analytic at $\l=\infty$.

By the well known Sokhotski-Plemelj formula,
\begin{equation}\label{gform2}
2\ggg(\l)={{\Rscr(\l)}\over{2\pi i}}\left[ 
\sum_{j=1}^g \oint_{\gt_j}{{\eta_jd\z}\over{(\z- \l)\Rscr(\z)}} +\oint_{\cup\gt_j}{{\th(\z)d\z}\over{(\z-\l)\Rscr(\z)}}\right],
\end{equation}
where the (Schwarz-symmetrical) radical $\Rscr(\l) \equiv \Rscr_g(\l)$ is given by \eqref{rsurf}, $\gt_j$ denotes the clockwise oriented loop around $\g_j$
and $\th(\l)$ is defined as $\th_{k}(\l)$ on $\gt_k$.
We assume the loops $\gt_j$ do not intersect each other and $\l$ is outside of any loop. 
Some mathematical details of the RHP derivations can be found, for example, in \cite{TVZ}, \cite{TVZ2008}. Here we 
proceed with the brief presentation of the  results relevant to the analysis of the semi-classical NLS box problem.

The requirement that $\ggg(z)$ is analytic at $z=\infty$ leads to the following $g$ real equations on $\eta_j$:
\begin{equation}\label{momentsNLS}
%\mbox{$k$th  moment condition for $g$}: \ \ \ \ \
\frac{1}{2\pi i}\oint_{\cup \gt_j}{\z^k {\th(\z)}\over{\Rscr(\z)}}d\z
+\frac{1}{2\pi i}\sum_{j=1}^g \oint_{\gt_{j}}{{\eta_j\z^k}\over{\Rscr(\z)}}d\z=0
\ \ \ \ \
k=0,1,\cdots, g-1.
\end{equation}

To shorten the notation, we will proceed by considering the case of genus $g=2$, with the understanding that  all the following 
calculations can be readily generalised  to an arbitrary genus $g\in\N$. Simple linear algebra shows that conditions \eqref{momentsNLS}
combined with \eqref{gform2} yield
\be\label{hn}
2\ggg(\l)=\frac{\Rscr(\l)}{|D|}K(\l),
\ee
where
\be\label{KNLS_box2}
K(\l)= \frac{1}{2\pi i}\times
\left| \begin{matrix}
  \oint_{\gt_{1}}\frac{ d\z}{\Rscr(\z) } &
 \oint_{\gt_{1}}\frac{\z d\z}{\Rscr(\z) } &  \oint_{\gt_{1}}\frac{d\z}{(\z-\l)\Rscr(\z) }\cr
%\cdots &\cdots & \cdots&
%\cdots & \cdots & \cdots  & \cdots\cr
  \oint_{\gt_{2}}\frac{ d\z}{\Rscr(\z) } &
  \oint_{\gt_{2}}\frac{\z d\z}{\Rscr(\z) }
 &\oint_{\gt_{2}}\frac{d\z}{(\z-\l)\Rscr(\z) }\cr
%\oint_{\gt_{m,0}}\frac{ d\z}{R(\z)} & \cdots & \oint_{\gt_{m,0}}\frac{\z^{N-1} d\z}{R(\z)} &\overline{ \oint_{\gt_{m,0}}\frac{ d\z}{R(\z)}}
%& \cdots &\overline{ \oint_{\gt_{m,0}}\frac{\z^{N-1} d\z}{R(\z)}}&\Im  \oint_{\gt_{m,0}}\frac{\z^{N} d\z}{R(\z)}&\oint_{\gt_{m,0}}\frac{d\z}{(\z-z)R(\z)}\cr
%
\oint_{\cup\gt_j}\frac{\th(\z)d\z}{\Rscr(\z) } 
&   \oint_{\cup\gt_j}\frac{\z \th(\z)d\z}{\Rscr(\z) } &
\oint_{\cup\gt_j}\frac{\th(\z)d\z}{(\z-\l)\Rscr(\z) }\cr
\end{matrix}\right|,~~~
%\ee
%\be\label{D-box}
D=\left( \begin{matrix}
  \oint_{\gt_{1}}\frac{ d\z}{\Rscr(\z) } &
 \oint_{\gt_{1}}\frac{\z d\z}{\Rscr(\z) }\cr
%\cdots &\cdots & \cdots&
%\cdots & \cdots & \cdots \cr
  \oint_{\gt_{2}}\frac{ d\z}{\Rscr(\z) } &
 \oint_{\gt_{2}}\frac{\z d\z}{\Rscr(\z) }
\end{matrix}\right).
\ee
The details of the calculations can be found in \cite{el_tovbis_2016}.
\subsection{RHP and modulation solution}
The Whitham modulation theory is concerned with the spatiotemporal evolution of the branch points $\a_j$, $\bar \a_j$ through which the physical parameters of the solution (amplitude, wavenumber, frequency etc) are expressed. The functions $\a_j(x,t)$, $\bar \a_j(x,t)$  are solutions of the Whitham modulation equations  (\ref{whitham-g}) with appropriate initial or boundary conditions derived from a given NLS initial-value problem. The RHP approach yields these dependencies  directly, by-passing the procedure of the integration of the Whitham modulation equations.  We stress that the modulation solution $\a_j(x,t)$, $\bar \a_j(x,t)$ is only part of the full RHP solution, and finding the full solution could be a rather demanding mathematical task.  Thus,  as long as the evolution of the branchpoints is concerned, one can take advantage of the appropriate part of the full RHP analysis for the derivation of the dependencies $\a_j(x,t)$ and then verify the formal result by checking its consistency with  the Whitham modulation equations and the corresponding initial or boundary conditions.  A detailed analysis of interrelations between the Whitham modulation theory for the focusing NLS equation and the RHP approach can be found in \cite{el_tovbis_2016}.

It is shown in \cite{TV2010}  that  the equations for the moving (non-constant) branchpoints $\a_j(x,t)$ follow from the condition
\be\label{mod_K2}
K(\a_j)=0, ~~~{\rm where}~~j=1,2~~~~~~~{\rm and}~~~c.c.
\ee
(In the case of box potential $\a_0=iq$ is a fixed branch point). 
We note that equation \eqref{mod_K2} for each particular $\a_j$ is equivalent to \cite{TV2010} 
\be\label{heqmod=}
\frac{\part}{\part \a_j}\ggg(\l)\equiv 0.~
\ee
Thus, on the solutions of the modulation equations \eqref{mod_K2},
\be\label{geneq0}
\frac{d}{dx}\ggg\equiv  \frac{\partial }{\partial x}\ggg,~~~\frac{d}{dt}\ggg \equiv  \frac{\partial }{\partial t}\ggg.
\ee
 In view of \eqref{f}, \eqref{KNLS_box2} and \eqref{geneq0}, the modulation solution \eqref{mod_K2} can be written in
the form 
\be\label{hodon}
x+\frac{K_t(\a_j)}{K_x(\a_j)}t=-\frac{K_0(\a_j)}{K_x(\a_j)} ~~~{\rm and ~~c.c}~~~~j=1,2,
\ee
where $K_x, K_t$ denote partial derivatives of $K$, and $K_0(z)$ is obtained from $K(z)$ by replacing $x,t$ with zero in 
 \eqref{f}, \eqref{KNLS_box2}.

Comparison of (\ref{hodon}) and the solution of the Whitham equations in the generalised hodograph form (\ref{w0}) suggests the identification
\be \label{ident}
V_j(\a, \bar \a) = \frac{K_t(\a_j)}{K_x(\a_j)} \, , \qquad w_j(\a, \bar \a) = -\frac{K_0(\a_j)}{K_x(\a_j)} \, , \quad j=1,2.
\ee
Direct substitution shows that Tsarev's equations (\ref{tsarev}) are indeed satisfied by (\ref{ident}).  Thus, the branchpoints $\a_j$ defined by \eqref{mod_K2} satisfy the Whitham equations \eqref{whitham-g}. The NLS initial conditions (the box potential (\ref{ic1}) in our case) enter the modulation via the functions $f_k(\la)$ (\ref{fk})  defining the jump functions $\theta_k$ (\ref{f}).  
The modulation solution (\ref{mod_K2}) (or, equivalently, (\ref{hodon})) constructed in such way automatically satisfies all the necessary matching conditions for $\a_j$, $\bar \a_j$ at the second breaking curve.  We shall later verify the matching explicitly for $x=0$.  

\subsection{Phases and characteristic speeds}
Our aim in this subsection is to obtain explicit expression (\ref{Vjg}) and (\ref{Upsilon})  for the characteristic speeds $V_j(\a, \bar \a) $  and phases $\Upsilon_j(\a, \bar \a)$ respectively. For that, we introduce the basis holomorphic differentials
on the Riemann surface $\Gamma$ by
\be \label{holomdiff}
d\Omega_j=\frac{p_j(\l)}{\Rscr(\a, \bar \a, \l)} d\l, \quad   j=1, 2, 
\ee 
where $p_j(z)=\k_{j,1}\l+\k_{j,2}$, and the coefficients $\k_{j,k}$ are found from the normalisation \eqref{holonorm}.
Then \eqref{holonorm} and {\eqref{KNLS_box2} imply
\be\label{coeffs}
\left( \begin{matrix}
  \k_{1,2} &
\k_{2,2}\cr
%\cdots &\cdots & \cdots&
%\cdots & \cdots & \cdots \cr
\k_{1,1} &
\k_{2,1}
\end{matrix}\right)=D^{-1}=\frac{1}{|D|}
\left( \begin{matrix} 
\oint_{\gt_{2}}\frac{\z d\z}{\Rscr(\z) } &
 -\oint_{\gt_{1}}\frac{\z d\z}{\Rscr(\z) }\cr
%\cdots &\cdots & \cdots&
%\cdots & \cdots & \cdots \cr
 - \oint_{\gt_{2}}\frac{ d\z}{\Rscr(\z) } &
  \oint_{\gt_{1}}\frac{ d\z}{\Rscr(\z) } 
\end{matrix}\right).
\ee
Note that the coefficients $\k_{i,j}$ determine the wavenumbers and frequencies of the multi-phase solution, see (\ref{kj}).
Making appropriate  linear combinations of the first two columns in the determinant representation (\ref{KNLS_box2}) for $K(\l)$, we obtain
\be\label{KNLS_box2norm}
K(\lambda)= \frac{|D|}{2\pi i}\times
\left| \begin{matrix}
  1 &
 0 &  \oint_{\gt_{1}}\frac{d\z}{(\z-\l)\Rscr(\z) }\cr
%\cdots &\cdots & \cdots&
%\cdots & \cdots & \cdots  & \cdots\cr
0 &
1
 &\oint_{\gt_{2}}\frac{d\z}{(\z-\l)\Rscr(\z) }\cr
%\oint_{\gt_{m,0}}\frac{ d\z}{R(\z)} & \cdots & \oint_{\gt_{m,0}}\frac{\z^{N-1} d\z}{R(\z)} &\overline{ \oint_{\gt_{m,0}}\frac{ d\z}{R(\z)}}
%& \cdots &\overline{ \oint_{\gt_{m,0}}\frac{\z^{N-1} d\z}{R(\z)}}&\Im  \oint_{\gt_{m,0}}\frac{\z^{N} d\z}{R(\z)}&\oint_{\gt_{m,0}}\frac{d\z}{(\z-z)R(\z)}\cr
%
\oint_{\cup\gt_j}\frac{p_1(\z)\th(\z)d\z}{\Rscr(\z) } 
&   \oint_{\cup\gt_j}\frac{p_2(\z) \th(\z)d\z}{\Rscr(\z) } &
\oint_{\cup\gt_j}\frac{\th(\z)d\z}{(\z-\l)\Rscr(\z) }\cr
\end{matrix}\right|,
\ee
which, together with \eqref{gform2} and \eqref{hn}
 implies that
\be\label{const-main}
\eta_j=-\oint_{\cup\gt_j}\frac{\th(\z)p_j(\z)d\z}{\Rscr(\z)}=4\pi i\left[\b_jt+\k_{j,1}x-\Upsilon_j\ri],~~~~j=1,2,
\ee
where (cf. \eqref{kj})
\be\label{coeff-const}
\b_j= - \frac{1}{4\pi i}\o_j, ~~~~\k_{j,1}=- \frac{1}{4\pi i}k_j,
\ee
and $\Upsilon_j$ is given by (\ref{Upsilon}).
Here the residue theory was used to calculate \eqref{const-main}, \eqref{coeff-const}, (\ref{Upsilon}).  Now, using (\ref{coeff-const}), (\ref{KNLS_box2norm}) it is not difficult to show that the determinant form (\ref{hodon}) of the modulation solution is equivalent to the hodograph solution (\ref{TsaUps})  obtained in terms of the phase $\Upsilon_j$. We recall that for the box potential the  functions $f_k(\l)$ in (\ref{coeff-const})  are given by (\ref{fk}). 

Equations \eqref{KNLS_box2norm}, {\eqref{geneq0} and  \eqref{f} yield convenient formulae
\be\label{K_xt}
\begin{split}
K_x(\l;x, t) =-2|D|\sum_{j=1}^2\k_{j,1}\oint_{\gt_j}\frac{d\z}{(\z-\l)\Rscr(\z)}d\z, \\
K_t(\l;x, t) =-2|D|\sum_{j=1}^2\left(\k_{j,1}\hbox{Re} \left[\sum_{k=1}^N\a_k \right]+\k_{j,2}\ri)\oint_{\gt_j}\frac{d\z}{(\z-\l)\Rscr(\z)}d\z,
\end{split}
\ee
so that  the characteristic speeds (\ref{ident}) are
\be\label{velos}
V_j=\frac{K_t(\a_j)}{K_x(\a_j)}= \hbox{Re} \left[\sum_{k=1}^2\a_k \right] +
\dfrac{\sum_{k=1}^2\k_{k,2}\oint_{\gt_k}\frac{d\z}{(\z-\a_j)\Rscr(\z)}d\z}{\sum_{k=1}^2\k_{k,1}\oint_{\gt_k}\frac{d\z}{(\z-\a_j)\Rscr(\z)}d\z} \, .
\ee
As it was mentioned in Section II, this expression can be readily generalized to the case of arbitrary genus $g$, see \eqref{Vjexp}.

\subsection{Modulation solution for the box problem ($g=2$, $x=0$) }

We now look closer at the modulation solution (\ref{hodon}) for the box potential. Specifically, we  consider behaviour of the solution at $x=0$ in the genus two region. 
First, we recall that, for the box potential (\ref{ic1}) we have $\a_0=iq$ so $\Rscr (\l)=R(\l)\nu(\l)$, where $R(\lambda)$, $\nu(\lambda)$ are given by (\ref{Rnu}).
%\be\label{Rnuapp}
%R(\l)=\sqrt{(\la - \a_1)(\la - \bar \a_1) \dots (\la - \a_{g})(\la - \bar \a_{g})}, \quad  \nu(\l)=\sqrt{\l^2+q^2}.
%\ee

In the case $x=0$ we have symmetry $\a_2=-\bar\a_1$. 
Then $R(\l)=\sqrt{(\l^2-\a^2)(\l^2-\bar\a^2)}$, where $\a=\a_2$.

By taking linear combinations of the first two rows in  
\eqref{KNLS_box2}, we can make $\gt_2\pm \gt_1$ to be the countours of integration for the first and for the second
row of \eqref{KNLS_box2}. 
Then
\be\label{KNLS_box2x=0}
K(\l)= 
4\left| \begin{matrix}
  \int_{[-iq,iq]}\frac{ d\z}{R(\z)\nu_+(\z)} &
 \int_{[-iq,iq]}\frac{\z d\z}{R(\z)\nu_+(\z)} &  \int_{[-iq,iq]}\frac{d\z}{(\z-\l)R(\z)\nu_+(\z)}\cr
%\cdots &\cdots & \cdots&
%\cdots & \cdots & \cdots  & \cdots\cr
  \int_{i\R\setminus[-iq,iq]}\frac{ d\z}{R(\z)\nu(\z)} &
  \int_{i\R\setminus[-iq,iq]}\frac{\z d\z}{R(\z)\nu(\z)}
 &\int_{i\R\setminus[-iq,iq]}\frac{d\z}{(\z-\l)R(\z)\nu(\z)}\cr
%\oint_{\gt_{m,0}}\frac{ d\z}{R(\z)} & \cdots & \oint_{\gt_{m,0}}\frac{\z^{N-1} d\z}{R(\z)} &\overline{ \oint_{\gt_{m,0}}\frac{ d\z}{R(\z)}}
%& \cdots &\overline{ \oint_{\gt_{m,0}}\frac{\z^{N-1} d\z}{R(\z)}}&\Im  \oint_{\gt_{m,0}}\frac{\z^{N} d\z}{R(\z)}&\oint_{\gt_{m,0}}\frac{d\z}{(\z-z)R(\z)}\cr
%
-2t +\frac{4L}{2\pi i}\oint_{\gt_2}\frac{ d\z}{R(\z)}
&   0 &\frac{4L}{2\pi i}\oint_{\gt_2}\frac{ d\z}{(\z-\l)R(\z)}
\cr
\end{matrix}\right|,
\ee
where the contours $\gt_2\pm \gt_1$ were deformed to $[-iq,iq]$ and $i\R\setminus[-iq,iq]$ respectively, and the subscript $`+'$ in $\nu_+$ indicates the limiting value of $\nu(z)$ on the positive side of the branchcut $[-iq, iq]$. 
Note that the function $R(\l)$ is even on $i\R$ whereas $\nu_+(\l)$ is even on $[-iq,iq]$ but odd on $i\R\setminus[-iq,iq]$.
Thus,
\be\label{KNLS_box2x=0_red}
\begin{split}
K(\a)= 
4\left| \begin{matrix}
  \int_{[-iq,iq]}\frac{ d\z}{R(\z)\nu_+(\z)} &
0 &  \int_{[-iq,iq]}\frac{d\z}{(\z-\a)R(\z)\nu_+(\z)}\cr
%\cdots &\cdots & \cdots&
%\cdots & \cdots & \cdots  & \cdots\cr
0&
  \int_{i\R\setminus[-iq,iq]}\frac{\z d\z}{R(\z)\nu(\z)}
 &\int_{i\R\setminus[-iq,iq]}\frac{d\z}{(\z-\a)R(\z)\nu(\z)}\cr
%\oint_{\gt_{m,0}}\frac{ d\z}{R(\z)} & \cdots & \oint_{\gt_{m,0}}\frac{\z^{N-1} d\z}{R(\z)} &\overline{ \oint_{\gt_{m,0}}\frac{ d\z}{R(\z)}}
%& \cdots &\overline{ \oint_{\gt_{m,0}}\frac{\z^{N-1} d\z}{R(\z)}}&\Im  \oint_{\gt_{m,0}}\frac{\z^{N} d\z}{R(\z)}&\oint_{\gt_{m,0}}\frac{d\z}{(\z-z)R(\z)}\cr
%
-2t +\frac{4L}{2\pi i}\oint_{\gt_2}\frac{ d\z}{R(\z)}
&   0 &\frac{4L}{2\pi i}\oint_{\gt_2}\frac{ d\z}{(\z-\a)R(\z)}
\cr \nonumber
\end{matrix}\right| \cr \\
=16\int_{i\R\setminus[-iq,iq]}\frac{\z d\z}{R(\z)\nu(\z)}
\left| \begin{matrix}
  \int_{[-iq,iq]}\frac{ d\z}{R(\z)\nu_+(\z)} &
 \int_{[-iq,iq]}\frac{d\z}{(\z-\a)R(\z)\nu_+(\z)}\cr
%\cdots &\cdots & \cdots&
%\cdots & \cdots & \cdots  & \cdots\cr
%\oint_{\gt_{m,0}}\frac{ d\z}{R(\z)} & \cdots & \oint_{\gt_{m,0}}\frac{\z^{N-1} d\z}{R(\z)} &\overline{ \oint_{\gt_{m,0}}\frac{ d\z}{R(\z)}}
%& \cdots &\overline{ \oint_{\gt_{m,0}}\frac{\z^{N-1} d\z}{R(\z)}}&\Im  \oint_{\gt_{m,0}}\frac{\z^{N} d\z}{R(\z)}&\oint_{\gt_{m,0}}\frac{d\z}{(\z-z)R(\z)}\cr
%
-\frac t2 +\frac{L}{2\pi i}\oint_{\gt_2}\frac{ d\z}{R(\z)}
&   \frac{L}{2\pi i}\oint_{\gt_2}\frac{ d\z}{(\z-\a)R(\z)}
\cr
\end{matrix}\right|=0,
\end{split}
\ee
so the complex modulation equation becomes
\be\label{com_modeq_2_x=0}
\frac{L}{2\pi i}\int_{-iq}^{iq}\frac{ d\z}{R(\z)\nu_+(\z)}\oint_{\gt_2}\frac{ d\z}{(\z-\a)R(\z)}=
\left[-\frac t2 +\frac{L}{2\pi i}\oint_{\gt_2}\frac{ d\z}{R(\z)}\right]\int_{-iq}^{iq}\frac{d\z}{(\z-\a)R(\z)\nu_+(\z)}.
\ee
We denote $\alpha= a+ ib$ and  make the change of variable $\z=i z$ to represent (\ref{com_modeq_2_x=0}) in the form (\ref{abx0}) containing only integrals over intervals of real line.
%\be \label{mod_sol_x=0_real}
%\frac{L}{2\pi}\int \limits_{-q}^{q} \frac{dz}{Q(z)\mu(z)} \int \limits^{\infty}_{-\infty} \frac{(z-b) - ia}{|z+i\a|^2}\frac{dz}{Q(z)} = \int \limits^q_{-q} \frac{(z-b) - ia}{|z+i\a|^2 } \frac{d z}{Q(z) \mu(z)}
%\left( -\frac{t}{2} + \frac{L}{2\pi} \int \limits^{\infty}_{-\infty} \frac{dz}{Q(z)}\right) ,
%\ee
%where
%\be \label{Qmuapp}
%Q(z) = \sqrt{[(z-b)^2 +a^2][(z+b)^2 +a^2]}, \qquad \mu(z) = \sqrt{q^2  - z^2} \, .
%\ee
%It is not difficult to see that at $t=L/2\sqrt{2}q$ (the moment of the collision  of two single-phase dispersive dam break flows at $x=0$,  corresponding to the tip of the $g=1$ region in the diagram in Fig. \ref{fig6}a) one has $a=q/\sqrt{2}$, $b=0$ as required.
%

As a by-product of our calculation we derive explicit expressions at $x=0$ for the coefficients $\k_{i,j}$ of the holomorphic differential (\ref{holomdiff}). These coefficients  determine, via (\ref{kj}),  the local wavenumbers and frequencies of the modulated multi-phase wave. 

For $g=2$ the values $\k_{i,j}$  are defined by the general formula (\ref{coeffs}) following from the normalisation conditions (\ref{holonorm}).  The crucial observation is that the determinant $|D|$ in (\ref{coeffs}) is the $(3,3)$ minor of the main determinant $K(z)$ (see  (\ref{KNLS_box2})). Then, at $x=0$ we take advantage of the representation (\ref{KNLS_box2x=0_red}) for $K(\a)$ to obtain: 
\be\label{Det}
|D| =  \int_{[-iq,iq]}\frac{ d\z}{R(\z)\nu_+(\z)}  \times \int_{i\R\setminus[-iq,iq]}\frac{\z d\z}{R(\z)\nu(\z)} \, .
\ee
Now, using (\ref{coeffs}) and taking into  account the symmetry $\a_2 = -\bar \a_1$ we obtain
 \begin{equation}\label{C12}
x=0: \quad \k_{1,1}(\a, \bar \a)=\left(2\int_{i\R\setminus[-iq,iq]}\frac{\z d\z}{R(\z)\nu(\z)}\right)^{-1}  \, , \quad \k_{1,2}(\a, \bar \a)= \left(2 \int_{[-iq,iq]} \frac{d\z}{R(\z)\nu(\z)}\right)^{-1} ,
\end{equation}
where, we recall, $\a= \a_2$. Introducing the change of variable $\z=i z$ we represent (\ref{C12}) in the form (\ref{kappa12}) containing only  integrals along the real axis,

\section{Numerical method}
Here we only present a brief description of the numerical method used in this paper for solving the small dispersion NLS equation (\ref{nls1}), leaving details to a separate publication. 
We first scale the
time variable in (\ref{nls1}) through $\tau=2t$ and write $\psi=\hat u+i\, \hat v$, where $\hat u$ and $\hat v$ are real-valued functions.
Then, Eq.(\ref{nls1}) can be written as the following system of equations:
\begin{equation} \label{nls2}
\begin{split}
\hat u_{\tau} =  
-\eps \hat v_{xx} - \frac{2}{\eps}\left(\hat u^2+ \hat v^2\right) \hat v, \\
\hat v_{\tau} =  
\eps \hat u_{xx} + \frac{2}{\eps}\left(\hat u^2+\hat v^2\right) \hat u.
\end{split}
\end{equation} 
The time derivatives $\hat u_{\tau}$ and $\hat v_{\tau}$ in (\ref{nls2}) are found by the 4th-order  
Adams-Bashforth-Moulton (ABM) predictor-corrector method \cite{ABM}. The first four time
steps are solved by another method (e.g. 4th-order Runge-Kutta) since the ABM method needs four 
initial values to be started. The spatial derivatives $\hat v_{xx}$ and $\hat u_{xx}$ are calculated
using a pseudo-spectral derivative approximation without any filtering.
Thus, the resulting algorithm is simple, totally explicit and can provide long-time
numerical evaluation without generating numerical artifacts for reasonably small
values of $\eps$, e.g. $\eps=1/33$. The stability region for the ABM method is narrower than that for the traditional Fourier split-step method, widely used for solving
the defocusing NLS equation. However, this latter method can easily yield wrong results when using
small values of $\eps$.

The above algorithm has been tested with both discontinous and  smoothed out initial data.  We have verified that the 
smoothing clearly preserves the structure of the solution  with the advantage of a better control, as   the value of $\varepsilon$ is decreased, over the round-off errors than in the discontinuous data case (the derivatives in the regions containing discontinuity are difficult to approximate numerically). As a result, one can avoid  the (numerically induced) effects of  modulational instability which are unavoidable with  the discontinuous initial data. The only drawback of smoothing the initial data is that the formation of the edge soliton is delayed and, as a result,  it has a different phase compared to the discontinuous  case.

\end{document}